\documentclass[dvips,a4paper,12pt,twoside]{book}
\usepackage[dvips]{graphicx,color}
\usepackage[dvips,pagebackref]{hyperref}
\hypersetup{
pdftitle={Search for CP symmetry violation in the decays of Ks mesons using the KLOE detector},
pdfauthor={Micha{\l} Silarski},
pdfkeywords={neutral kaons, meson, CP symmetry},
unicode=false,
pdfstartview={Fit},
pdffitwindow=false,
colorlinks=true,
linkcolor=black,
citecolor=black,
breaklinks=true,
}
\usepackage[english]{babel}
\usepackage[numbers,sort&compress]{natbib}
\usepackage{hypernat}
\usepackage[T1]{fontenc}
\usepackage[utf8]{inputenc}
\usepackage{times}
\usepackage{geometry}
\usepackage{fancyhdr}
\usepackage{amsfonts}
\usepackage{amsmath}
\usepackage{mathrsfs}
\usepackage{wasysym}
\fancypagestyle{plain}{%
\fancyhead{} 
\fancyhf{}%
}
\begin{document}
\pdfbookmark[-1]{Title pages}{title_en}
\cleardoublepage
\pdfbookmark[0]{Title page}{title_en}
\begin{titlepage}
\begin{center}
\vspace{0.7cm}
\begin{Huge}
JAGIELLONIAN UNIVERSITY\\
INSTITUTE OF PHYSICS\\
\end{Huge}
\begin{figure}[!h]
\begin{center}
\includegraphics[width=0.17\textwidth]{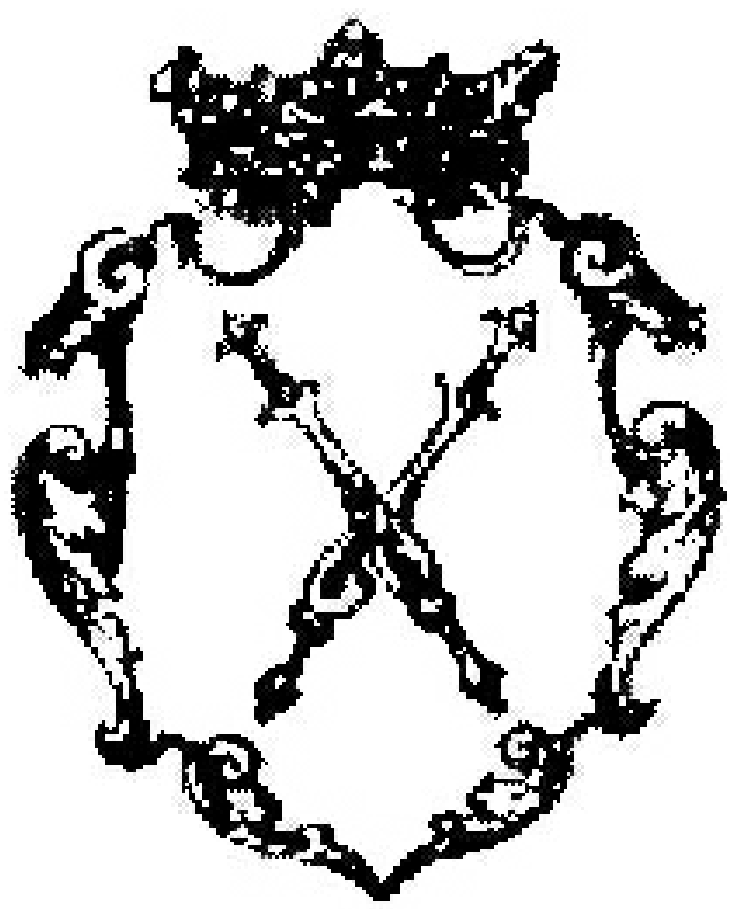}
\end{center}
\end{figure}
\vspace{1cm}
\begin{LARGE}
Search for the $\mathcal{CP}$ symmetry violation in the decays of $K_S$ mesons using the KLOE detector\\
\end{LARGE}
\vspace{1.5cm}
\begin{Large}
Micha{\l} Silarski
\end{Large}
\vspace{6.0cm}
\begin{Large}
\\
PhD thesis prepared in the Department of Nuclear Physics of the Jagiellonian University
under supervision\\ of Prof. Pawe{\l} Moskal\\
\end{Large}
\begin{large}
\vspace{1.5cm}
Cracow 2012
\end{large}
\end{center}
\end{titlepage}

\thispagestyle{plain}
\cleardoublepage
\pdfbookmark[0]{Strona tytu{\l}owa}{title_pl}
\begin{titlepage}
\begin{center}
\vspace{0.7cm}
\begin{Huge}
UNIWERSYTET JAGIELLO{\'N}SKI\\
INSTYTUT FIZYKI\\
\end{Huge}
\begin{figure}[!h]
\begin{center}
\includegraphics[width=0.17\textwidth]{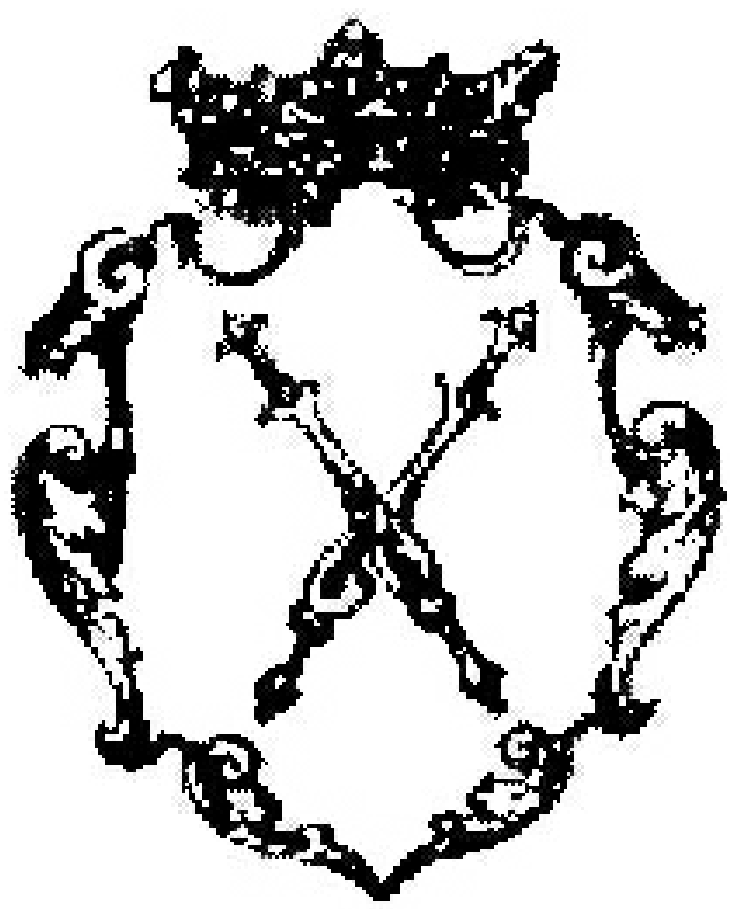}
\end{center}
\end{figure}
\vspace{1cm}
\begin{LARGE}
Poszukiwania {\l}amania symetrii CP w rozpadach mezonu $K_S$ za pomoc\k{a} detektora KLOE\\
\end{LARGE}
\vspace{1.5cm}
\begin{Large}
Micha{\l} Silarski
\end{Large}
\vspace{6.0cm}
\begin{Large}
\\Praca doktorska wykonana w Zak{\l}adzie Fizyki J\k{a}drowej
Uniwersytetu Jagiello{\'n}skiego pod kierunkiem\\
Prof. dr hab. Paw{\l}a Moskala\\
\end{Large}
\vspace{1.5cm}
\begin{large}
Krak{\'o}w 2012
\end{large}
\end{center}
\end{titlepage}

\thispagestyle{plain}
\cleardoublepage
\thispagestyle{plain}
\hspace{1cm}\\
\vfill
\begin{flushright}
,, \textit{Nur ein Leben f\"ur die anderen ist lebenswert} ''\\
Albert Einstein
\end{flushright}

\thispagestyle{plain}
\cleardoublepage
\pdfbookmark[-1]{Abstracts}{abstract_en}
\pdfbookmark[0]{Abstract}{abstract_en}
\begin{titlepage}
\begin{center}
\bf{Abstract}
\end{center}
The aim of this work was to determine the $K_S \to 3\pi^0$ decay branching ratio
and a modulus of the $\eta_{000}$ parameter, defined as the ratio of amplitudes
for $K_S \to 3\pi^0$ to $K_L \to 3\pi^0$ decays, which characterizes the $\mathcal{CP}$ symmetry
violation in this decay.\\
The measurement has been carried out with the KLOE detector operating at the $\phi$~--~factory
DA$\Phi$NE in the Italian National Center for Nuclear Physics in Frascati.
DA$\Phi$NE
collides the $e^+$ and $e^-$ beams at the center of mass energy $\sqrt{s} = 1019.45$ MeV.
The $e^+e^-$ collisions result in the $\phi$ meson creation which is almost at rest and
decay predominantly to kaon pairs.
The decay products are registered using the KLOE detection setup, which consists of large
cylindrical drift chamber surrounded by the electromagnetic calorimeter.
The detectors are placed in a magnetic field of $B \approx 0.52$ T generated by
superconducting solenoid. The $K_S$ mesons were identified with high efficiency
via registration of these $K_L$ mesons which crossed the drift chamber without decaying
and then interacted with the KLOE electromagnetic calorimeter. The $K_S$
four -- momentum vector was then determined using the registered position of the $K_L$ meson
and the known momentum of the $\phi$ meson. Next, the search for the $K_S \to 3\pi^0 \to 6\gamma$ decay
was carried out by the selection of events with six gamma quanta which momenta were
reconstructed using time and energy measured by the electromagnetic calorimeter.
To increase the signal over background ratio after identification of the $K_S$ meson and requiring six
reconstructed photons a discriminant analysis is performed. It is based on kinematical fit, testing of
the signal and background hypotheses and exploiting of the differences in kinematics of the $K_S$ decays
into 2$\pi^0$ and 3$\pi^0$.\\
The search for the $K_S \to 3\pi^0$ decay presented in this work failed to detect a signal of
sufficient statistical significance. Hence, we have obtained the upper limit on the
$K_S \to 3\pi^0$ branching ratio at the 90$\%$ confidence level:
\begin{equation}
\nonumber
BR(K_S \to 3\pi^0) \leq 2.7\cdot 10^{-8}~,
\end{equation}
which is almost five times lower than the latest published result.
This upper limit can be translated into a limit on the modulus of the $\eta_{000}$
parameter amounting to: \mbox{$|\eta_{000}| \leq 0.009$} at the 90$\%$
confidence level. This corresponds to an improvement of the $|\eta_{000}|$ uncertainty
by a factor of two with respect to the latest direct measurement.\\
The upper limit on the $K_S \to 3\pi^0$ branching ratio determined in this work is still
about one order of magnitude larger than the prediction based on the Standard Model. Hence, the
search for this decay will be continued with the upgraded KLOE detector, which has realistic
chances to observe the $K_S \to 3\pi^0$ decay for the first time in the near future.
\end{titlepage}

\thispagestyle{plain}
\cleardoublepage
\pdfbookmark[0]{Streszczenie}{abstract_pl}
\begin{titlepage}
\begin{center}
\textbf{Streszczenie}
\end{center}
Celem tej pracy by{\l}o wyznaczenie stosunku rozga{\l}\k{e}zie{\'n} dla rozpadu
$K_S \to 3\pi^0$, oraz modu{\l}u parametru $\eta_{000}$, zdefiniowanego jako stosunek
amplitud na rozpad $K_S \to 3\pi^0$ i $K_L \to 3\pi^0$, kt{\'o}ry charakteryzuje
niezachowanie symetrii $\mathcal{CP}$ w tym  procesie.\\
Pomiary wspomnianego rozpadu by{\l}y prowadzone za pomoc\k{a} detektora KLOE
dzia{\l}aj\k{a}cego na akceleratorze wi\k{a}zek przeciwbie{\.z}nych DA$\Phi$NE we W{\l}oskim
Narodowym Centrum Fizyki J\k{a}drowej we Frascati.
DA$\Phi$NE zderza wi\k{a}zki $e^+$ $e^-$ przy energii w centrum masy $\sqrt{s} = 1019.45$~MeV
r{\'o}wnej masie mezonu $\phi$. W wyniku zderze{\'n} $e^+e^-$ powstaj\k{a} mezony $\phi$.
Produkowane s\k{a} one  praktycznie w spoczynku i rozpadaj\k{a} si\k{e} g{\l}{\'o}wnie na pary
kaon{\'o}w. Do pomiaru powstaj\k{a}cych cz\k{a}stek wykorzystywany jest uk{\l}ad detekcyjny KLOE.
Zbudowany jest on z cylindrycznej komory dryfowej otoczo\-nej kalorymetrem elektromagnetycznym.
Ca{\l}o{\'s}{\'c} umieszczona jest w polu magnetycznym nadprzewodz\k{a}cego solenoidu o 
indukcji $B \approx 0.52$~T.
Mezony $K_S$ iden\-ty\-fi\-ko\-wa\-ne by{\l}y z du{\.z}\k{a} wydajno{\'s}ci\k{a} poprzez rejestracj\k{e} tych
mezon{\'o}w $K_L$, kt{\'o}re nie rozpad{\l}y si\k{e} w komorze dryfowej i zareagowa{\l}y z materia{\l}em
kalorymetru elektromagnetycznego. Wektor czterop\k{e}du mezonu $K_S$
okre{\'s}lany by{\l} na podstawie zarejestrowanej w kalorymetrze pozycji mezonu $K_L$ i znanego
wektora czterop\k{e}du mezonu $\phi$. Poszukiwania rozpad{\'o}w $K_S \to 3\pi^0 \to 6\gamma$ 
prowadzone by{\l}y nast\k{e}pnie poprzez wy\-bra\-nie zdarze{\'n} z sze{\'s}cioma zrekonstruowanymi
kwantami gamma. Ich p\k{e}dy okre{\'s}lane by{\l}y na podstawie czasu i energii mierzonych za pomoc\k{a}
kalorymetru.
Aby zwi\k{e}kszy{\'c} stosunek sygna{\l}u do t{\l}a zdarzenia ze zidentyfikowanym mezonem $K_S$
i sze{\'s}cioma zrekonstruowanymi kwantami gamma zosta{\l}y poddane dalszej analizie, opartej na dopasowaniu
kinematycznym, testowaniu hipotez t{\l}a i sygna{\l}u oraz wykorzystaniu r{\'o}{\.z}nic w kinematyce
rozpad{\'o}w mezonu $K_S$ na dwa i trzy mezony $\pi^0$.\\
W wyniku przeprowadzonych poszukiwa{\'n} nie zarejestrowano {\.z}adnego zdarzenia
od\-po\-wia\-da\-j\k{a}\-ce\-go
rozpadowi $K_S \to 3\pi^0$. Dlatego okre{\'s}lono g{\'o}rn\k{a} granic\k{e}
stosunku rozga{\l}\k{e}zie{\'n} dla rozpadu $K_S \to 3\pi^0$ na poziomie ufno{\'s}ci 90$\%$:
\begin{equation}
\nonumber
BR(K_S \to 3\pi^0) \leq 2.7\cdot 10^{-8}~.
\end{equation}
Otrzymana warto{\'s}{\'c} jest oko{\l}o pi\k{e}{\'c} razy ni{\.z}sza od ostatniej opublikowanej
g{\'o}rnej gra\-ni\-cy na ten stosunek rozga{\l}\k{e}zie{\'n}.
Otrzymana na poziomie ufno{\'s}ci 90$\%$ warto{\'s}{\'c} g{\'o}rnej granicy dla modu{\l}u parametru
$\eta_{000}$ wynosi: \mbox{$|\eta_{000}| \leq 0.009$}, co odpowiada zmniejszeniu jego niepewno{\'s}ci
dwa razy w stosunku do ostatniego bezpo{\'s}redniego pomiaru.\\
Otrzymana warto{\'s}{\'c} g{\'o}rnej granicy na $BR(K_S \to 3\pi^0)$ jest wi\k{e}ksza od teoretycznych
przewidywa{\'n} o rz\k{a}d wielko{\'s}ci, dlatego poszukiwania tego rozpadu b\k{e}d\k{a} kontynuowane za pomoc\k{a}
systemu detekcyjnego KLOE wyposa{\.z}onego w nowe detektory.
\end{titlepage} 
\thispagestyle{plain}
\frontmatter
\cleardoublepage
\pdfbookmark[-1]{Contents}{tableofcontents}
\tableofcontents
\mainmatter
\chapter{Introduction}
In 1918 Amalie Emmy Noether proved a theorem connecting the symmetries of the physical
systems and conservation laws~\cite{noether} which became one of the greatest achievements
of the twentieth century theoretical physics.
It shows for instance that a system invariant under translations of time, space, or rotation
will obey the laws of conservation of energy, linear momentum, or angular momentum, respectively.
From that time on the symmetries have become an essential part of almost all physics theories
and models, especially in the particle physics. And so for example every quantum field theory
describing the interaction and properties of elementary particles are formulated requiring
the Lorentz invariance. Furthermore, the discrete symmetries of Parity $\mathcal{P}$, Charge Conjugation
$\mathcal{C}$ and Time Reversal $\mathcal{T}$ proved to be very useful in the calculation of
the cross sections and decay rates, especially for the processes governed by the strong interaction.
These symmetries became also an important tool in the Standard Model formulation.\\

Among the known elementary forces the weak interaction has appeared to be very peculiar mainly
because it violates $\mathcal{P}$ and $\mathcal{C}$ symmetries~\cite{Wu,Garwin} as well as their
combination: $\mathcal{CP}$.
The $\mathcal{CP}$ violation was discovered unexpectedly in 1964 by Christenson, Cronin, Fitch and Turlay
during the regeneration studies of the neutral $K$ mesons~\cite{Christensen}. In the framework of Standard Model
the $\mathcal{CP}$ violation mechanism is introduced by the quark mixing described by the complex Cabibbo -- Kobayashi -- Maskawa
matrix
with one nonzero phase~\cite{cabibo,ckm}, which requires the existence of three generation of quarks.
Parameters describing the $\mathcal{CP}$ violation in the neutral kaon system were measured with a good precision
by several experiments~\cite{na48,ktev,passeri}, and at present the main experimental effort is focused on studies
of the neutral $B$ and $D$ meson systems~\cite{babar,belle,Wilson,lhcb}. However, there are still
several interesting open issues in the kaon physics. One of them is the $K_S \to 3\pi^0$ decay which,
assuming the $\mathcal{CPT}$ invariance, allows one to investigate the direct $\mathcal{CP}$
symmetry violation~\cite{sozzi}.
Despite several direct searches~\cite{snd,Matteo} and $K_S K_L$ interference stu\-dies~\cite{cplear,na48b},
this decay remains undiscovered and the best upper limit on
the bran\-ching ratio $BR(K_S \to 3\pi^0)<1.2\cdot 10^{-7}$~\cite{Matteo,pdg2010} is still two orders of magnitude larger
than the predictions based on the Standard Model~\cite{sozzi}.\\

This work is focused on the measurement of the $K_S \to 3\pi^0$ decay branching ratio based on
the data sample gathered in 2004 -- 2005 with the KLOE detector operating at the $\phi$ -- factory
DA$\Phi$NE in the Italian National Center for Nuclear Physics in Frascati.
DA$\Phi$NE collides the $e^+$ and $e^-$ beams at the center of mass energy of $\sqrt{s}$ = 1019.45 MeV near
the $\phi$ meson mass~\cite{kloe2008}. The $e^+e^-$ collisions result in $\phi$ meson creation which is almost at rest
($\beta_{\phi} \approx 0.015$) and decay predominantly to $K^+K^-$ (49$\%$), $K_SK_L$ (34$\%$), $\rho\pi$ (15$\%$) and $\eta\gamma$ (1.3$\%$)
final states~\cite{pdg2010}. The decay products are registered using the KLOE detection setup, which consists
of large cylindrical drift chamber surrounded by the electromagnetic calorimeter.
The detectors are placed in a magnetic field of $B \approx 0.52$ T generated by superconducting solenoid.
Since the $\phi$ mesons are produced almost at rest, kaons arising from the decay move with the relative
angle close to 180$^0$, and as a consequence, their decay products are registered in
the well separated parts of the detector.
The $K_S$ mesons are identified with high efficiency ($\sim$~34$\%$) via registration of these $K_L$ mesons
which cross the drift chamber
without decaying and then interact with the KLOE electromagnetic calorimeter (so called $K_S$ tag).
The $K_S$ four -- momentum vector is then determined using the registered position of the $K_L$ meson
and the known momentum of the $\phi$ meson, which is estimated as an average of the momentum distribution
measured using large angle $e^+e^-$ scattering.
The search for the $K_S\to 3\pi^0\to 6\gamma$ decay is then carried out by the selection of events
with six $\gamma$ quanta which momenta are reconstructed using time and energy measured by the electromagnetic
calorimeter.
Background for the searched decay originates mainly from the $K_S \to 2\pi^0$ events with two spurious
clusters from fragmentation of the electromagnetic showers (so called splitting) or accidental activity,
or from false $K_L$ identification~\cite{HyperF}.
To increase the signal over background ratio after identification of the $K_S$ meson and requiring six
reconstructed photons a discriminant analysis is performed. It is based on kinematical fit, testing of
the signal and background hypotheses and exploiting of the differences in kinematics of the $K_S$ decays
into 2$\pi^0$ and 3$\pi^0$.\\

This thesis is divided into nine chapters. The detailed description of the $\mathcal{CP}$ violation
mechanism in the neutral kaon system is presented in the second chapter together with the motivation
to search for the $K_S \to 3\pi^0$ decay.\\
The third chapter provides the description of experimental tools used for the measurement:
the DA$\Phi$NE collider, the KLOE detector as well as the trigger and data acquisition system.\\
The method used to identify the $K_S$ mesons based on the detection of the $K_L$ interactions in the
electromagnetic calorimeter is presented in chapter four.\\
Chapter five is devoted to the identification of the $K_S \to 2\pi^0$ events used for the normalization
of the measured branching ratio.\\
The discriminant analysis used to increase the signal over background ratio is described in chapter six,
where also the background estimation based on the Monte Carlo simulations and the
final result of the $K_S \to 3\pi^0$ identification are presented.\\
The seventh chapter is devoted to the estimation of the systematic uncertainties of the measurement.\\
The determined upper limit on the $K_S \to 3\pi^0$ branching ratio is given in chapter eight together
with the estimation of the modulus of the $\eta_{000}$ parameter, defined as the ratio of amplitudes
for $K_S \to 3\pi^0$ to $K_L \to 3\pi^0$ decays.\\
Finally, the ninth chapter comprises the summary and perspectives. In particular we discuss
the possibility of the first observation of the $K_S \to 3\pi^0$ decay in the next KLOE--2 data~--~taking campaign
during 2013~--~15. It will be conducted with the KLOE detector upgraded by Inner Tracker and with improved photon
acceptance brought about by new calorimeters installed in the final focusing region.
\chapter{Neutral Kaon system and $\mathcal{CP}$ violation}
Discrete symmetries as parity $\mathcal{P}$, charge conjugation $\mathcal{C}$
and time reversal $\mathcal{T}$, as well as their combinations such as $\mathcal{CP}$
and $\mathcal{CPT}$ play a fundamental role in particle physics.
The parity transformation $\mathcal{P}$ changes the signs of the three space
coordinates, while $\mathcal{C}$ changes particle
to its antiparticle and vice versa changing its internal quantum numbers.
The strong and electromagnetic interaction preserve eigenvalues of both,
parity and charge conjugation operators, as well as eigenvalues of the $\mathcal{CP}$ operator.
The weak interaction instead do not preserve these quantum numbers which result
in a far -- reaching consequences, especially in case of the $\mathcal{CP}$ operator.
In the framework of the Standard Model the $\mathcal{CP}$ violation implies the existence of the third
generation of quarks. Moreover, it is a very important mechanism which could have an essential
contribution to the asymmetry between matter and antimatter in the Universe.
In 1967 A. Sakharov laid out three conditions that would enable a Universe containing initially
equal amounts of matter and antimatter to evolve into a matter dominated universe,
which we see today~\cite{Sakharov}. The first condition was a violation of the baryon number
conservation $B$, for which there is still no experimental evidence. However, simple baryon number
violation would not be enough to explain matter -- antimatter asymmetry if $\mathcal{C}$ and
$\mathcal{CP}$ were exact symmetries\footnote{In that case there would be a perfect equality
between rates of different $B$ violating processes and no asymmetry could be generated from the initially
symmetric state~\cite{Farrar}.}. As the third condition Sakharov proposed that the Universe
should undergo a phase of extremely rapid expansion~\cite{Sakharov}.\\
Since the first discovery of the $\mathcal{CP}$ -- violating neutral kaon decay in 1964, there
have been made a big effort to describe the $\mathcal{CP}$ symmetry breaking within the Standard Model.
The favoured theoretical framework was provided in 1973 by Kobayashi and Maskawa, who pointed out that
$\mathcal{CP}$ violation would follow automatically if there were at least six quark flavours.
At present the main experimental effort is focused on the neutral $B$ and $D$ meson system
studies~\cite{babar,belle,Wilson,lhcb}. However, there are still several interesting open
issuses in the kaon physics which, as it will be shown in this chapter, can contribute to our
better understanding of the $\mathcal{CP}$ violation mechanism.
\section{The neutral kaon system}
\begin{figure}
\centering
\includegraphics[width=0.5\textwidth]{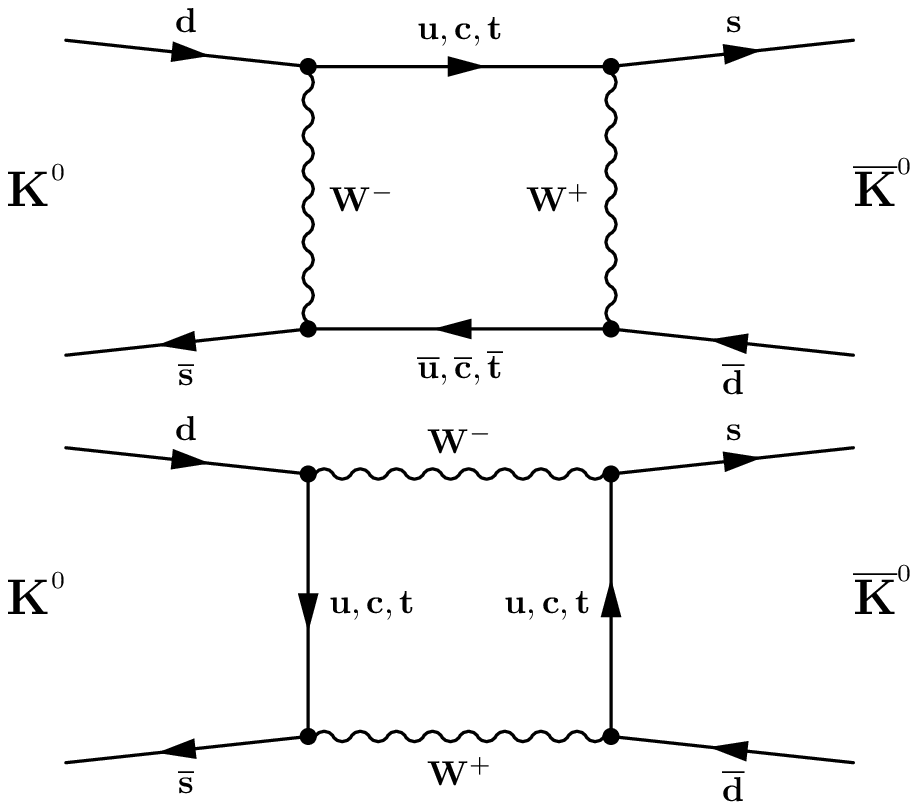}
\caption{Quark diagrams for the $K^0$ -- $\overline{K}^{0}$ transitions.}
\label{fig:boxmix}
\end{figure}

Kaons were discovered in 1947 by G. D. Rochester and Clifford C. Butler
while studying cosmic ray showers with a cloud chamber~\cite{Rochester}.
The contrast between the production and decay times  of these new particles entailed in 1953
introduction of a new quantum number called ,,strangeness'' $S$~\cite{GellMann,Nakano}.
Strangeness is conserved by both electromagnetic and strong processes while first order weak
interaction can induce transitions  with $\Delta S = 1$.\\
Kaons appear in isospin $I = \frac{1}{2}$ doublets: $(K^{+},K^{0})$ with $S = 1$ and
$(\overline{K}^{0},K^{-})$ with $S = -1$. They can be produced via strong interactions
in processes e.g. like:
\begin{description}
\centering
\item[$\pi^+ p \to K^{+} \overline{K}^{0} p$]
\item[$\pi^- p \to K^{0} \Lambda$]
\item[$p\overline{p} \to K^- \pi^+ K^0$]
\item[$p\overline{p} \to K^+ \pi^- \overline{K}^0~.$]
\end{description}
	From the point of view of strong interactions the $K^0$ meson is a particle with a corresponding
	antiparticle $\overline{K}^{0}$. Violation of strangeness conservation by weak interaction allows for
	transitions like $K^0 \to 2\pi \to \overline{K}^{0}$ or $K^0 \to 3\pi \to \overline{K}^{0}$. Thus,
	the two strangeness eigenstates can oscillate one into another via the $\Delta S = 2$, second order
	weak interactions, i.e., via virtual 2-pion and 3-pion states. The corresponding quark diagrams for
these transitions are presented in Fig.~\ref{fig:boxmix}.
Therefore, in the evolution of a kaon in a free space states with well defined mass and width are
mixtures of $K^0$ and $\overline{K}^{0}$~\cite{Winstein}.
The time evolution of the neutral kaon system,
which may be given in the $K^0$ -- $\overline{K}^{0}$ rest frame is determined by Hamiltonian $\mathbf{H}$
and the following equation:
\begin{equation}
  {i \frac{\partial}{\partial t}\left( \begin{array}{c}
    K^0\\
    \overline{K}^{0}
  \end{array} \right)}
  = \mathbf{H}
  \left( \begin{array}{c}
    K^0\\
    \overline{K}^{0}
  \end{array} \right)
= \left( \mathbf{M} -i\frac{\mathbf{\Gamma}}{2} \right)
\left( \begin{array}{c}
    K^0\\
    \overline{K}^{0}
  \end{array} \right)~,
\label{eq:kkevol1}
\end{equation}
where $\mathbf{M}$ and $\mathbf{\Gamma}$ are 2 x 2 hermitian mass and decay matrices, respectively.
In the Weisskopf -- Wigner approximation the elements of the mass matrix $\mathbf{M}$ can be
expressed as a sum of contributions due to strong and weak interactions~\cite{Winstein}:
\begin{equation}
 M_{ij} = m_{k}\delta_{ij} +\langle i|\mathbf{H_W} | j \rangle + \sum_{n \neq K^0,\overline{K}^0}
 \frac{\langle i| \mathbf{H_W} | n \rangle\langle n | \mathbf{H_W} | j \rangle}{m_{K} - E_n}~,
\end{equation}
where $m_K$ is the neutral kaon mass and $\mathbf{H_W}$ denotes the effective weak Hamiltonian.
The sum $\sum_{n} |n\rangle\langle n|$ runs over all virtual and real states connecting $K^0$ and $\overline{K^0}$.  
The decay matrix elements $\Gamma_{ij}$ related to the kaon decay width by unitarity, originate instead
only from  $\mathbf{H_W}$~\cite{Maiani}:
\begin{equation}
\Gamma_{ij} = 2\pi \sum_{n \neq K^0,\overline{K}^0} \delta(E_{n} - m_{K})\langle i|
\mathbf{H_W} | n\rangle\langle n|\mathbf{H_W}| j \rangle~.
\label{eq:kkevol2}
\end{equation}
If the Hamiltonian of the system is invariant under $\mathcal{T}$, $\mathcal{CP}$ and $\mathcal{CPT}$
$M_{ij}$ and $\Gamma_{ij}$ have to satisfy the following relations:
\begin{description}
	\item[$|M_{12} -i\Gamma_{12}/2| = |M^{*}_{12} -i\Gamma^{*}_{12}/2|$ $\mathrm{(\mathcal{T}~conservation),}$]
	\item[$|M_{12} -i\Gamma_{12}/2| = |M^{*}_{12} -i\Gamma^{*}_{12}/2|$ $\mathrm{and}$ $M_{11} = M_{22},$
	$\Gamma_{11} = \Gamma_{22}$ $\mathrm{(\mathcal{CP}~invariance)}$]
	 \item[$M_{11} = M_{22},$ $\Gamma_{11} = \Gamma_{22}$ $\mathrm{( \mathcal{CPT}~conservation).}$]
\end{description}
Without any assumption about symmetry invariance the Hamiltonian eigenstates of the neutral kaon system seen in nature
can be written in the following form~\cite{Aslanides}:
\begin{align}
\nonumber
|K_{S}\rangle	& = \frac{1}{\sqrt{2(1+|\epsilon_S|)^2}} \left[(1+\epsilon_S)|K^0\rangle + (1-\epsilon_{S})|\overline{K}^0\rangle\right]\\
|K_{L}\rangle	& = \frac{1}{\sqrt{2(1+|\epsilon_L|)^2}} \left[(1+\epsilon_L)|K^0\rangle - (1-\epsilon_{L})|\overline{K}^0\rangle\right],
\label{eq:kkk}
\end{align}
where $\epsilon_S$ and $\epsilon_L$ are complex parameters expressing possible $\mathcal{CP}$ and $\mathcal{CPT}$ violation.
In particular, if $\mathcal{CPT}$ invariance holds: $\epsilon_S = \epsilon_L = \epsilon$.
It is important to stress, that $K_S$ and $K_L$ are kaon states which preserve their identity during  the evolution in free space.\\
The experimental values for the lifetimes of these two particles differ by three orders of magnitude.
The lifetime of the ,,short'' state $|K_S\rangle$ amounts to\\
$\tau_S = (8.953 \pm 0.005)\cdot 10^{-11}$~s, while the ,,long'' living
particle $|K_L\rangle$ has a lifetime $\tau_L = (5.116 \pm 0.020)\cdot 10^{-8}$~s~\cite{pdg2010}.
This large difference was explained by assuming $\mathcal{CP}$ to be an exact symmetry of the weak
interactions. In this case the mass eigenstates defined in Eq.~\ref{eq:kkk} reduce to the
$\mathcal{CP}$ eigenstates\footnote{
Here we assume a phase convention where $\mathcal{CP}|K^{0}\rangle = |\overline{K^{0}}\rangle$ and 
$\mathcal{CP}|\overline{K^{0}}\rangle = |K^{0}\rangle$~\cite{Aslanides}.}
($\epsilon_L = \epsilon_S = 0$):
\begin{align}
\nonumber
|K_{1}\rangle	& = \frac{1}{\sqrt{2}} \left[|K^0\rangle + |\overline{K}^0\rangle\right] \mathrm{with~\mathcal{CP} =1}\\
|K_{2}\rangle	& = \frac{1}{\sqrt{2}} \left[|K^0\rangle - |\overline{K}^0\rangle\right] \mathrm{with~\mathcal{CP} =-1.}
\label{eq:kkk1}
\end{align}
Neutral kaons decay mainly to the two -- and three -- pion final states with a well defined $\mathcal{CP}$ eigenvalues
~\cite{Perkins}:
\begin{description}
\centering
\item[$\mathcal{CP} |\pi^+ \pi^-\rangle = |\pi^+ \pi^-\rangle$]
\item[$\mathcal{CP} |\pi^0 \pi^0\rangle = |\pi^0 \pi^0\rangle$]
\item[$\mathcal{CP} |\pi^+ \pi^- \pi^0\rangle = (-1)^{l+1}|\pi^+ \pi^- \pi^0\rangle$]
\item[$\mathcal{CP} |\pi^0 \pi^0 \pi^0\rangle = -|\pi^0 \pi^0 \pi^0\rangle ~.$]
\end{description}
For the $|\pi^+ \pi^- \pi^0\rangle$ final state the eigenvalue depends on the total angular momentum $l$. However, since the three
pions from the kaon decay are mainly in the relative $s$ -- wave state we can assume with a good approximation that
the $(\pi^+, \pi^-, \pi^0)$ system is $\mathcal{CP}$ -- odd.
Thus, $\mathcal{CP}$ conservation would imply that $|K_1\rangle$ state is allowed to decay only to two pions while the ,,long'' living
$|K_2\rangle$ decays only to three pions state. Moreover, the large phase space difference between these two decay modes manifests
itself in the difference between observed lifetimes.\\
However, as it is presented in the next section, the $\mathcal{CP}$ invariance is violated by the weak interaction which
entails big consequences for the whole particle physics and cosmology.
\section{$\mathcal{CP}$ violation in kaon decays}
\begin{figure}
\centering
\includegraphics[width=0.3\textwidth]{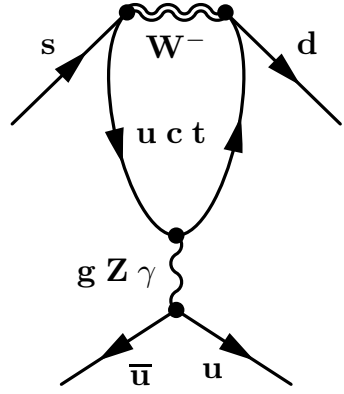}
\caption{,,Pengiun'' diagram for the $s$ quark decay.}
\label{fig:pingwin}
\end{figure}

In 1964 an experiment by Christenson, Cronin, Fitch and Turlay, unexpectedly exhibited that the long -- lived kaon
can decay also to the two -- pion final states with branching ratio of about $2\cdot 10^{-3}$~\cite{Christensen}.
Thus, the neutral kaons states seen in nature are not $\mathcal{CP}$ eigenstates defined in Eq.~\ref{eq:kkk1}. However,
they still can be expressed in the ($|K_1\rangle$, $|K_2\rangle$) basis:
\begin{align}
\nonumber
|K_{L}\rangle	& = \frac{1}{\sqrt{1+|\epsilon|^2}} \left(|K_{2}\rangle + \epsilon|K_{1}\rangle\right)\\
|K_{S}\rangle	& = \frac{1}{\sqrt{1+|\epsilon|^2}} \left(|K_{1}\rangle - \epsilon|K_{2}\rangle\right)~.
\label{kskldef}
\end{align}
Since up to now there is no signs of the $\mathcal{CPT}$ symmetry violation from now on we assume
\footnote{Although there are some theoretical predictions for the $\mathcal{CPT}$ violation
~\cite{AmelinoCamelia,Balwierz:2012uk,Bernabeu1,Bernabeu2}, all the tests done so far resulted in
the confirmation that it is an exact symmetry~\cite{pdg2010,DiDomenico,Ambrosino:2006,Thomson,Abouzaid}.}
$\epsilon_S = \epsilon_L = \epsilon$.\\
We can understand the $\mathcal{CP}$ symmetry breaking within the scope of two distinct mechanisms referred
to as ,,direct'' and ,,indirect'' breaking.
The ,,indirect'' violation corresponds to the statement that the true
eigenstates of both the strong and electroweak interactions are not exactly $\mathcal{CP}$
eigenstates but have small admixtures of the state with opposite $\mathcal{CP}$~\cite{Perkins}.
It is also possible that $\mathcal{CP}$ violation occurs ,,directly'' in the weak decays themselves.
In the following the ,,direct'' violation will be explained on the example of kaon decays to two pions.
The two -- pion systems originating from decays of kaons can be produced with isospin $I = 0$
or $I = 2$ (~isospin equal one is forbidden by Bose symmetry~\cite{Perkins}):
\begin{align}
\nonumber
|\pi^0 \pi^0\rangle &= \sqrt{\frac{1}{3}}|\pi^0 \pi^0;I=0\rangle - \sqrt{\frac{2}{3}}|\pi^0 \pi^0;I=2\rangle\\
\nonumber
|\pi^+ \pi^-\rangle &= \sqrt{\frac{2}{3}}|\pi^+ \pi^-;I=0\rangle + \sqrt{\frac{1}{3}}|\pi^+ \pi^-;I=2\rangle.
\end{align}
The corresponding weak decay amplitudes of kaons can be expressed as~\cite{Maiani}:
\begin{align}
\nonumber
\langle \pi \pi; I|\mathbf{H_W}|K^0\rangle &= A_{I}e^{i\delta_{I}}\\
\nonumber
\langle \pi \pi; I|\mathbf{H_W}|\overline{K^0}\rangle& = A^{*}_{I}e^{i\delta_{I}},
\end{align}
where we have explicitly exhibited the final state phases $\delta_{I}$, which arise
from the final state strong interactions of the pions.
Direct $\mathcal{CP}$ violation, occurring at the decay vertices, appears as a phase difference
between the $A_{0}$ and $A_{2}$ amplitudes\footnote{In general $A_{0}$ and $A_{2}$ could be real
if $\mathcal{CP}$ would be conserved.}~\cite{Maiani}. This phase difference is generated by a class
of so called ,,pengiun'' diagrams for $s$ quark decay, one of which is presented
in Fig.~\ref{fig:pingwin}.
\\Typically the $\mathcal{CP}$ violation in the neutral kaon sector is characterized in terms of
the following parameters:
\begin{align}
\nonumber
\eta_{+-} &= \frac{A( K_{L}\rightarrow \pi^{+}\pi^{-})}{A(K_{S}\rightarrow \pi^{+}\pi^{-})} =
|\eta_{+-}|e^{\phi_{+-}} \cong \epsilon + \epsilon'\\
\eta_{00} &= \frac{A( K_{L}\rightarrow \pi^0\pi^0)}{A(K_{S}\rightarrow \pi^0\pi^0)} =
|\eta_{00}|e^{\phi_{00}} \cong \epsilon - 2\epsilon',
\label{eq:eta00}
\end{align}
where $\epsilon$ is the mixing parameter defined before in Eq.~\ref{kskldef} and $\epsilon^{'}$
accounts for
the direct $\mathcal{CP}$ violation and can be expressed in terms of the weak amplitudes~\cite{Maiani}:
\begin{align}
\nonumber
\epsilon' &= \frac{\langle \pi \pi;0|\mathbf{H_W}|K_{S}\rangle \langle \pi \pi;2|
\mathbf{H_W}|K_{L}\rangle - \langle \pi \pi;0|\mathbf{H_W}|K_{L}\rangle \langle\pi \pi;2|
\mathbf{H_W}|K_{S}\rangle} {\sqrt{2} \langle \pi \pi;0|\mathbf{H_W}|K_{S}\rangle^2}\\
&\approx i\frac{e^{i(\delta_{2} - \delta_{0})}}{\sqrt{2}}Im\left(\frac{A_2}{A_0}\right).
\label{eq:epsilp}
\end{align}
Above defined parameters were measured many times and are known with a good precision~\cite{pdg2010}:
\begin{align}
\nonumber
|\eta_{+-}| &= (2.232 \pm 0.011)\cdot 10^{-3};~&\phi_{+-} = (43.51 \pm 0.05)^{\circ}\\
|\eta_{00}| &= (2.221 \pm 0.011)\cdot 10^{-3};~&\phi_{00} = (43.52 \pm 0.05)^{\circ}\\
\nonumber
|\epsilon| &= (2.228 \pm 0.011)\cdot 10^{-3};~&\phi_{\epsilon} = (43.51 \pm 0.05)^{\circ}.\\
\nonumber
\label{eq:params}
\end{align}
Moreover, measurements of the double ratio of the two pion decay rates $|\eta_{+-}|/|\eta_{00}|$
have proved that $\epsilon'$ is different from zero indicating occurrence of the direct
$\mathcal{CP}$ violation~\cite{Abouzaid}.\\
Analogous $\mathcal{CP}$ invariance breaking should appear also in the $K_S$ decays.
As before we can define the following amplitude ratios;
\begin{align}
\eta_{+-0} &= \frac{A( K_{S}\rightarrow \pi^{+}\pi^{-} \pi^0)}{A(K_{L}\rightarrow \pi^{+}\pi^{-}\pi^0)} =
|\eta_{+-0}|e^{\phi_{+-0}} \cong \epsilon + \epsilon'_{+-0}\\
\eta_{000} &= \frac{A( K_{S}\rightarrow \pi^0\pi^0 \pi^0)}{A(K_{L}\rightarrow \pi^0\pi^0 \pi^0)} =
|\eta_{000}|e^{\phi_{000}} \cong \epsilon + \epsilon'_{000}~.
\label{eq:eta000} 
\end{align}
As in the case of two -- pion decays the ratios contain direct $\mathcal{CP}$ violation parameters related
in the lowest order of the Chiral Perturbation Theory by the following equations:
$\epsilon'_{+-0} = \epsilon'_{000} = -2\epsilon'$~\cite{MPaver}.
The possible $|\pi^+ \pi^- \pi^0\rangle$ final state originating from the neutral kaon decays can
be produced with  isospin $I = 0, 1, 2,$ or 3. The $I = 0$ and $I = 2$ states have $\mathcal{CP} = 1$,
and $K_S$ can decay into them without violation of the $\mathcal{CP}$ symmetry. However, they are expected
to be strongly suppressed by centrifugal barrier effects~\cite{pdg2010}.
For the $I = 1$ and $I = 3$ states there is no centrifugal barrier and $\mathcal{CP} = -1$ so $K_S$
decay requires violation of this symmetry. Anyhow the two kinds of final states can be separated
by the analysis of the $\pi^+ \pi^- \pi^0$ Dalitz plot~\cite{pdg2010}. 
In the case of $|\pi^0 \pi^0 \pi^0\rangle$ final state, only isospin $I = 1$ or $I = 3$ is allowed,
for which $\mathcal{CP} = -1$. Therefore, the $K_S \to 3\pi^0$
decay is a purely $\mathcal{CP}$ violating process~\cite{pdg2010}.\\
The present knowledge about $\eta_{+-0}$ and $\eta_{000}$ is poor mainly due to very low decay
rates for the $K_S \to 3\pi$ decays. The current value of the $K_S \to \pi^+ \pi^- \pi^0$ branching
ratio amounts to $BR(K_S \to \pi^+ \pi^- \pi^0) = (3.5^{+1.1}_{-0.9})\cdot 10^{-7}$, and
the $K_S \to 3\pi^0$ has been never observed~\cite{pdg2010}. The best upper limit on this decay
branching ratio was set by KLOE collaboration and amounts to $BR(K_S \to 3\pi^0) < 1.2 \cdot 10^{-7}$~\cite{Matteo},
while the prediction based on Standard Model is equal to about $2\cdot 10^{-9}$~\cite{MPaver}. The corresponding
knowledge about the amplitude ratios can be summarized as follows~\cite{pdg2010}:
\begin{align}
\nonumber
Re(\eta_{+-0}) &= -0.002 \pm 0.007^{+0.004}_{-0.001}\\
\nonumber
Im(\eta_{+-0}) &= -0.002 \pm 0.009\\
\nonumber
Im(\eta_{000}) &= -0.001 \pm 0.016\\
\nonumber
|\eta_{000}| &< 0.018. 
\end{align}
Therefore, it is clear, that the full understanding of the $\mathcal{CP}$ violation in the neutral kaon system
demands new high statistics measurements, in particular of the $K_S \to 3\pi^0$ decay which is a subject of this
work. One of the high precision experiments, which has been greatly contributed to this quest, is the KLOE detection
setup which will be presented in the next chapter. 
\chapter{The KLOE experiment at DA$\Phi$NE}
In this chapter the characteristics of the DA$\Phi$NE collider and the KLOE detector
are briefly described. More detailed description can be found in
Ref.~\cite{kloe2008,Antonelli:1996hn,Adinolfi:2002hs,Finocchiaro:1995fu,Flavio,Ambrosino:2004qx}. 
\section{The DA$\Phi$NE Collider}
DA$\Phi$NE is an $e^{+}e^{-}$ collider, optimized to work with a center of mass
energy around the $\phi$ mass, $M_{\phi} = (1019.418 \pm 0.008)$ MeV~\cite{Flavio}.
The ,,heart'' of the collider are two storage rings in which 120 bunches of both, electrons and positrons, are stored.
Each bunch collides with its counterpart once per turn, minimizing the mutual perturbations of colliding beams.
\begin{figure}[h!]
\centering
\includegraphics[width=0.8\textwidth]{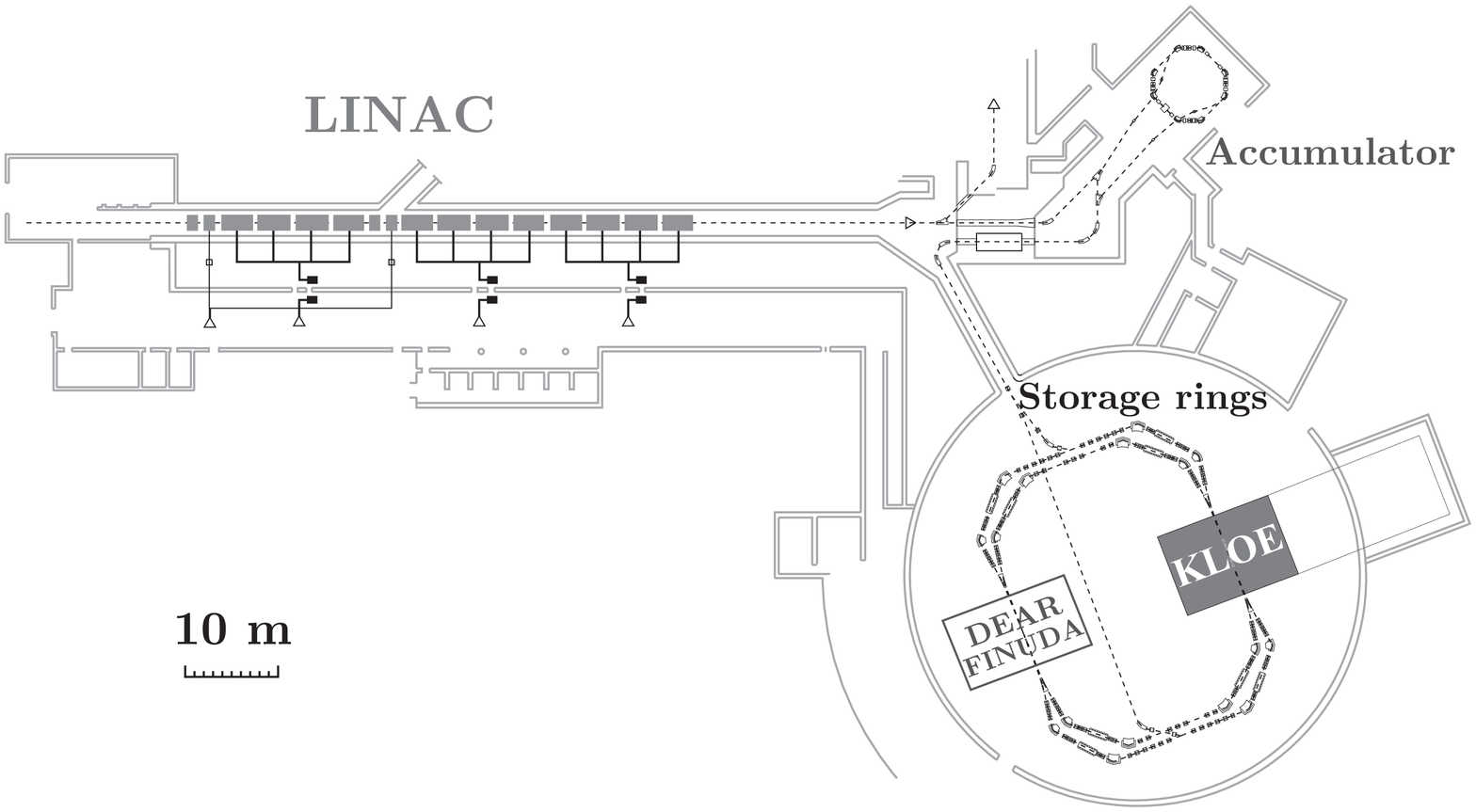}
\caption{Schematic view of the DA$\Phi$NE collider. The picture was adapted from~\cite{kloe2008}.}
\label{fig:dafne}
\end{figure}
Electrons are
accelerated to final energy in the Linac (see Fig.~\ref{fig:dafne}), accumulated
and cooled in the accumulator and transferred to a single bunch in the ring~\cite{kloe2008}.
Positrons are created in an intermediate station in the Linac, and then follow the same
procedure as electrons. Both, electrons and positrons are injected into the rings at final energy of about 510 MeV.
The beams collide in two interactions regions, with a frequency
up to 356 MHz, corresponding to a minimum bunch crossing period of $T_{rf} = 2.7$ ns.
The $e^+e^-$ collisions result in $\phi$ meson creation which is almost at rest
($\beta_{\phi} \approx 0.015$) and decay predominantly to $K^+K^-$ (49$\%$), $K_SK_L$ (34$\%$),
$\rho\pi$ (15$\%$) and $\eta\gamma$ (1.3$\%$) final states~\cite{pdg2010}.
The decay products are registered using the KLOE detection setup, which consists
of large cylindrical drift chamber surrounded by the electromagnetic calorimeter.
The components of KLOE will be briefly described in the next section.
\section{The KLOE detector}
\begin{figure}[b!]
\centering
\includegraphics[width=0.6\textwidth]{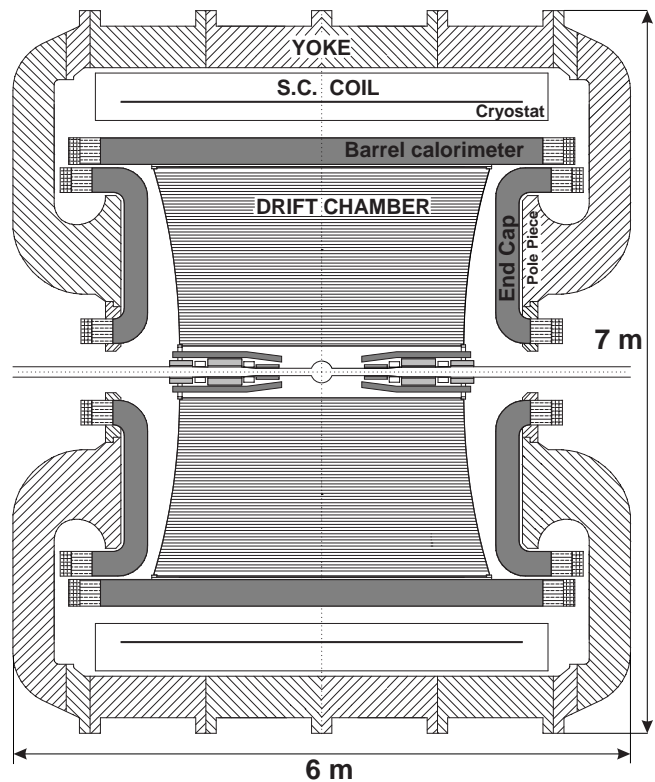}
\caption{Vertical cross section of the KLOE detector. The figure is adapted from~\cite{kloe2008}.}
\label{fig:kloe}
\end{figure}

The KLOE detector consists of a large cylindrical drift chamber and a hermetic
electromagnetic calorimeter. A superconducting coil and an iron yoke (see Fig.~\ref{fig:kloe})
surrounding the calorimeter provides a 0.52 T magnetic field.
The beam pipe at the interaction region is a beryllium sphere with 10 cm of radius and 0.5 mm
thick. This structure minimizes both the multiple scattering and the energy loss of the charged
particles from $K_S$ decays, as well as the probability of $K_L$ regeneration~\cite{Flavio}.
\subsection{The Drift Chamber}
\begin{figure}
\centering
\includegraphics[width=0.7\textwidth]{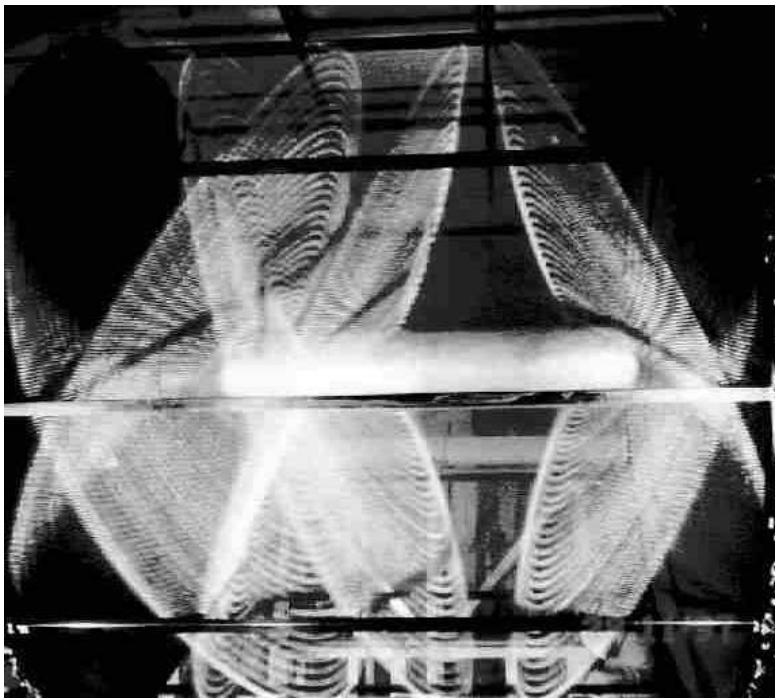}
\caption{Drift chamber stereo wires geometry. The figure is adapted from~\cite{Balwierz}.}
\label{fig:d}
\end{figure}

\begin{figure}
\centering
\includegraphics[width=0.49\textwidth]{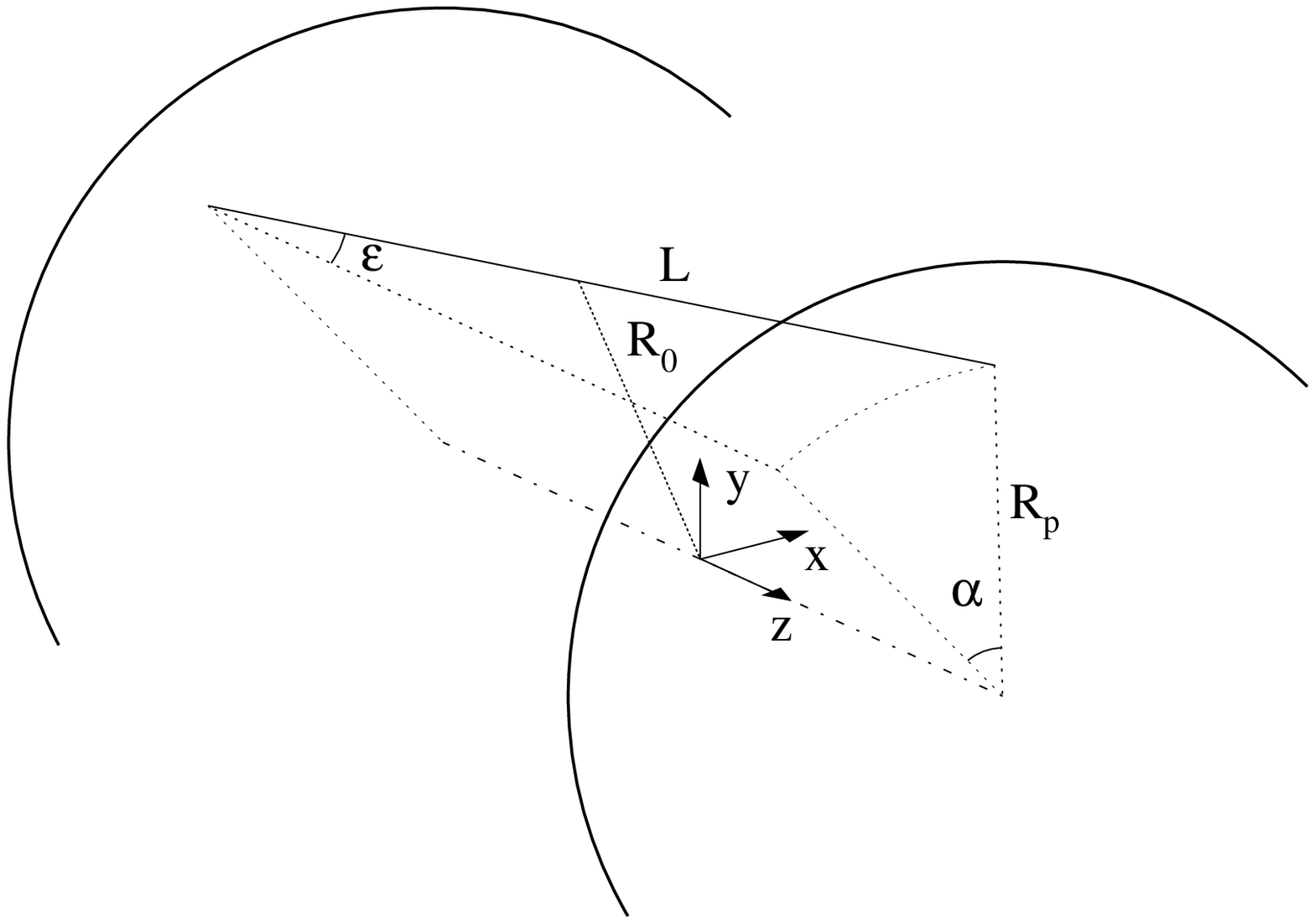}
\includegraphics[width=0.49\textwidth]{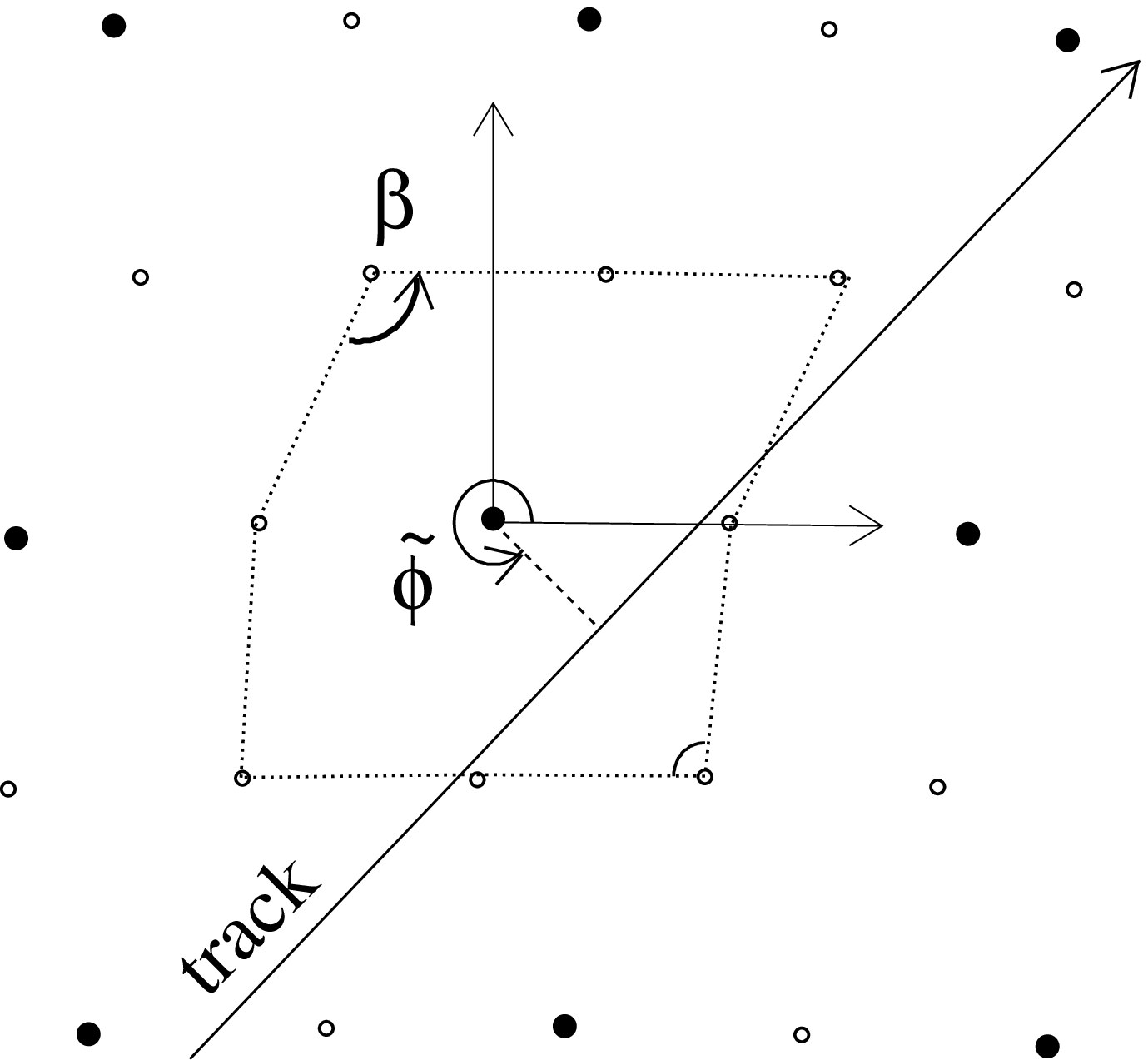}
\caption{Left: Wire geometry with the definition of stereo angle $\epsilon$ between the wire of length $L$
and the z -- axis. Right: Definition of $\beta$ and $\widetilde{\phi}$ angles characterizing the shape of the
cell and the angle of the incident track~\cite{kloe2008}. The figures are adapted from~\cite{kloe2008}.}
\label{fig:a}
\end{figure}

The KLOE drift chamber has a cylindrical shape 3.3 m long, with internal and external
radii of 25 cm and 2 m, respectively~\cite{Flavio}. It was designed to register all charged secondary
products from the $K_L$ decay and measure their properties with great precision~\cite{kloe2008}.
Thus, its size was dictated by a long lifetime of 
this particle\footnote{The mean decay path of the $K_L$ meson produced in the $\phi$ decay
amounts to about 3.4~m~\cite{kloe2008}.}.
To minimize the $K_L$ regeneration, multiple Coulomb scattering and photon absorption
KLOE drift chamber is constructed out of carbon fiber composite with low-Z and low
density, and uses a gas mixture of helium (90$\%$) and isobutane (10$\%$)~\cite{HyperF}.
The radiation length of the gas amounts to about 900 m, including the contribution
of the 52140 wires~\cite{Flavio}.
In order to obtain high and uniform track and vertex reconstruction efficiencies,
wires are strung in an all -- stereo geometry, with stereo angles varying with the radius
from 50 mrad to 120 mrad going outward~\cite{Flavio}.
This design results in a uniform filling of the sensitive volume with almost square drift
cells, with shape slowly changing along z axis\footnote{ The z axis of the KLOE reference
frame is defined as the bisector of the angle between colliding $e^+$ and $e^-$ beams~\cite{kloe2008}.}.
Fig.~\ref{fig:d} shows the wire geometry during the drift chamber construction as illuminated by light.
Particles from the $\phi$ decays are produced with small momenta and therefore track density is much
higher at small radii~\cite{Flavio}. Thus, dimensions of the cells were designed to be of about 2 x 2 cm$^2$
for the 12 innermost wire layers, and to of about 3 x 3 cm$^2$ for the remaining 48 layers~\cite{Finocchiaro:1995fu}.\\
To extract the space position from the measured drift time of the incident particle,
232 space -- to -- time relations are used. They are parametrized in terms of two angles $\beta$ and $\widetilde{\phi}$
defined in Fig.~\ref{fig:a}. The $\beta$ angle characterizes the geometry of the cell directly related
to the electric field responsible for the avalanche multiplication mechanism. $\widetilde{\phi}$ instead
gives the orientation of the particle trajectory in the cell's reference frame, defined in the
transverse plane and with origin in the sense wire of the cell~\cite{kloe2008}.\\
Using the wire geometry, space -- to -- time relations and known magnetic field one can reconstruct
the tracks and vertices of charged particles. The reconstruction procedure starts with pattern
recognition and is followed by track and vertex fitting.
The pattern recognition associates hits\footnote{As a hit we consider a presence of a signal on
a sense wire.} close in space to form track candidates and gives a first estimate of
the track parameters. Then track fitting provides the final values of these parameters
by minimization procedure based on the difference between the fitted and the expected
drift distances (so called residuals), as evaluated from measured drift times and
space -- to -- time relations.
Finally the vertex fit procedures search for possible primary and secondary vertices, on the basis
of the distance of closest approach between tracks~\cite{kloe2008}.\\
To ensure the stability in time of the KLOE drift chamber performance, the system is calibrated
periodically by acquiring samples of cosmic ray events suitable for the measurement of
about 200 different space -- to -- time relations~\cite{kloe2008}. The calibration is performed at the beginning
of each KLOE run and selects about 80000 cosmic ray events~\cite{Finocchiaro:1995fu}.
These events are tracked using the existing space -- to -- time relations and the average
value of the residuals for hits in the central part of the cells is monitored.
If the residuals exceed 40 $\mu$m
additional 3$\cdot 10^5$  cosmic ray events are collected, and a new set of calibration
constants is obtained. Finally, during data taking the drift chamber performances are monitored
using selected samples of events~\cite{Finocchiaro:1995fu}.\\
The KLOE drift chamber provides tracking in three dimensions with a resolution in the
transverse plane of about 200~$\mu$m, resolution in the z-co\-or\-di\-na\-te measurement
of about 2 mm and of 1 mm on the decay vertex position.
The momentum of the particle is determined from the curvature of its trajectory in the magnetic
field with a fractional accuracy $\sigma_p/p~=~0.4\%$ for polar angles larger
than 45$^{\circ}$~\cite{kloe2008}.
\subsection{The Electromagnetic Calorimeter}
\begin{figure}
\centering
\includegraphics[width=0.2\textwidth,angle=-90]{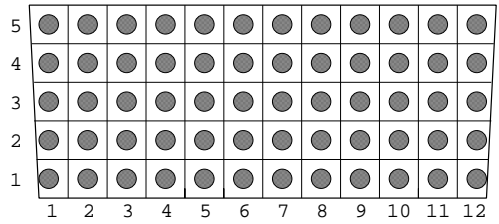}
\caption{Schematic view of the readout structure on one side of the barrel module~\cite{Zdebik}.
60 defined cells form 5 planes and 12 columns of the calorimeter module. 
Filled circles represents photomultipliers.
The figure is adapted from~\cite{Zdebik}.}
\label{fig:bb}
\end{figure}

The KLOE electromagnetic calorimeter was designed to provide hermetic  detection of low energy
gamma quanta with high efficiency, good energy resolution and excellent time resolution for
the neutral vertex reconstruction and to trigger the events~\cite{kloe2008}. It consists
of a barrel built out of 24 trapezoidal shaped modules and side detectors (so called endcaps)
read out from both sides by a set of photomultipliers.
The barrel is a cylinder with an inner diameter of 4 m, made of 24 modules 4.3 m long and 23 cm thick.
Each endcap consists of 32 vertical C -- shaped modules. This structure covers 98$\%$ of the full solid
angle. Each module consists of a mixture of lead (48$\%$ of the volume), scintillating fibers
(42$\%$), and glue (10$\%$)~\cite{Flavio}. Fibers, each with a diameter of 1 mm, are embedded in 0.5 mm
lead foils accelerating the showering processes. The special care in design and assembly
of the Pb -- fiber composite ensures that the light propagates along the fiber in a single mode
with velocity $\sim$17 cm/ns, which greatly reduces spread of the light arrival time at the fiber ends~\cite{kloe2008}.
Calorimeter modules are read out at both ends viewed by light guides of area of 4.4 x 4.4~cm$^2$
coupled to the photomultipliers transforming the light into electric impulses.
This defines so called ,,calorimeter cells'' which form five larger
structures (see Fig.~\ref{fig:bb}): planes 4.4~cm wide\footnote{The last plane of cells is 5.2~cm wide.}.
\\When a particle hits the calorimeter for each cell both, the charge as well as time of arrival
of the photomultiplier signals are registered. The cell energy is taken as the average of
the energy registered at both sides, after correcting for the light attenuation along the fiber~\cite{Flavio}.
The energy calibration starts by a first equalization in cell response to minimum
ionizing particles at calorimeter center, and by determining the attenuation length
of each single cell using cosmic rays acquired in dedicated runs.
This is done before the start of each long data taking period~\cite{kloe2008}.
\begin{figure}
\begin{center}
\includegraphics[width=0.7\textwidth]{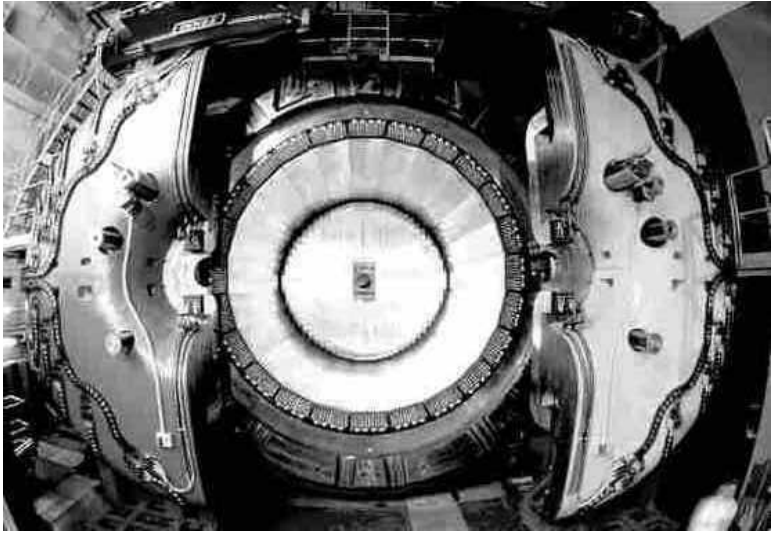}
\end{center}
\caption{Photograph of the KLOE calorimeter. One can see 24 modules of the barrel and
the inner plane of one of the endcaps. The figure is adapted from~\cite{Zdebik}.}
\label{fig:b}
\end{figure}
The energy determined from the measured amplitudes of signals for both sides $A$ and $B$
of a cell $S^{AB}$ amounts to:
\begin{equation}
E^{A,B}(\mathrm{MeV}) = \frac{S^{A,B} - S_{0}^{A,B} }{S_{M}} \cdot K ,  
\end{equation}
where 
$ S_{0}^{A,B} $ are the zero -- offsets of the amplitudes scale,    
$ S_{M} $ corresponds to the response for the minimum ionizing particle crossing the calorimeter center and 
K factor gives the energy scale in MeV~\cite{Zdebik}.  
The total energy deposited in a cell is calculated as the mean of values determined at both ends for each cell.
The determination of the absolute energy scale $K$ relies on a monochromatic source
of 510 MeV photons: the $e^+e^- \to \gamma\gamma$ sample. The  latter calibration
is routinely carried out each 200 -- 400 nb$^{-1}$ of collected luminosity~\cite{kloe2008}.\\

For each cell two time signals $T^A$ and $T^B$ (digitized by the Time to Digital Converter TDC )
are recorded. The arrival time $t$ and position $s$ of the impact point along the fiber direction
can be determined as\footnote{$s = 0$ is assumed
to be at the fiber center.}:
\begin{align}
  t (\mathrm{ns}) & = \frac{1}{2} (t^{A} + t^{B} - t^{A}_{0} - t^{B}_{0}) - \frac{L}{2v} ~, \\
  s (\mathrm{cm}) & = \frac{v}{2} (t^{A} - t^{B} - t^{A}_{0} + t^{B}_{0}) ~,
\end{align}
with $ t^{A,B} = c^{A,B} \cdot T^{A,B} $, where $ c^{A,B} $ are the TDC calibration constants, 
$ t^{A,B}_{0} $ denotes overall time offsets,    
 $L$ stands for length of the cell (cm) and  
$v$ is the light velocity in fibers (cm/ns)~\cite{Antonelli:1996hn}.\\
Based on the reconstructed energies, times and positions cells are merged into calorimeter
clusters. First the adjacent cells are grouped into so called
,,preclusters''\footnote{A cell is added into a precluster only if its times and energies
were reconstructed at both sides of the calorimeter module.}.
The time spread of cells forming the precluster has to be smaller than 2.5 ns~\cite{Zdebik}.
Moreover, cells are merged in one cluster if a distance between them and the center of the
precluster is less than 20 cm~\cite{Zdebik}.
The cluster energy is evaluated as the sum of the cells energies: 
\begin{align}
E_{cl} = \sum_{i} E_{i}~,
\end{align}
while the time and position centroids are obtained as weighted averages:
\begin{align}
T_{cl} & = \frac{\sum_{i}E_{i}\cdot t_{i}}{\sum_{i} E_i}\\
\mathbf{R_{cl}} &= \frac{\sum_{i}E_{i}\cdot \mathbf{r_{i}}}{\sum_{i} E_i}~.
\end{align}
$i$ denotes the $i$th cell belonging to the cluster and $\mathbf{r_{i}}$ stands for the cell's
position vector with respect to the interaction point.
$T_{cl}$ is next related to the time of flight of particle from the interaction point to the
cluster position. It is done subtracting the event global time offset, common
to all channels and depending on the trigger formation time with respect
to the real $e^+e^-$ interaction time.
Due to the spread of the particle's arrival times, the KLOE trigger is not able to identify
the bunch crossing related to each event, which has to be determined offline~\cite{kloe2008}. 
The common ,,Start'' signal to the calorimeter TDC boards is provided by the
first level trigger, which will be described in the next section. The ,,Stop'' instead 
is given by the photomultiplier signals delayed  because of the electronics and light
propagation in the fibers~\cite{Flavio}. Time measured by the calorimeter can be expressed as:
\begin{align}
T_{cl} & = T_{tof} + \delta_{c} - N_{bc}\cdot T_{rf}~,
\end{align}
where $T_{tof}$ is the time of flight of a particle from the interaction point to
the cluster position, $\delta_{c}$ is a single number accounting
for the overall electronic offsets and cable delays, and $N_{bc}$ is the number of
bunch -- crossing periods needed to generate the TDC start. The values of
$\delta_{c}$ and $N_{bc}$ are determined for each data taking run with
$e^+e^- \to \gamma\gamma$ events by looking at the $\Delta_{TOF} = T_{cl} -R_{cl}/c$
distributions ($c$ denotes the speed of light )~\cite{kloe2008}.
For such events this distribution shows well separated peaks corresponding to
different values of $N_{bc}$.
We define $\delta_{c}$ as the position of the largest peak in the 
distribution, and obtain $T_{rf}$ from the distance between 
peaks~\cite{Ambrosino:2004qx}.
During offline processing, to allow the cluster times to be related to the particle
time of flight, we determine for each event the corrected cluster times:
\begin{align}
t_{cl} & = T_{cl} - (\delta_{c} - N_{bc}\cdot T_{rf})~.
\end{align}
\\
The KLOE electromagnetic calorimeter allows for measurements of particle energies and flight time
with accuracies of
$\sigma_E = \frac{5.7\% }{\sqrt{E[\mathrm{GeV}]}}E$
and $\sigma(t) = \frac{57 \mathrm{ps}}{\sqrt{E[\mathrm{GeV}]}} \oplus 140$~ps, respectively~\cite{Antonelli:1996hn}.
Analysis of the signal amplitude distributions allows to determine the location where the
particle hits the calorimeter module with accuracy of about 1 cm in the plane transverse to
the fiber direction. The longitudinal coordinate precision is energy dependent:
$\sigma_z = \frac{1.2~\mathrm{cm}}{\sqrt{E[\mathrm{GeV}]}}$~\cite{HyperF}.
\subsection{The Trigger system}
The KLOE trigger system is based on local energy deposits in the electromagnetic calorimeter and 
hit multiplicity information from the drift chamber.
It has been optimized to retain almost all $e^{+}e^{-} \to \phi$ decays, and provide
efficient rejection on the two main sources of background: small angle $e^{+}e^{-} \to e^{+}e^{-}$
scattering
and particle lost from the DA$\Phi$NE beams~\cite{kloe2008}. Moreover, all $e^{+}e^{-}$ scattering and $\gamma\gamma$
events produced at large polar angles are gathered for detector monitoring and calibration.
Since the DA$\Phi$NE bunch crossing period amounts to $T_{rf} = 2.7$ ns, KLOE trigger must
operate in continuous mode. A two level scheme was chosen. A first level trigger T1 is produced
with a minimal delay ($\sim$ 200 ns) and is synchronized with the DA$\Phi$NE master clock~\cite{Adinolfi:2002hs}.
The T1 signal initiates conversion in the
front -- end electronics modules, which are subsequently read out following a fixed time
interval of about 2.6 $\mu$s. This corresponds to the typical drift time of electrons
travelling in the drift chamber cells~\cite{kloe2008}.
After the arrival of a first level trigger, additional information is collected from the drift
chamber, which is used together with the calorimeter information as a second level trigger T2.
It confirms the first level trigger, initializes digitisation of the drift chamber electronics
and starts the data acquisition readout. If no T2 signal arrives before the end of 2.6 $\mu$s
dead time, all readout is reset~\cite{Adinolfi:2002hs}.\\
T1 and T2 triggers are based on the topology of energy deposits in the KLOE electromagnetic
calorimeter and on the number and spatial distribution of the drift chamber hits.
Since $\phi$ decay events have a relatively high multiplicity, they can be efficiently
selected by the calorimeter trigger by requiring two isolated energy deposits above a
threshold of 50 MeV in the barrel and 150 MeV in the endcaps. Events with only two
fired sectors in the same endcap are rejected, because this topology is dominated by
machine background. Moreover, we require about 15 hits in the drift chamber within a time
window of 250 ns from beam crossing~\cite{kloe2008}. The trigger identifies
$e^{+}e^{-} \to e^{+}e^{-}$ events requiring clusters with energy of about 350 MeV. 
An event which satisfies at least one of the two above
conditions and is not recognized as $e^+e^-$ scattering, generates a first level
trigger T1\footnote{As it was mentioned a part of the $e^{+}e^{-} \to e^{+}e^{-}$ events
are gathered for detector monitoring and calibration.}.
The level -- 2 trigger T2, requires further multiplicity or geometrical conditions
for the electromagnetic energy deposits, or about 120 drift chamber wire signals within
a 1.2 $\mu$s time window. At the level 2 trigger recognizes also the cosmic ray events
by the presence of two energy deposits above 30 MeV in the outermost calorimeter
layers~\cite{kloe2008}. A fraction about 80$\%$ of the cosmic ray events are identified
and rejected at the trigger level with this technique. Further suppression of the DA$\Phi$NE
background events and cosmic rays is performed by an off -- line filter called FILFO
(\textit{FILtro FOndo}: background filter). FILFO identifies  background events at a very
early stage of the data reconstruction using only information from the calorimeter~\cite{memo288}.\\
For the search of the $K_S \to 3\pi^0$ decay only the calorimeter signals are used to
trigger the event. Two energy deposits above threshold about 50 MeV for
the barrel and about 150 MeV for the endcaps are required~\cite{Matteo}. 
\chapter{First stage of the event selection}
\label{rozdz4}
The $\phi$ meson produced in the $e^+e^-$ collision at DA$\Phi$NE is in a pure $J^{PC} = 1^{- -}$ state.
Since the $\phi \to K_S K_L$ decay is driven by the strong interaction, the initial $K_S K_L$
state is antisymmetric with the same quantum numbers and can be written in the $\phi$ rest frame as:
\begin{equation}
\left | i \right \rangle =N \cdot \left[\left|K_S(\vec{p})\right\rangle\left|
K_L(-\vec{p})\right\rangle -\left|K_L(\vec{p})\right\rangle\left|K_S(-\vec{p})\right\rangle\right]~,
\label{ksklstate}
\end{equation}
where $\vec{p}$ denotes the momentum of each kaon and $N$ is a normalization factor~\cite{DiDomenico}.
Since the $\phi$ resonance is moving with a small momentum in the horizontal plane
$P_{\phi} \approx 13~\mathrm{MeV/c}$ $K_S$ and $K_L$ mesons are produced almost back -- to -- back
in the laboratory frame.
Therefore, observation of a $K_L$ ($K_S$ ) decay ensures the presence of a $K_S$ ($K_L$~)
meson travelling in the opposite direction\footnote{
We refer to the process of defining a $K_S$ or $K_L$ sample as tagging:
observation of a $K_L$ ($K_S$ ) decay tags the presence of \mbox{a $K_S$} ($K_L$ ) meson
and allows for the determination of its momentum~\cite{flavor_KLOE}.}
~\cite{flavor_KLOE}. Thus, at DA$\Phi$NE we obtain pure $K_S$ and $K_L$ ,,beams'' with
precisely known momenta and flux, which can be
used to measure absolute branching ratios~\cite{kloe2008}. In this chapter the $K_S$ tagging
technique with the detection of the $K_L$ interaction in the KLOE calorimeter is described.
%
\section{Identification of $K_S$ via detection of $K_L$}
Neutral kaons produced at KLOE have a velocity in the $\phi$ rest frame equal to $\beta \approx 0.22$.
This corresponds to the $K_L$ time of flight from the interaction point to the calorimeter
equal to about 31 ns, which means that about 60$\%$ of produced $K_L$ mesons reach the calorimeter without
decaying~\cite{kloe2008}. $K_L$ mesons interact in the calorimeter with an energy release up to $\sim$
497 MeV (so called ,,$K_L$ -- crash''). Thanks to the exceptional timing capabilities of the KLOE
calorimeter\footnote{
For an energy release of 100 MeV the resolution of time measured by the calorimeter amounts to
about 0.3 ns, which corresponds to about 1$\%$ accuracy in the determination of
the $K_L$ velocity~\cite{kloe2008}.}
and the low velocity of kaons one can use the Time of Flight technique to tag the $K_S$ meson,
as described in the next section. Adding the information about the position of the energy release
($K_L$ cluster), the direction of the $K_L$ flight path can be determined with $\sim$ 1$^\circ$
angular accuracy~\cite{kloe2008}. This allows to estimate the $K_L$ momentum vector and as a
consequence, knowing the $\phi$ four -- momentum, to determine the four -- momentum of the
tagged $K_S$ meson.

\subsection{Identification of the $K_L$ meson}
The identification of the $K_L$ interaction in the calorimeter is performed after tracks reconstruction
and association to the clusters\footnote{
The track -- to -- cluster association procedure establishes correspondence between tracks in the
drift chamber and clusters in the calorimeter.}, and after the preselection aiming at the rejection of events
with $K_L$ meson decay inside the drift chamber. Events for which there is one reconstructed vertex
with two tracks having opposite curvature are rejected. Moreover, an event is discarded if there are two
reconstructed tracks, having opposite curvature, associated to two vertices reconstructed less than 30 cm
away from the interaction point in the transverse plane. These cuts reject most of the
background events with $K_L$ decaying before reaching the calorimeter~\cite{km146}.  
For each surviving event we look for the $K_L$ clusters in the calorimeter taking into account only
clusters not associated to any track. For each that kind of clusters we calculate velocity of the
contributing particle defined in the laboratory frame as:
\begin{equation}
\beta_{cl} = \frac{R_{cl}}{c\cdot t_{cl}}~,
\label{eq:betakl}
\end{equation}
where $R_{cl}$ denotes the distance from the $e^+e^-$ interaction point to the reconstructed position
of the cluster center, $t_{cl}$ stands for the measured time of flight of the particle and $c$ is
the speed of light. Since in the $\phi$ rest frame kaons have a well known velocity
$\beta \approx 0.22$ it is convenient to transform $\beta_{cl}$ to this reference frame:
\begin{equation}
\beta_{cr} = \frac{\sqrt{\beta_{\phi}^2+\beta_{cl}^2+2\beta_{\phi} \beta_{cl} \mathrm{cos}\alpha}}
{1+\beta_{\phi}\beta_{cl} \mathrm{cos}\alpha}~,
\label{eq:betaklstar}
\end{equation}
where $\beta_{\phi}$ denotes the velocity of the $\phi$ meson in the laboratory frame and $\alpha$
stands for the angle between the $\phi$ momentum vector and a direction vector connecting the
interaction point with the cluster position.
\begin{figure}
\centering
\includegraphics[width=0.49\textwidth]{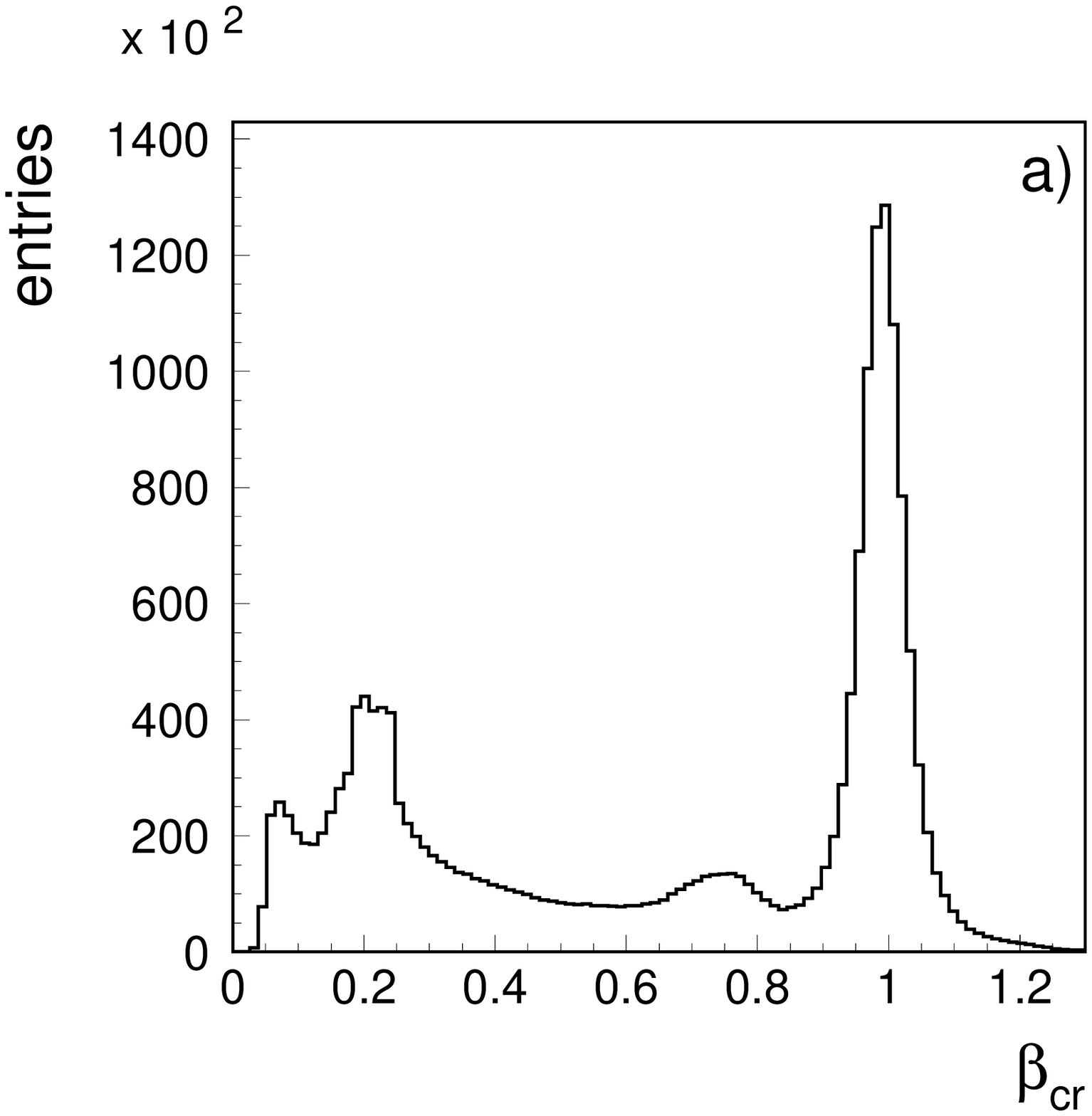}
\includegraphics[width=0.49\textwidth]{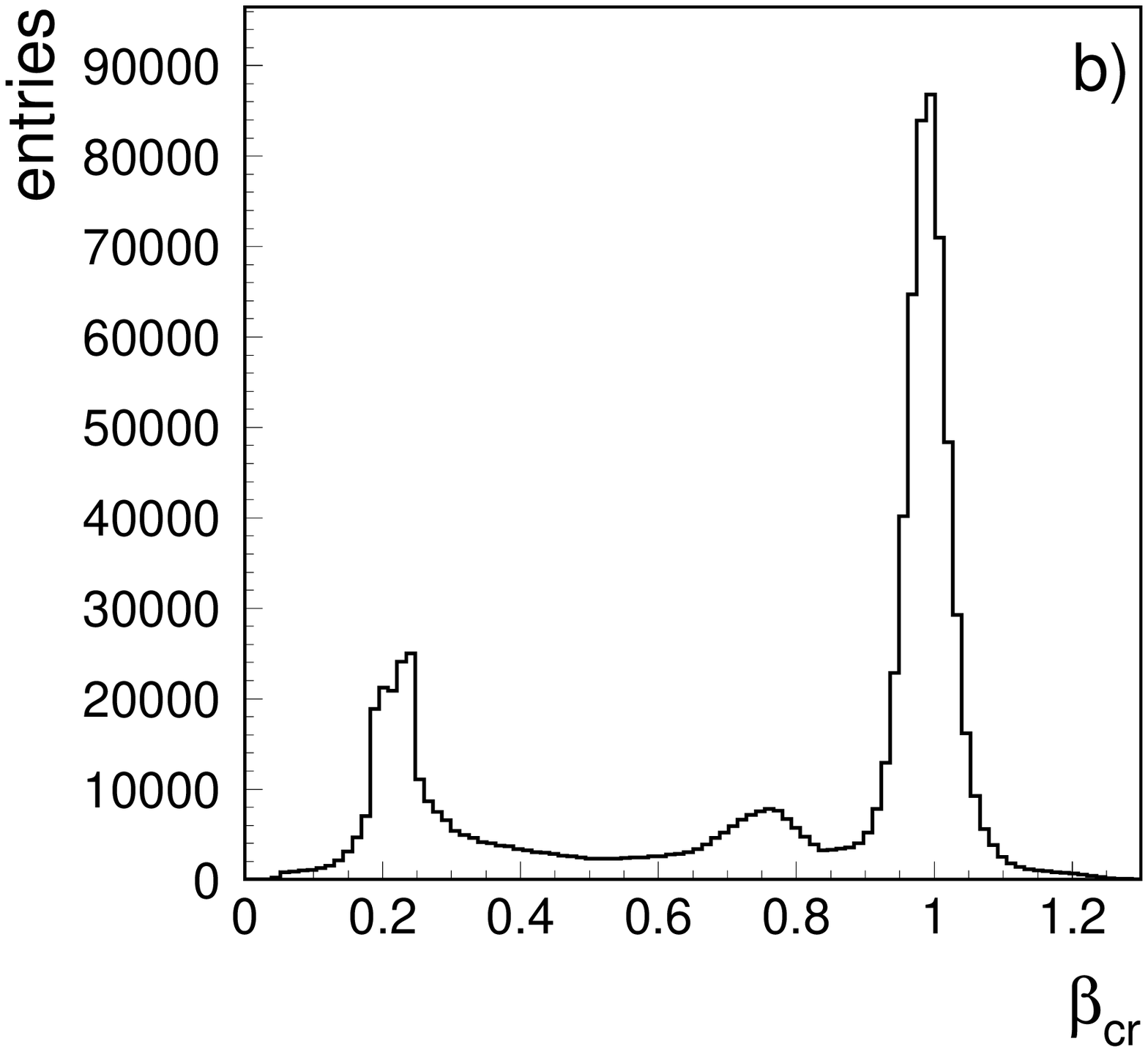}
\caption{Distribution of the $\beta_{cr}$ velocity reconstructed for clusters not associated
to the track for a sample of events before identification of the $K_L$ interaction in the calorimeter.
The spectra are made for all clusters before (a) and after (b) cut on energy $E_{cr}>100$ MeV.}
\label{fig:betacl}
\end{figure}
The distribution of $\beta_{cr}$ is presented in Fig.~\ref{fig:betacl}a. A big peak around
$\beta_{cr}~=~1$ corresponds mainly to clusters formed by gamma quanta from $K_S$ mesons
which decay very close to the interaction point. Clusters originating from the decay products
of remaining $K_L$ decays are instead characterized by smaller velocities distributed in the range from
$\beta_{cr} \approx 0.28$ to $\beta_{cr} \approx 1$~\cite{km146}. In Fig.~\ref{fig:betacl}a one can
also see a structure around $\beta_{cr} \approx 0.22$ corresponding to $K_L$ mesons and a smaller
peak for very low velocities. The latter peak originates mainly from the charged particles (e.g. pions)
for which the track -- to -- cluster association procedure failed. Additionally, most of
these clusters are characterized by energy deposits smaller than \mbox{100 MeV}~\cite{km146}.
The $\beta_{cr}$ distribution after the cut on $E_{cr} > 100$ MeV is
shown in Fig.~\ref{fig:betacl}b, where one can see a big suppression of the clusters with lowest velocity. 
Therefore, clusters originating from the $K_L$ interaction in the calorimeter are defined
with the following conditions:
\begin{align}
\nonumber
0.17 < \beta_{cr} < 0.28\\
E_{cr} > 100~\mathrm{MeV}~,
\label{eq:consKlcr}
\end{align}
where $E_{cr}$ is the energy of the $K_L$ cluster.\\
The main remaining background sources to this tagging algorithm are the cosmic muons
entering KLOE through the intersection between the barrel and endcap calorimeters.
Such muons may give a signal in the calorimeter without a track in the KLOE drift chamber.
The other contributions to the background originate from DA$\Phi$NE activity
and $\phi \to K^+K^-$ decays~\cite{Flavio}.
The angular momentum of the $K_L$$K_S$ system is equal to the spin of the $\phi$ meson $s=1$.
Therefore, kaons from the $\phi$ decay are mostly emitted in the direction perpendicular
to the beam axis and the background can be additionally suppressed selecting
only ,,$K_L$ -- crash'' clusters in the barrel~\cite{Flavio}.\\
$K_L$ meson interacting in the calorimeter usually induces more than one cluster,
therefore to estimate the direction of the tagging $K_L$ meson we consider the ,,fastest''
candidate cluster which was produced as the first one.
\subsection{$K_L$ -- momentum estimate}
In the $\phi \to K_S K_L$ decay the $K_L$ four -- momentum $\mathbb{P_{K_L}}$ can be
determined completely knowing the center of mass energy $\sqrt{s}$, the $\phi$ momentum
$P_{\phi}$ vector and $\alpha$ angle between the $\phi$ momentum and the $K_L$ flight
direction determined in the laboratory frame from the reconstructed center of the cluster.
For $\sqrt{s}$  and $P_{\phi}$
we use the mean values measured for each running period using the gathered sample
of $e^+e^-$ scattered at large angles. Determination of $\mathbb{P_{K_L}}$ allows to
calculate the four -- momentum of the tagged $K_S$ meson:
$\mathbb{P_{K_S}} = \mathbb{P_{\phi}} - \mathbb{P_{K_L}}$.
\chapter{Normalization sample}
\label{rozdz5}
\begin{figure}
\centering
\includegraphics[width=0.5\textwidth]{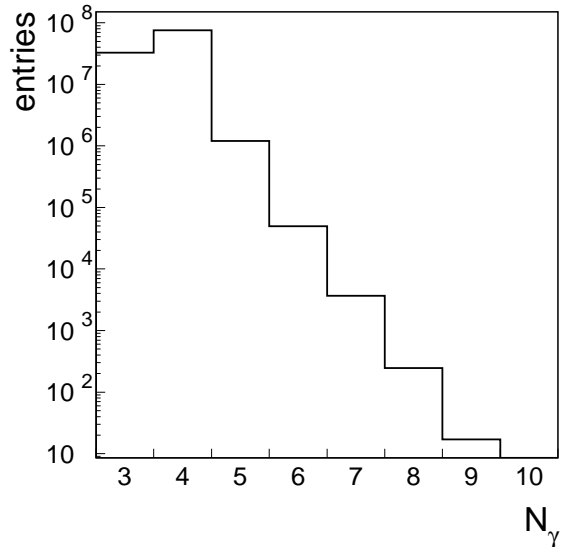}
\caption{
The experimental distribution of the reconstructed $\gamma$ quanta multiplicities
after imposing the tight $K_S$ tag requirements \mbox{( $E_{cr} > 150~\mathrm{MeV}$}
and \mbox{$0.200 <\beta_{cr} < 0.225$ )}
and acceptance cuts defined by Eqs.~\ref{acc1},
\ref{acc2} and \ref{acc3}.
}
\label{fig:multi}
\end{figure}

Registration of the $K_L$ interactions in the calorimeter allows for the simultaneous
identification of the $K_S$ meson. Since one of the goals of this measurement is
to determine the $K_S \to 3\pi^0$ decay branching ratio, the number of events selected as
the signal has to be normalized to the number of all $K_S$ decays. To this end the $K_S \to 2\pi^0\to 4\gamma$
events were also counted (further on they will be referred to as the normalization sample).
This process is one of the main
$K_S$ decay channel with well -- known branching ratio $BR(K_S \to 2\pi^0) = 0.3069 \pm 0.0005$~\cite{pdg2010}.
The number of events produced for both the signal and the normalization sample can be expressed as:
\begin{eqnarray}
N_{2\pi}~=~L\cdot\sigma_{\phi}\cdot BR(\phi \to K_{S}K_{L})\cdot BR(K_{S}\to 2\pi^{0})\cdot\epsilon_{2\pi}\cdot\epsilon_{cr}~\\
\nonumber
N_{3\pi}~=~L\cdot\sigma_{\phi}\cdot BR(\phi \to K_{S}K_{L})\cdot BR(K_{S}\to 3\pi^{0})\cdot\epsilon_{3\pi}\cdot\epsilon_{cr}~,
\end{eqnarray}
where $L$ is the integrated luminosity, $\sigma_{\phi}$ denotes the total cross section for $\phi$ production,
$\epsilon_{cr}$ stands for the tagging efficiency and $\epsilon_{3\pi}$ and $\epsilon_{2\pi}$ are the
identification efficiencies for the appropriate channel. The ratio:
\begin{equation}
\frac{N_{3\pi}}{N_{2\pi}}~=~\frac{BR(K_{S}\to 3\pi^{0})\cdot\epsilon_{3\pi}}{BR(K_{S}\to 2\pi^{0})\cdot\epsilon_{2\pi}}
\label{eq:1roz5}
\end{equation}
allows for the $BR(K_{S}\to 3\pi^{0})$ determination independently of $L$, $\sigma_{\phi}$, $\epsilon_{cr}$
and the $\phi \to K_{S}K_{L}$
branching fraction avoiding all the systematic effects originating from measurements of these quantities.\\
After identification of the $K_L$ meson interacting in the calorimeter the preselection is based on the number of
reconstructed $\gamma$ quanta
in each event. To this end we consider only calorimeter clusters not associated to any
track reconstructed in the drift chamber. Moreover, the reconstructed time of the cluster $t_{cl}$ should be
compatible with the time of flight of photon equal to $R_{cl}/c$, where $R_{cl}$ is the distance from the cluster
position to the interaction point\footnote{ Since the mean free path of the $K_S$ originating from the $\phi$
decay amounts to about 6~mm (the kaon velocity $\beta$~$\sim$ 0.215) , which corresponds to negligible time delay equal to
$\sim$ 100 ps,
we assume that it decays exactly in the interaction point.}
and $c$ denotes the speed of light. We assume the two times to be consistent if:
\begin{equation}
|t_{cl} - R_{cl}/c| \le \mathrm{MIN}(3.5\cdot \sigma_{t}(E_{cl}), 2~\mathrm{ns})~, 
\label{acc1}
\end{equation}
where $\sigma_t$ is the calorimeter time resolution  parametrized as a function of the cluster energy $E_{cl}$:
\begin{equation}
\sigma_{t}(E_{cl})~=~\frac{57~\mathrm{ps}}{\sqrt{E_{cl}(\mathrm{GeV})}}~\oplus~140~\mathrm{ps}~. 
\end{equation}
The cutoff on 2 ns is used to reduce the number of the machine background clusters accidentally overlapping with
the event. To this end we apply also cuts  on the minimal cluster energy and polar angle:
\begin{equation}
E_{cl} > 7~\mathrm{MeV}
\label{acc2}
\end{equation}
\begin{equation}
|\cos(\theta_{cl})| \le 0.915~\Longleftrightarrow~23.8^{\circ} \le \theta_{cl} \le 156.2^{\circ}~.
\label{acc3}
\end{equation} 
Distribution of the $\gamma$ quanta multiplicities is shown in Fig.~\ref{fig:multi}.
At this stage of analysis we select two data subsets: the signal sample which consists of events with
six reconstructed
photons and the normalization sample of $K_S\to 2\pi^0$ candidates with N$_{\gamma}$~=~4.\\
For both channels the expected background as well as the detector acceptance and the analysis efficiency
is estimated using the Monte Carlo simulations based on the GEANT3 package~\cite{geant3}. The simulations incorporate
a detailed geometry and material composition of the KLOE apparatus and all the conditions of the experiment
e.g. DA$\Phi$NE background rates, position of the interaction point and beam parameters\footnote{The detailed
description of the KLOE Monte Carlo simulation program GEANFI can be found in Ref.~\cite{Ambrosino:2004qx}.}.
\section{Comparison between data and simulations results for 4$\gamma$ events}
\label{4gammacomp}
\begin{figure}
\includegraphics[width=0.5\textwidth]{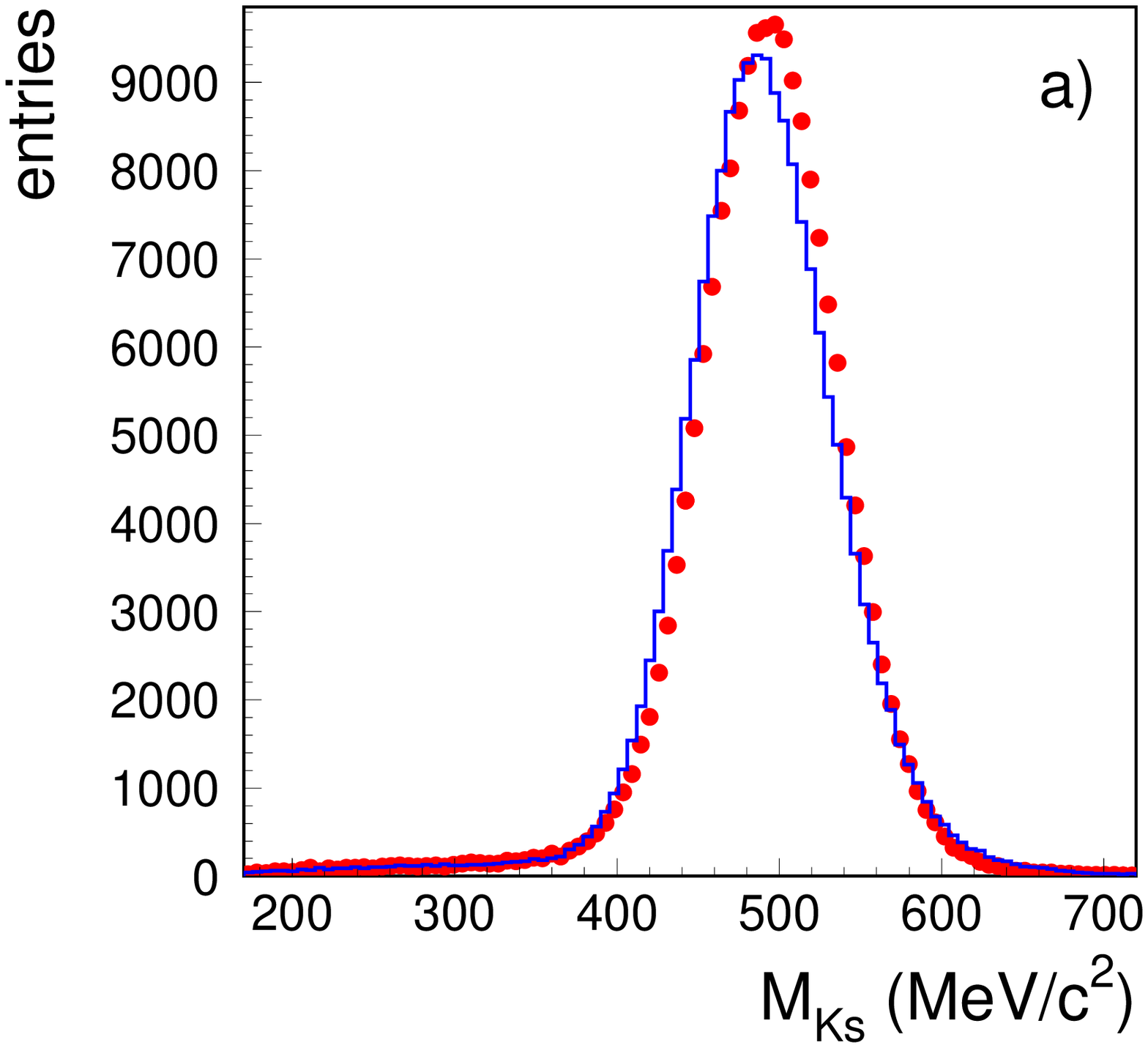}
\includegraphics[width=0.5\textwidth]{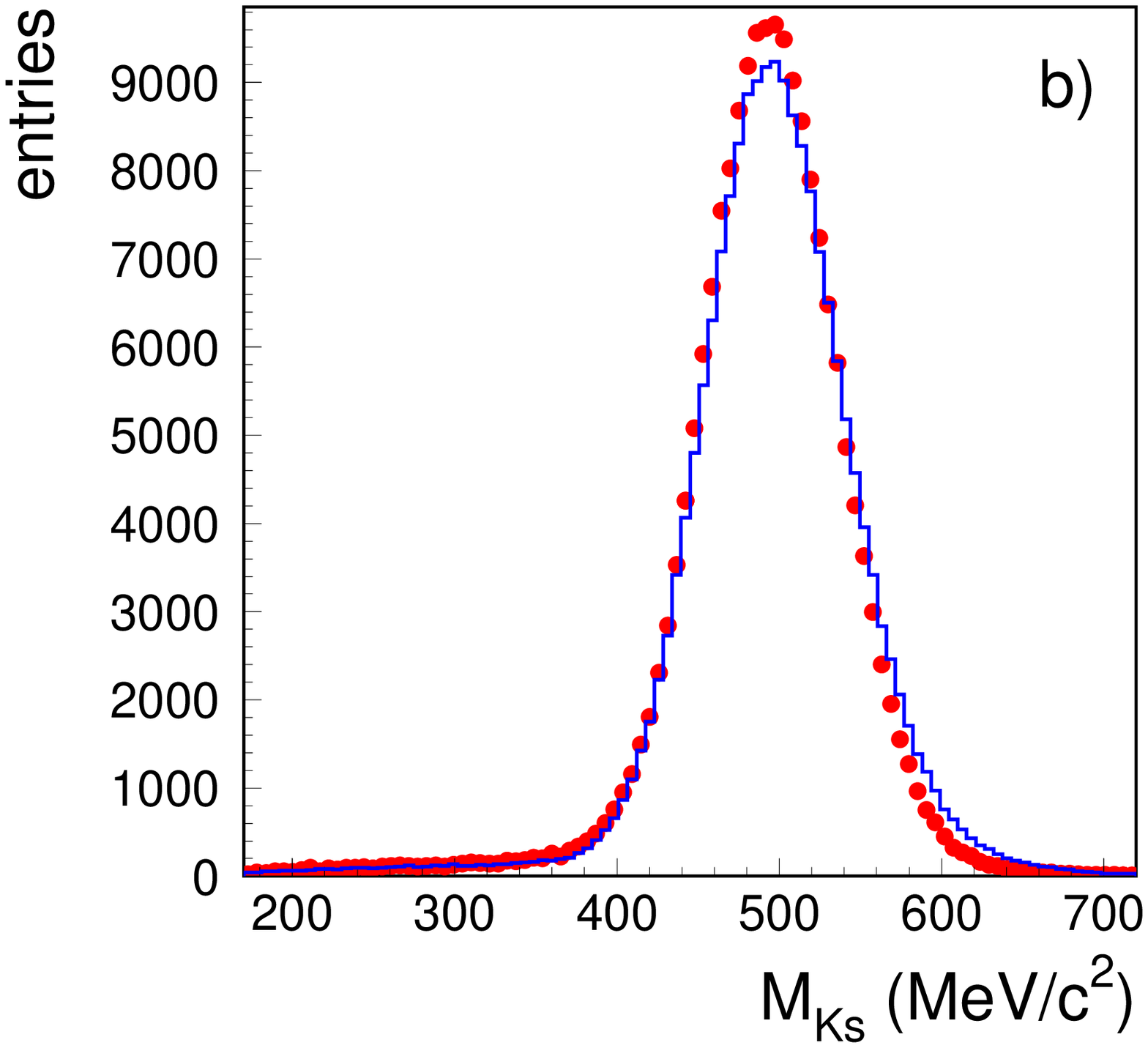}
\caption{Comparison between the reconstructed $K_S$ mass distributions  for data (red points) and simulations
(blue histogram) before a) and after b) the energy scale correction. The $K_S$ mass is reconstructed from the
$K_S \to 2\pi^0$ events.}
\label{figMks}
\end{figure}

For the search of rare processes like the $K_S \to 3\pi^0$ the estimation of the background has to be as
precise as possible. In our research to this end we use Monte Carlo simulations described briefly in the
introduction to this chapter. Moreover, the determination of the efficiencies of cuts
and discriminant analysis is also based on the simulated samples of events.
Therefore, we have checked the reliability of the KLOE Monte Carlo simulations and optimized them for
the best  possible description of the experimental data.\\
Since the reconstruction efficiency of clusters in the calorimeter is slightly higher for simulations
compared to the measured data we apply a correction determined based on the $\phi \to \pi^+\pi^-\pi^0$ 
sample\footnote{The detailed description of the cluster reconstruction efficiency studies for
data and simulations can be found in Ref.~\cite{note201}.}.
The efficiency for both data and simulations were parametrized as a function of the $\gamma$ quanta energy
and polar angle. The correction is then applied deleting randomly photons from the simulated events with a
probability equal to the ratio of efficiencies for data and simulations~\cite{note,note201}.\\
Apart from the cluster reconstruction efficiency the simulations were corrected also for the energy scale
of the reconstructed gamma quanta. The necessity of this additional correction is justified in
Fig.~\ref{figMks}a where we observe a small shift between the distributions of the reconstructed $K_S$
mass for data and simulations before the correction.
\begin{figure}
\includegraphics[width=0.5\textwidth]{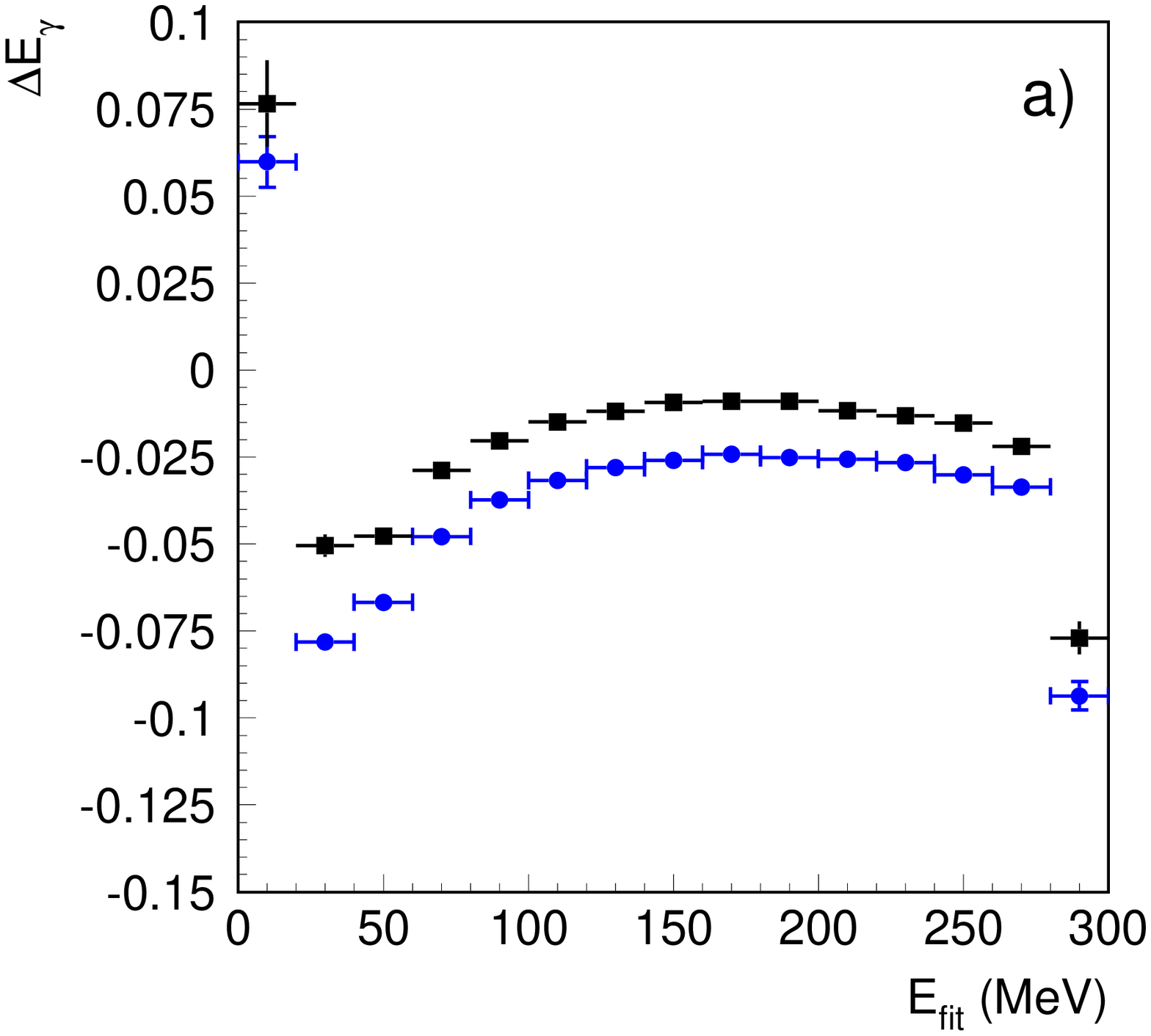}
\includegraphics[width=0.5\textwidth]{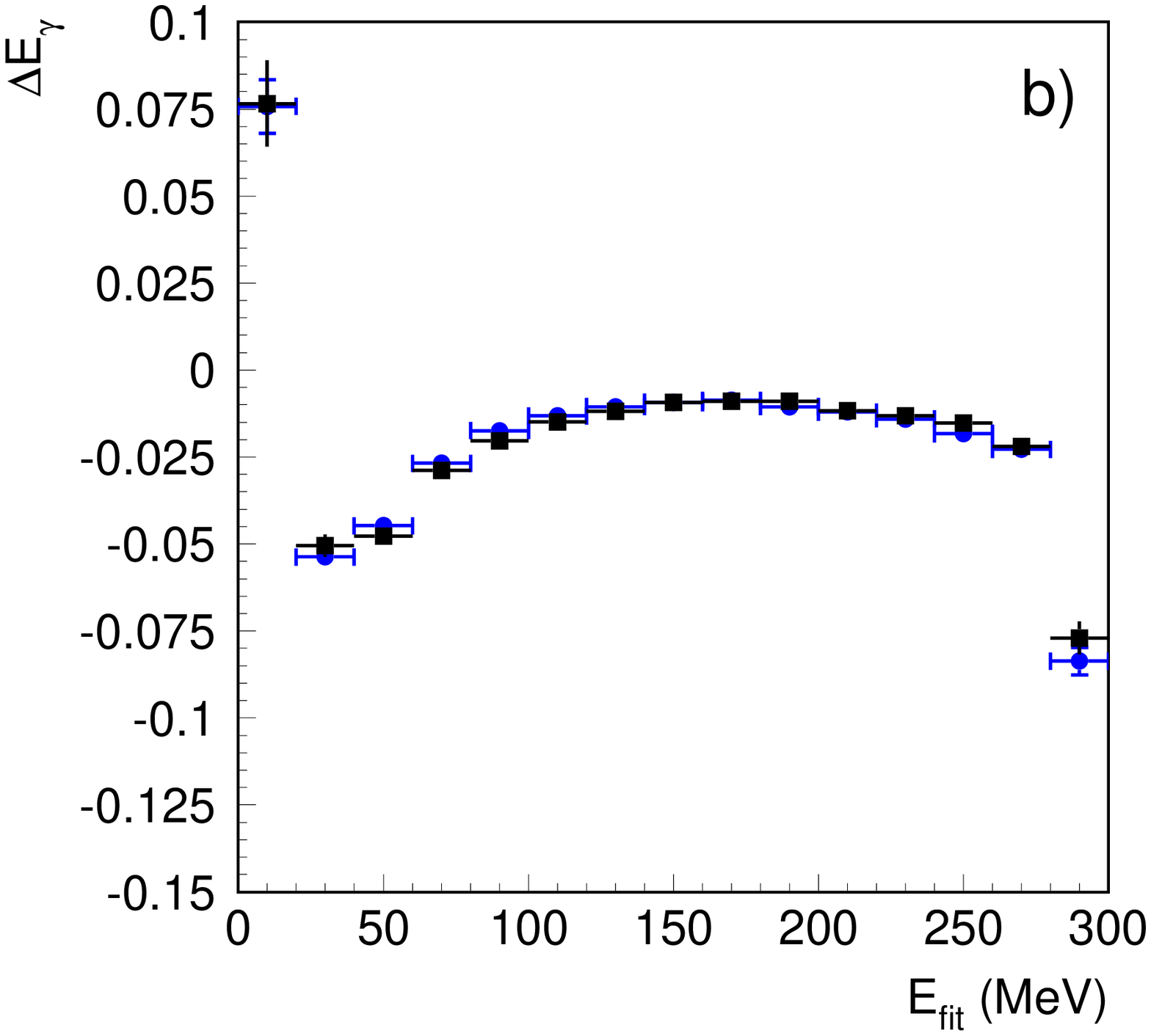}
\caption{Distributions of the mean $\Delta E_{\gamma}$ for data (black squares) and simulations (blue circles)
without energy scale correction a) and after the correction b). The values are obtained using the
fit described in the text.}
\label{figMks1}
\end{figure}
The procedure of the energy scale correction for Monte Carlo events is based on the $K_S \to 2\pi^0 \to 4\gamma$
sample which is almost background free. For both, data and simulations the following variable has been constructed:
\begin{equation}
\Delta E_{\gamma} = \frac{E_{\gamma} - E_{fit}}{E_{fit}}~,
\label{eqDEcl}
\end{equation}
where $E_{\gamma}$ is the energy of reconstructed gamma quantum and $E_{fit}$ denotes energy of the same gamma quantum
corrected by the kinematical fit procedure. The gamma quanta were then divided into groups of 20 MeV with respect to the
$E_{fit}$. For each group the $\Delta E_{\gamma}$ distribution was fitted with the Gauss function.
The mean values of the fitted Gauss distributions are shown in Fig.~\ref{figMks1}a. As it can be seen Monte Carlo
simulations systematically underestimate the data. The energy scale correction was applied by increasing the cluster
energies by a factor parametrized as a function of $E_{fit}$. For the first 20 MeV bin we started with 2.4$\%$ shift
while for every next group of clusters the correction was decreasing by a factor of 0.1$\%$. The result is presented
in Figs.~\ref{figMks}b and \ref{figMks1}b where one can see much better agreement with the data.
\\For further validation and tuning of the simulations, after applying the cluster efficiency correction,
we have determined relative fractions of number of events with given $\gamma$ quanta multiplicity $k$ with respect to
the total number of events with 3 -- 6 reconstructed photons:
\begin{equation}
F_{k} = \frac{N_{ev}(k)}{\displaystyle\sum_{i=3}^6 N_{ev}(i)}~,
\end{equation}
and compared experimental values of $F_k$ with results of simulations.
The distributions of the relative fractions as a function of the KLOE running period are presented
in Fig.~\ref{figfk}.
\begin{figure}
\includegraphics[width=0.5\textwidth]{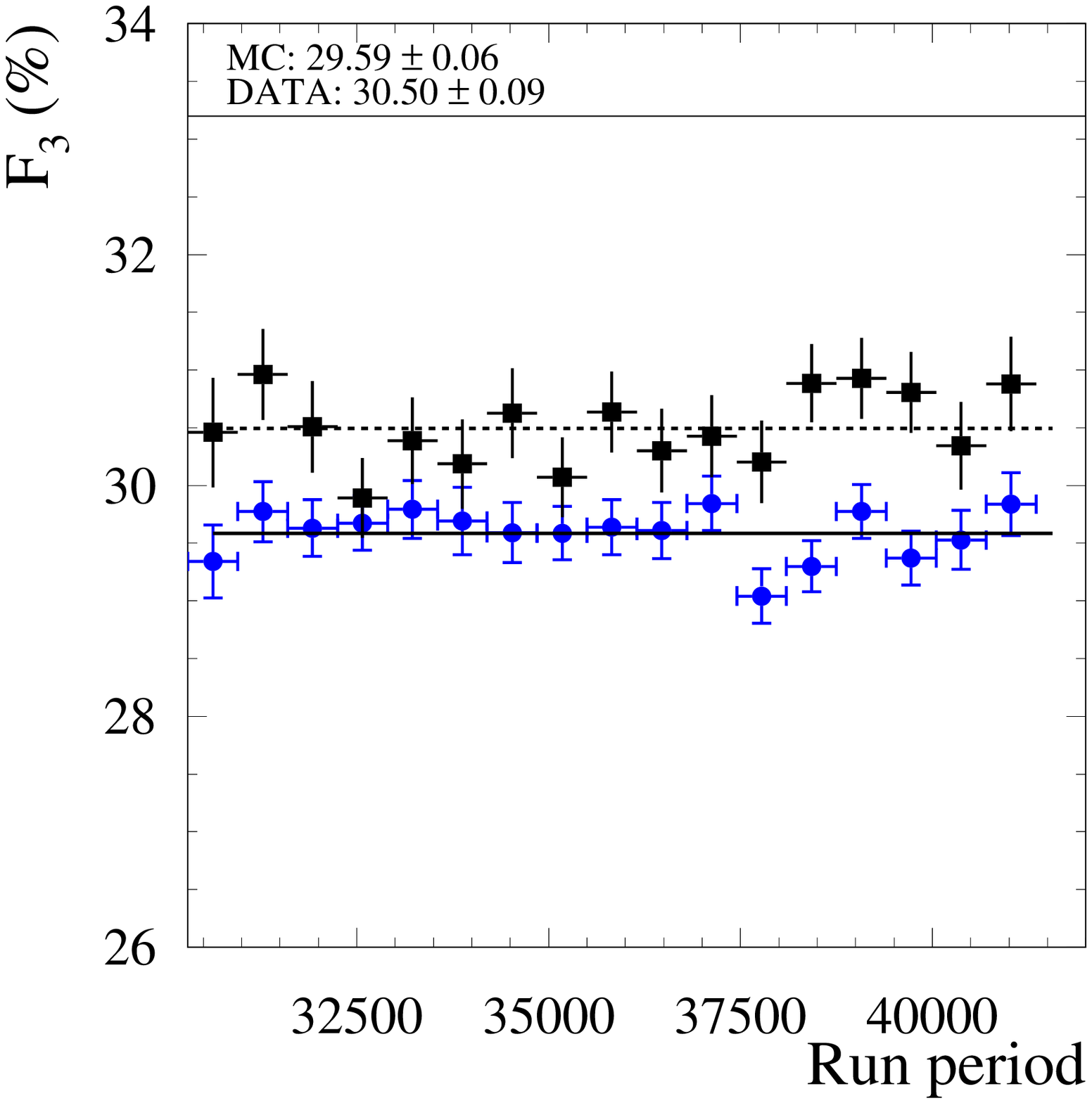}
\includegraphics[width=0.5\textwidth]{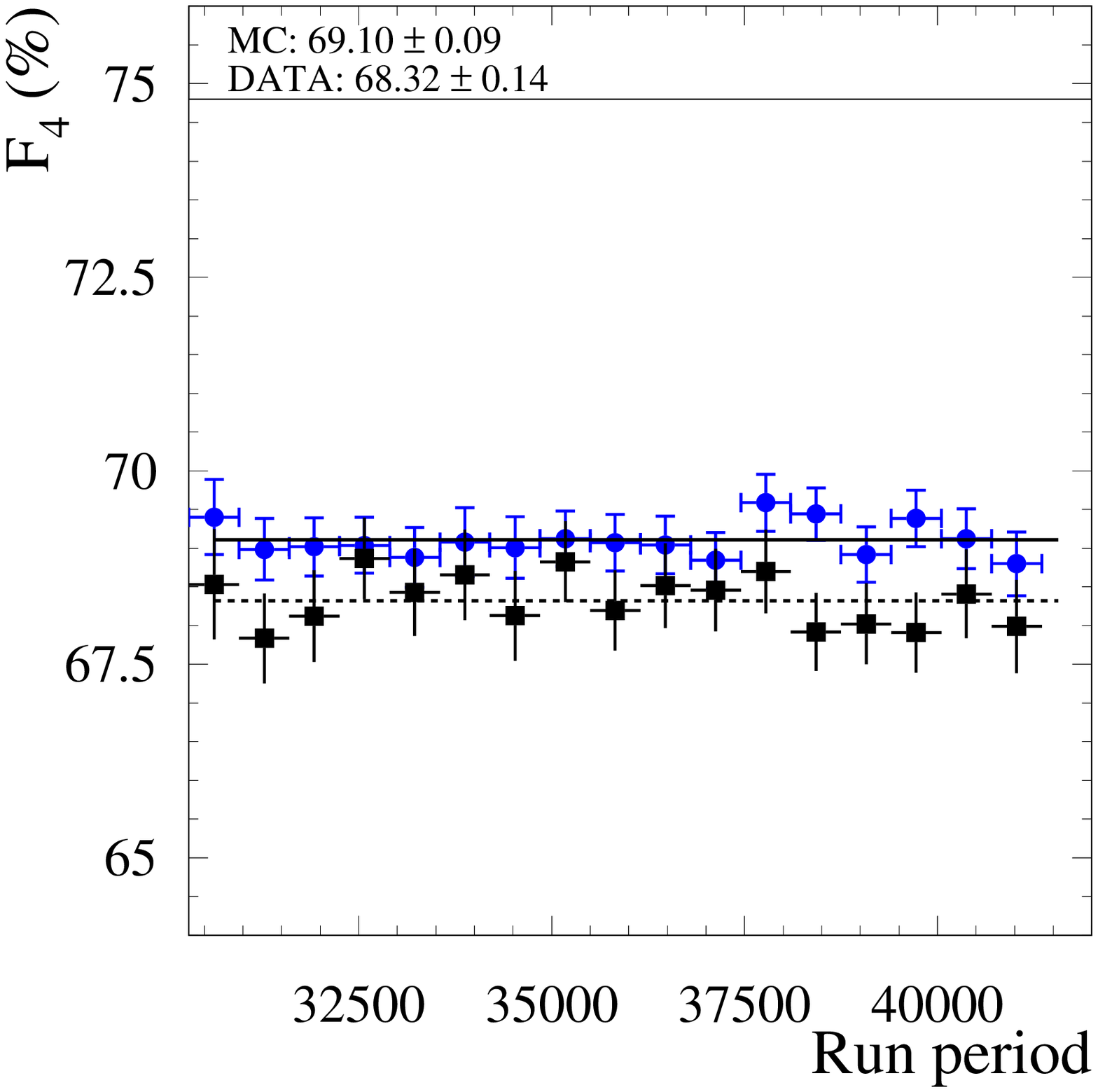}
\includegraphics[width=0.5\textwidth]{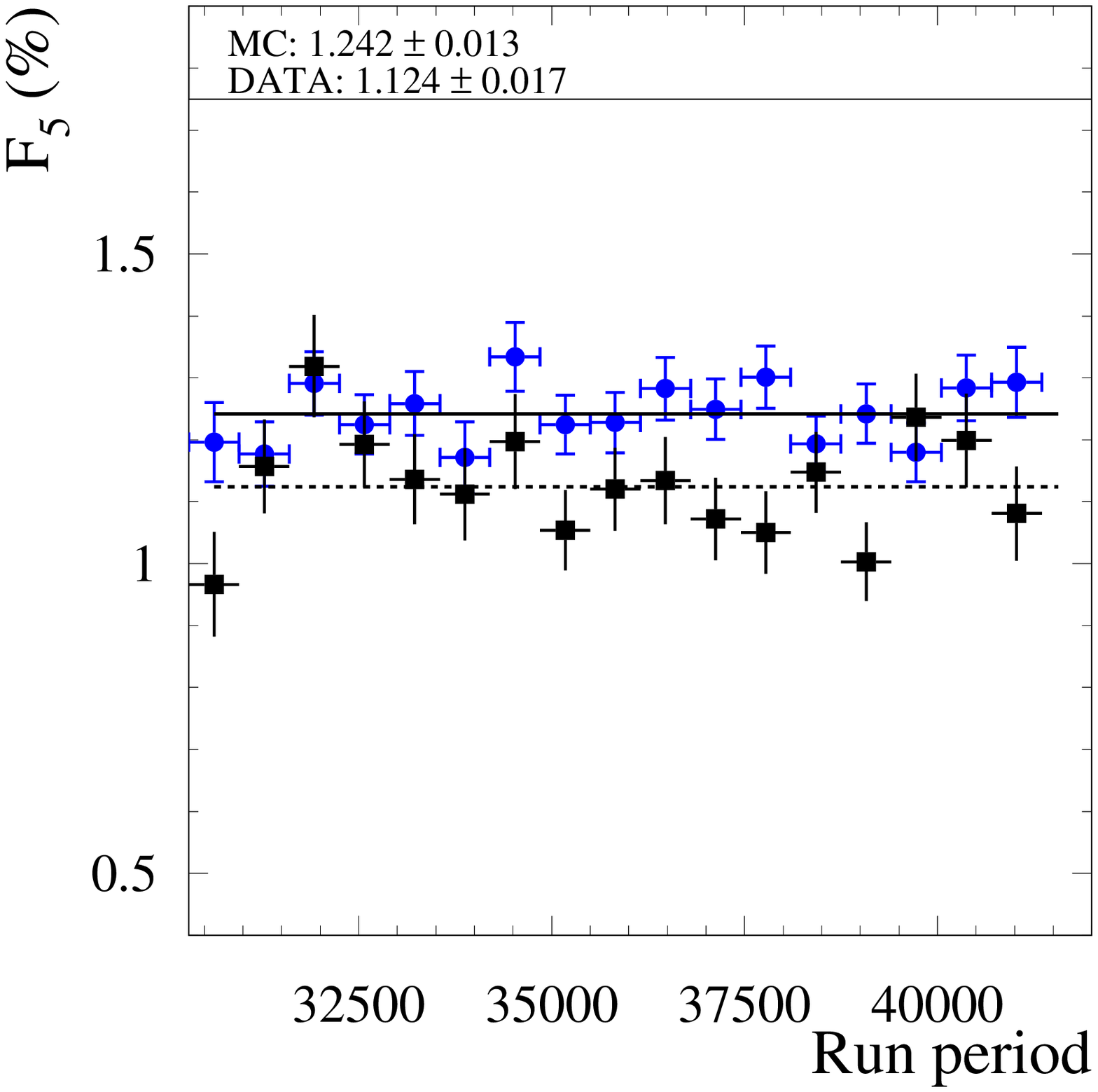}
\includegraphics[width=0.5\textwidth]{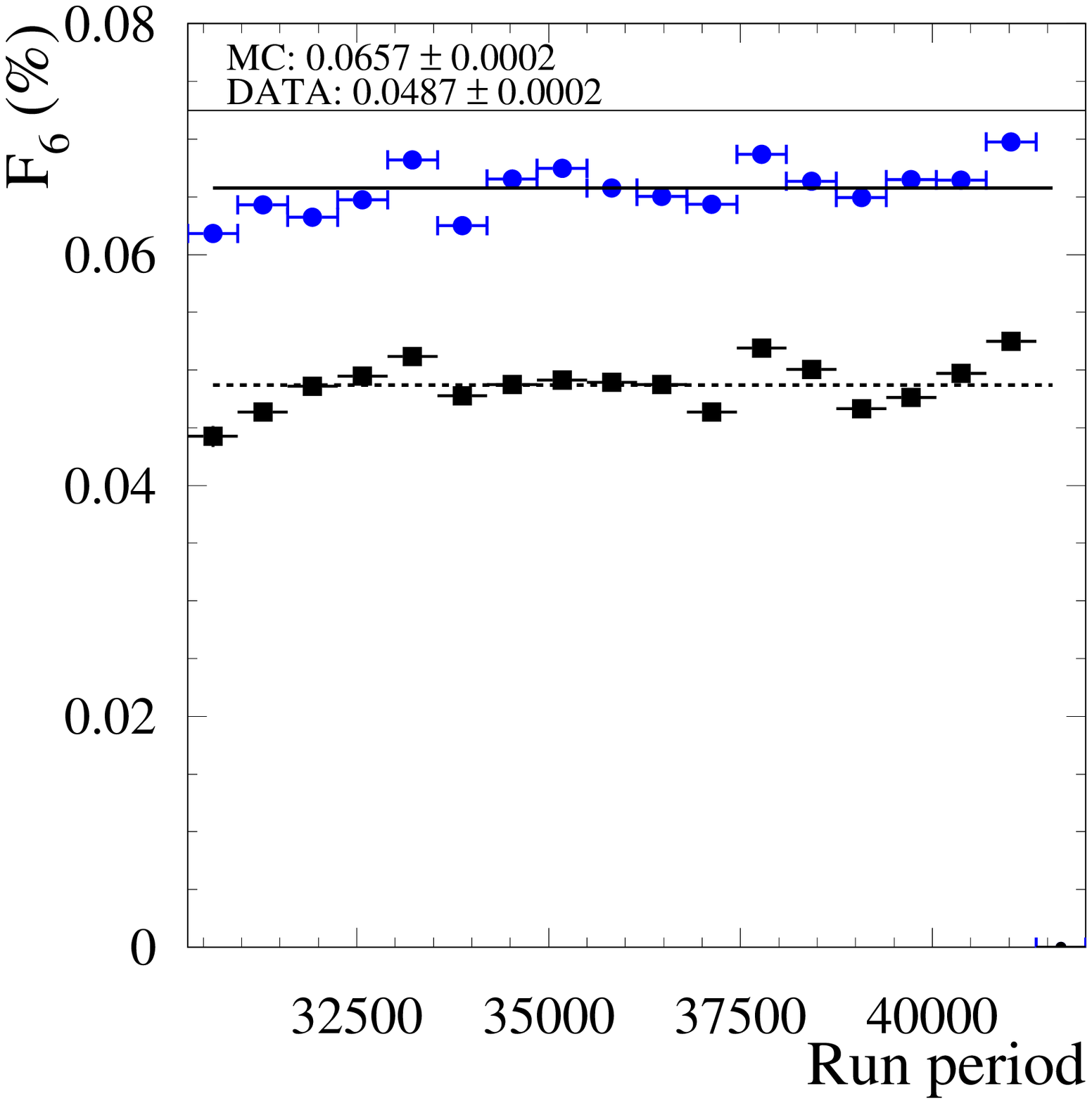}
\caption{Comparison of the $F_k$ distributions for data (black squares) and background simulations (blue circles).
The solid and dashed lines denote linear fits ($F_k = const.$) to the spectra of simulations and data, respectively.
The results of the fit are reported on the top of each distribution.}
\label{figfk}
\end{figure}
The agreement between data and simulations is reasonable apart from the most important multiplicity $F_6$ where results
of the simulations systematically overestimate the data during the whole data taking period. To understand the source of
this discrepancy we have determined the probabilities to find one ( $P_{A1}$ ) or two \mbox{( $P_{A2}$ )} accidental clusters
in the prompt time window defined in Eq.~\ref{acc1} for both data and simulations. To this end we have considered
clusters in so called early time window i.e. background clusters originating from earlier bunch crossing fulfilling
the condition:
\begin{equation}
(t_{cl} - R_{cl}/c) \in [-54,-14]~\mathrm{ns}~,
\end{equation}
which corresponds to about fifteen groups of accidental clusters sources\footnote{The minimum bunch crossing period
of DA$\Phi$NE is equal to T$_{rf}$~=~2.715 ns.}.
The times of these clusters were then shifted by a number of bunch crossing periods T$_{rf}$ obtained for
each event from the time of the earliest cluster $t_{cl}^{f}$ as follows:
\begin{equation}
t_{cl}^{n}(i)~=~t_{cl} + \mathrm{T}_{rf} \cdot \mathrm{INT}(\frac{t_{cl}^{f} - R_{cl}^{f}/c}{\mathrm{T}_{rf}})~,
\end{equation}   
and next the acceptance cuts defined in Eq.~\ref{acc1} were imposed.
This allowed to determine the fraction of events with one or more accidental clusters
in the acceptance and, as a consequence, to calculate $P_{A1}$ and $P_{A2}$ (see Tab.~\ref{tabprob}).\\
The values of $P_{A1}$ and $P_{A2}$ were next used to estimate the probability for a cluster to produce one ( $P_{S1}$ )
or more fragments ( $P_{S2}$ ) reconstructed as an additional cluster. In order to do that
the true relative photon multiplicities $F^{true}_{k}$ not affected by the accidental
activity or cluster splitting\footnote{ In the simulations the accidental clusters can be ignored
referring to the GEANT KINE indices for particles contributing to each cluster.}
were determined based on the information about the decay chain of the simulated events.
Knowing the $F^{true}_{k}$ values and the determined probabilities of accidental coincidence $P_{A1}$ and $P_{A2}$ we can fit
the measured $F_{k}$ distributions treating the $P_{S1}$ and $P_{S2}$ as the free unknown
parameters\footnote{The detailed description of the fit procedure and used probabilistic model is described in Ref.~\cite{note201},
where the technique to measure the $P_{A1}$ and $P_{A2}$ probabilities are also presented more detailed.}.
The obtained splitting probabilities are presented for both data and simulations in Tab.~\ref{tabprob}.
These results show that for the simulations there is about 50$\%$ more events with two splitted clusters and about
30$\%$ more events with one cluster originating from the machine background, which explains the discrepancy for
the 6 -- gamma events. The technique used to account for this difference is presented in chapter~\ref{rozdz6}.
\begin{table}
\begin{center}
\begin{tabular}{|c|c|c|c|c|}
\hline
 {}& $\boldsymbol{P_{A1}}$ [\%] & $\boldsymbol{P_{A2}}$ [\%] & $\boldsymbol{P_{S1}}$ [\%] & $\boldsymbol{P_{S2}}$ [\%]\\
\hline
\textbf{DATA} & 0.378 $\pm$ 0.004 & 0.025 $\pm$ 0.001 & 0.30 $\pm$ 0.01 & 0.0103 $\pm$ 0.0001\\
\hline
\textbf{SIMULATIONS}& 0.492 $\pm$ 0.004& 0.027 $\pm$ 0.001 & 0.31 $\pm$ 0.01 & 0.0156 $\pm$ 0.0002\\
\hline
\end{tabular}
\end{center}
\caption{
\label{tabprob}
The probabilities to find one ( $P_{A1}$ ) or two ( $P_{A2}$ ) accidental clusters
and to reconstruct one ( $P_{S1}$ ) or more ( $P_{S2}$ ) splitted clusters
estimated using events in out-of-time window and fit to the $F_{k}$ distributions, as it is described
in the text.}
\end{table}
\\As described in chapter~\ref{rozdz4} at the preselection stage we cut on the velocity $\beta_{cr}$
and ener\-gy $E_{cr}$ of the $K_L$ meson, therefore the simulations of its interaction in
the calorimeter should be also realistic and precise.
\begin{figure}
\centering
\includegraphics[width=0.4\textwidth]{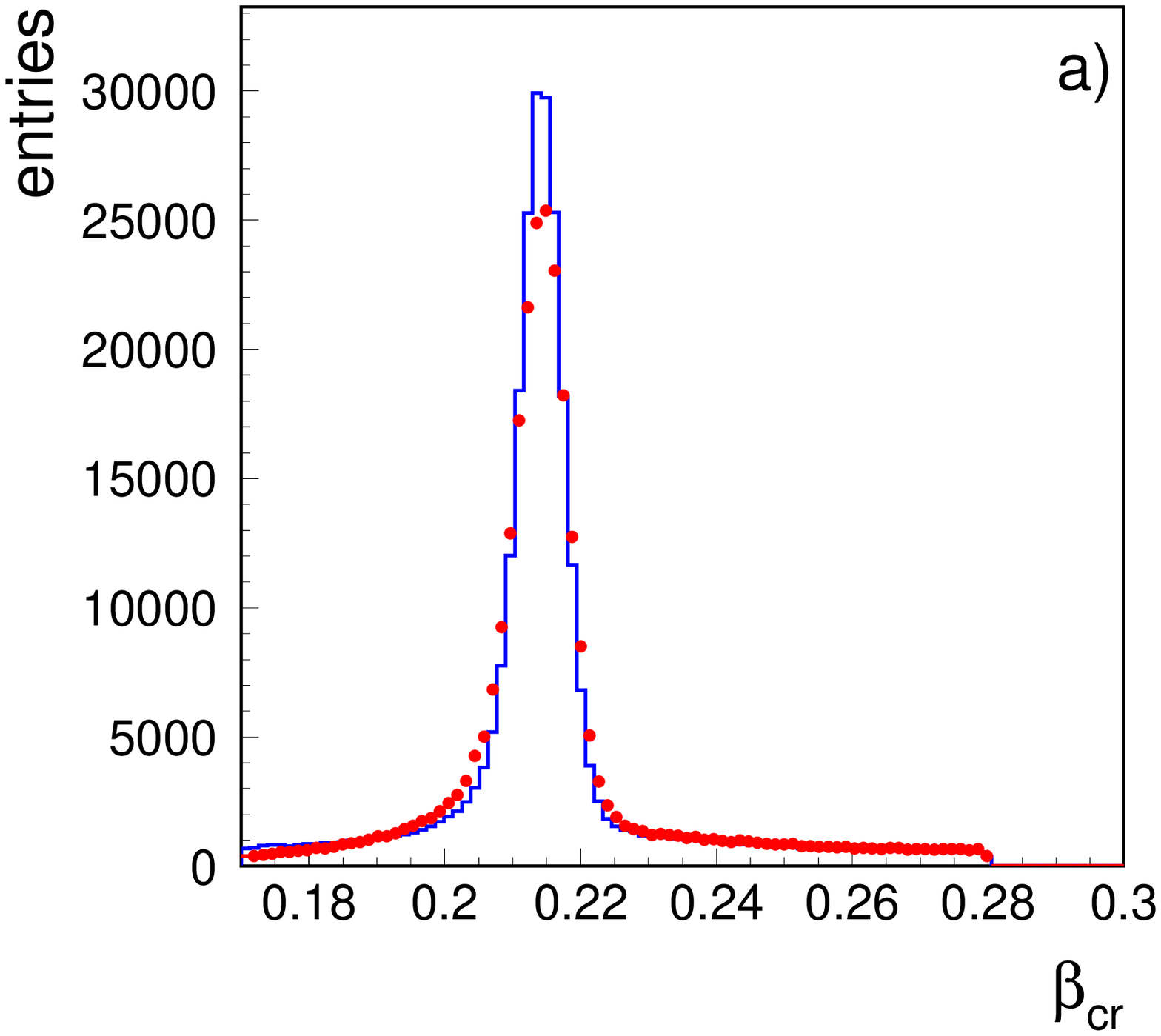}
\includegraphics[width=0.4\textwidth]{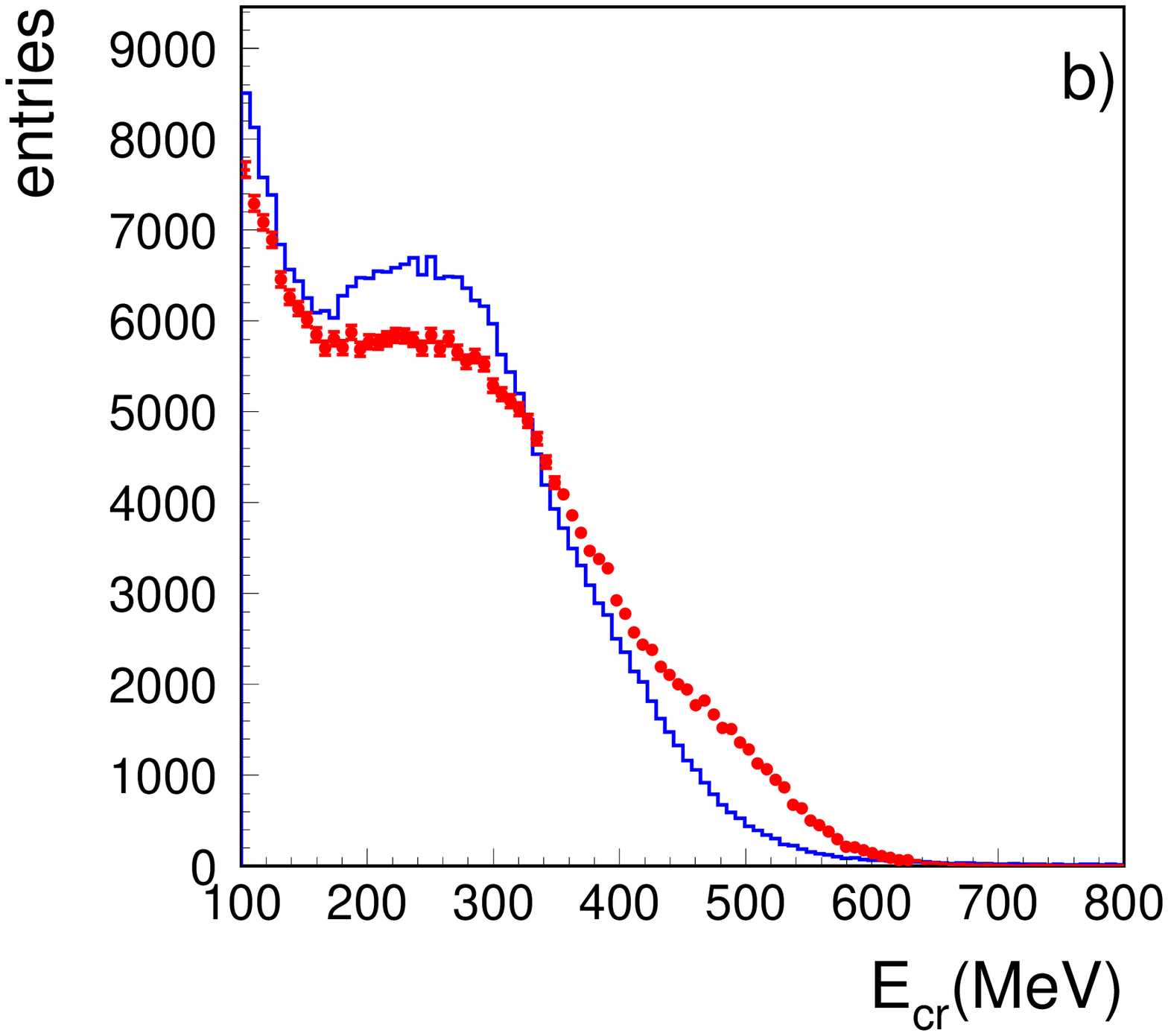}
\includegraphics[width=0.4\textwidth]{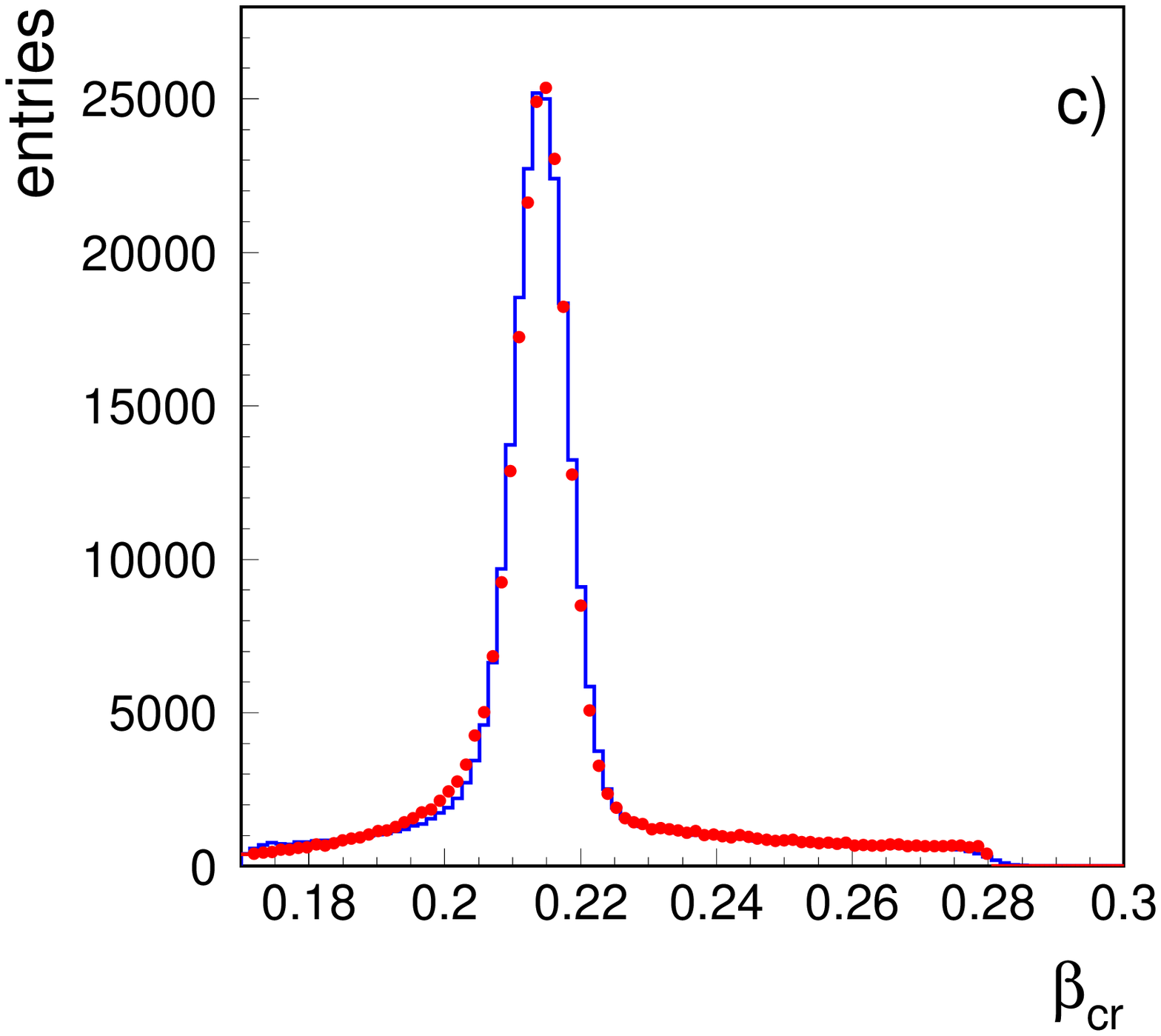}
\includegraphics[width=0.4\textwidth]{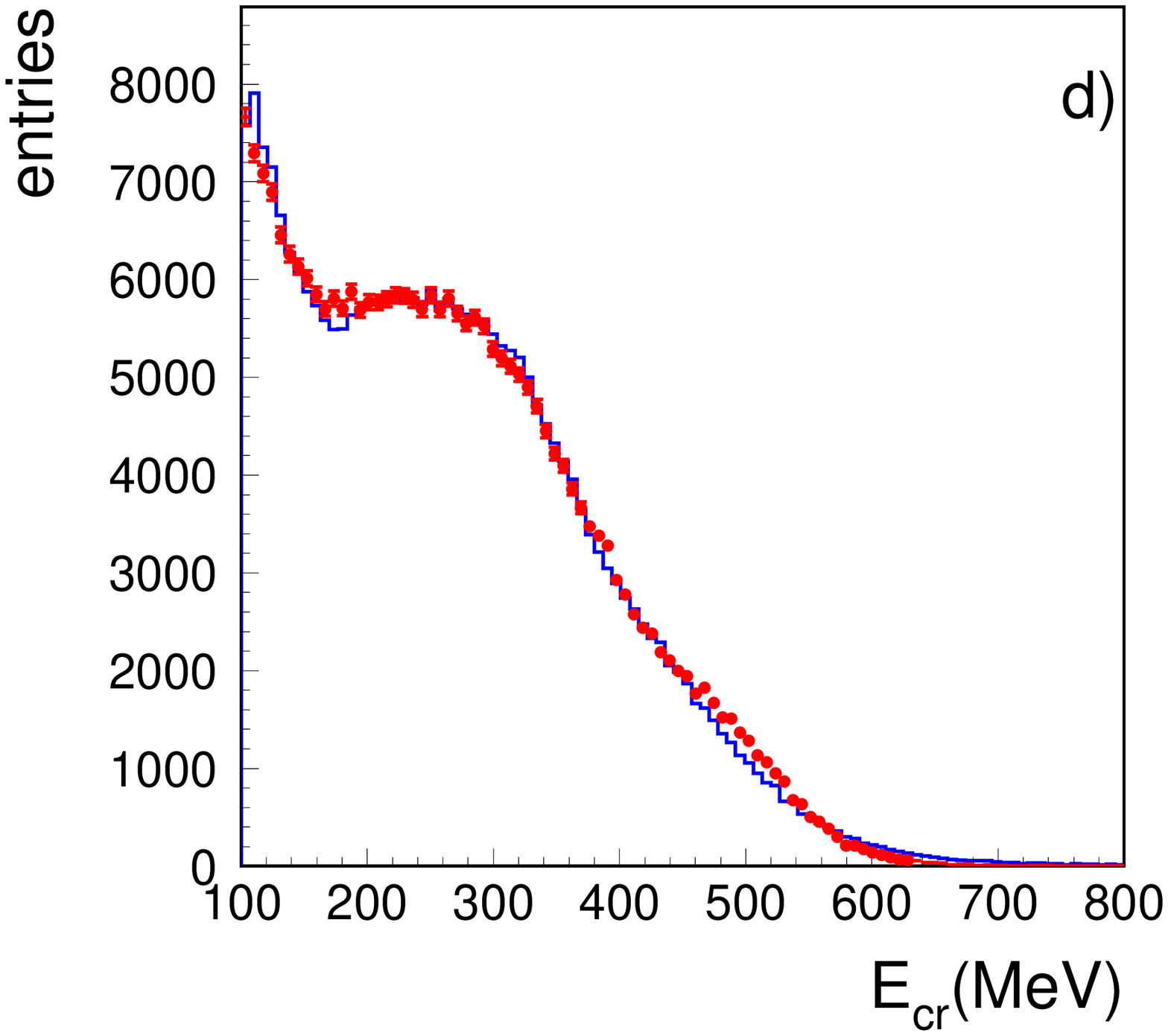}
\caption{Distributions of energy and velocity of the tagging $K_L$ meson for data (red points)
and simulations (blue histogram) before (~a and b~) and after the corrections
(~c and d~). Here only events with four gamma quanta are taken into account.}
\label{fKlcor}
\end{figure}
The comparison of the $\beta_{cr}$ and $E_{cr}$ distributions for data and Monte Carlo simulations
is presented in Fig.~\ref{fKlcor}a and ~\ref{fKlcor}b. It is clear, that the simulated $K_L$ velocity is in
a reasonable agreement with data while there is a big discrepancy in the $E_{cr}$ distribution.
Therefore before the cuts optimization another correction for the Monte Carlo simulated events had to be
applied. A small correction to the $\beta_{cr}$ was made adding a shift of 1$\%$ multiplied by a
Normal -- distributed random number.
The $E_{cr}$ was instead modified by 2.5$\%$ correction increasing every 1 MeV by 0.03$\%$.
The resulting distributions are shown in Fig.~\ref{fKlcor}c and ~\ref{fKlcor}d where one can see a much
better agreement with data.\\
Finally applying all the corrections described before we have compared some other simulated inclusive
distributions for the 4 -- gamma sample with the experimental ones. As it can be seen in
Fig.~\ref{fig4gamma} the agreement is reasonable so we can proceed with further analysis and
counting of the $K_S \to 2\pi^0$ events.
\begin{figure}
\centering
\includegraphics[width=0.4\textwidth]{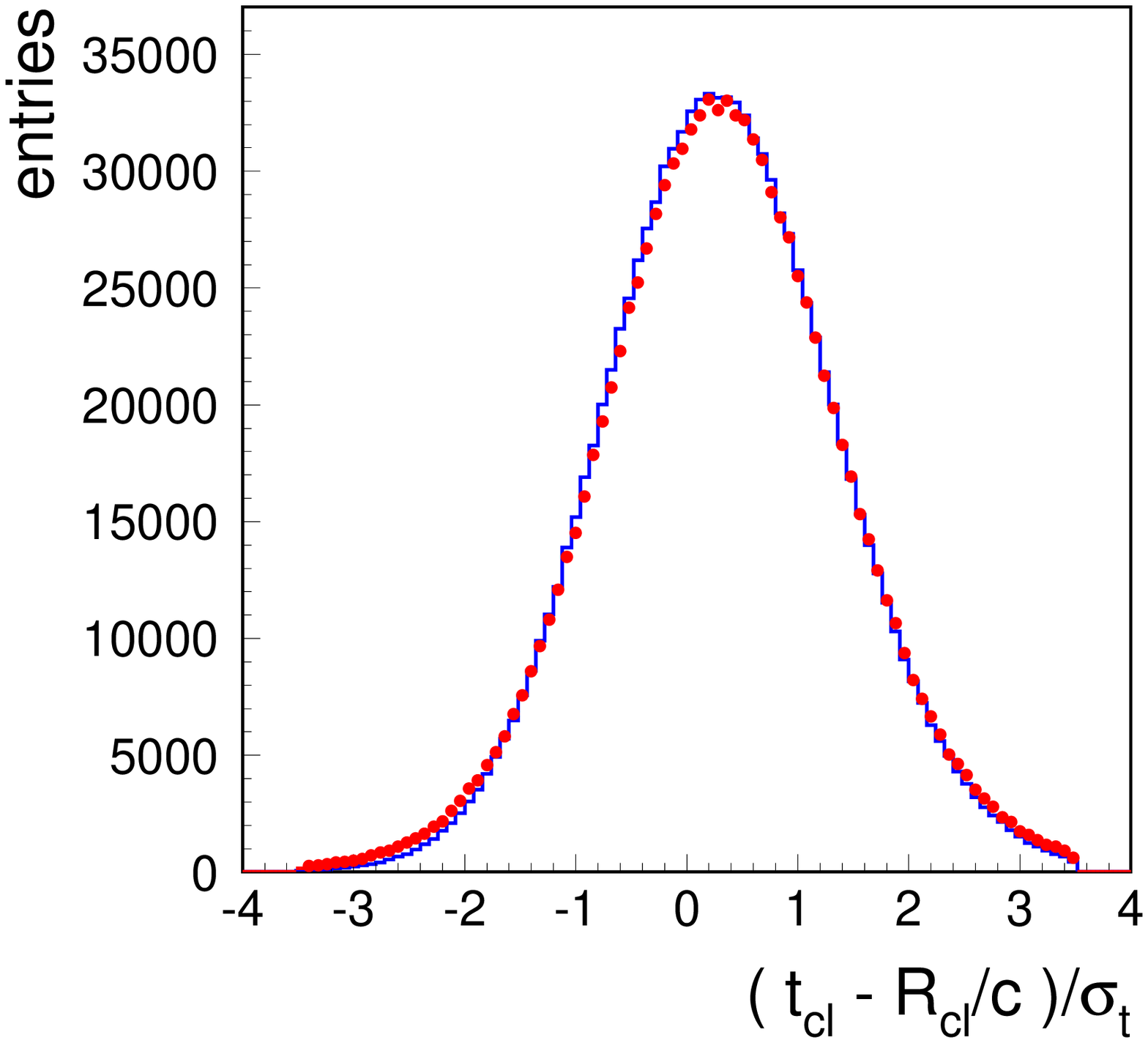}
\includegraphics[width=0.4\textwidth]{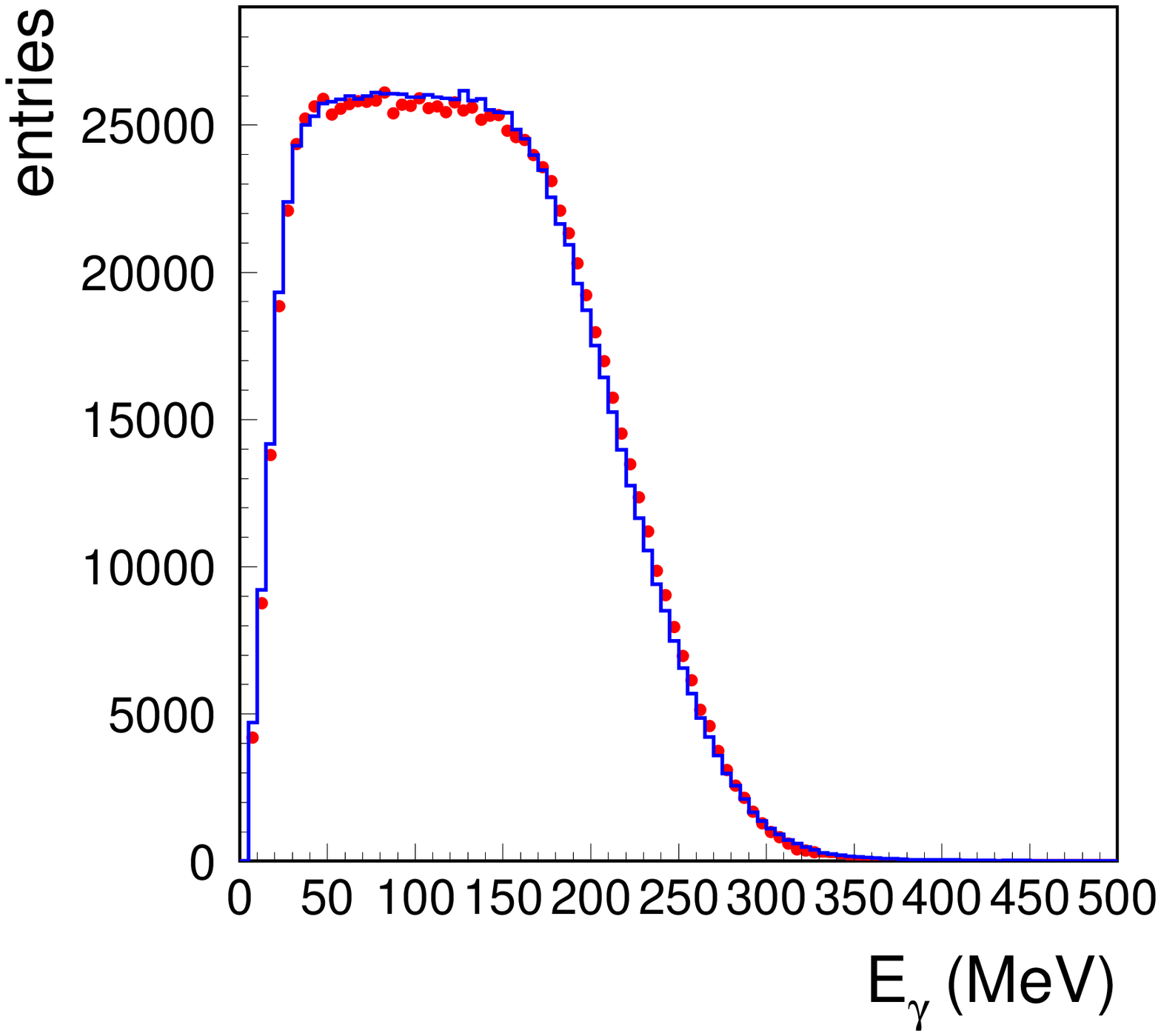}
\caption{Inclusive distributions for the $K_s \to 2\pi^0 \to 4\gamma$ decay after all corrections
applied to the Monte Carlo simulations (blue histogram). The red points denote the experimental result.}
\label{fig4gamma}
\end{figure}
\section{Background estimation and counting of the $K_S\to 2\pi^0 \to 4\gamma$ events}
Before the final counting of the $K_S \to 2\pi^0$ sample one has to estimate
the number of events originating from processes different than $K_S \to 2\pi^0$
for which we have found four reconstructed gamma quanta. To this
end we have used the Monte Carlo simulated events of $\phi$ decays for which the true decay chain is
known. Based on this information we have found that the background events constitute
a negligible fraction of the 4 -- gamma sample amounting to about 0.1$\%$\footnote{These are
above all the $\phi \to K^+K^-$ events.}. This allows for
counting of the $K_S \to 2\pi^0$ candidates without any further cuts. With the tight
requirements for reconstructed $K_L$ energy \mbox{( $E_{cr} > 150~\mathrm{MeV}$} and velocity
\mbox{($0.200 <\beta_{cr} < 0.225$ )}
we have found the following number of events with four reconstructed gamma quanta\footnote{
The justification of using this set of cuts will be presented in chapter~\ref{rozdz6}.}:
\begin{equation}
N_{2\pi}~=~(7.533 \pm 0.018) \times 10^7~.
\end{equation}
The Monte Carlo simulations allow also for the determination of the $K_S \to 2\pi^{0} \to 4\gamma$
selection efficiency. To this end we have used a simulated sample of $N_{tot}~=~59141$ $K_S \to 2\pi^0$
events fulfilling above mentioned tight conditions for $\beta_{cr}$ and $E_{cr}$.
Next we apply the acceptance cuts and count events with a given number of reconstructed gamma quanta
(see Tab.~\ref{tab:eff2pi}). The efficiency is then defined as:
\begin{equation}
\epsilon_{2\pi}~=~\frac{N^{rec}_{4\gamma}}{N_{tot}}~.
\label{eq:ep2pi}
\end{equation}
$N^{rec}_{4\gamma}$ denotes the number of events with four reconstructed gamma
quanta and $N_{tot}$ stands for the total number of simulated $K_S \to 2\pi^0$ events.
In both cases we count only events fulfilling the tight $K_L$ tag conditions.
\begin{table}[h]
\begin{center}
\begin{tabular}{|c|c|c|c|c|c|c|c|c|c|}
\hline
$\boldsymbol{k}$ & 0 & 1 & 2 & 3 & \textbf{4} & 5 & 6 & 7 & 8\\
\hline
$\boldsymbol{N^{rec}_{k\gamma}}$ & 25 & 181 & 3059 & 16185 & \textbf{39012} & 636 & 40 & 1 & 2\\
\hline
$\boldsymbol{N^{rec}_{k\gamma}/N_{tot}} [\%]$ & 0.043 & 0.31 & 5.17 & 27.37 & \textbf{65.97} & 1.08 & 0.068 & 0.0017 & 0.0034\\
\hline
\end{tabular}
\end{center}
\caption{
\label{tab:eff2pi}
The number of events $N^{rec}_{k\gamma}$ with a given multiplicity of reconstructed clusters $k$
for $N_{tot}$ = 59141 simulated $\phi \to K_SK_L \to 4\gamma K_L$ events with 0.200 < $\beta_{cr}$ < 0.225 and
$E_{cr}$ > 150 MeV.}
\end{table}
Taking into account the estimated selection efficiency: $\epsilon_{2\pi} = 0.660 \pm 0.002_{stat}$ we can
determine the final number of produced events:
\begin{equation}
N_{norm} = \frac{N_{2\pi}}{\epsilon_{2\pi}} = (1.142 \pm 0.005) \cdot 10^8.
\end{equation}
This number will be used for the normalization of the $K_S \to 3\pi^0$ branching ratio.
\chapter{Search for the $K_{S}\to 3\pi^{0} \to 6\gamma$ signal}
\label{rozdz6}
After the $K_S$ tagging via the $K_L$ interactions in the calorimeter the selection
of the $K_{S}\to 3\pi^{0}$ decay is performed by searching for six photons from the decay of the pions.
After the preselection with the conservative\footnote{
Such requirements have been used in the previous analysis of the 
KLOE data~\cite{Matteo}.}
$K_{S}$ tag requirements\\
\mbox{( $E_{cr} > 100~\mathrm{MeV}$ and $0.17 <\beta_{cr} < 0.28$ )}
we have found 76689 events with six reconstructed $\gamma$ quanta. For these events we perform
further discriminant analysis to increase the signal to background ratio. As it was in the case
of the $K_S \to 2\pi^0$ normalization sample we have considered only calorimeter clusters not
associated to any track reconstructed in the drift chamber and imposed acceptance cuts defined
with Eqs.~\ref{acc1}, \ref{acc2} and \ref{acc3}.\\
As it was mentioned in chapter~\ref{rozdz5}, the expected number of background events as well as
the analysis efficiency is estimated using the KLOE Monte Carlo simulations.
All the processes contributing to the background were simulated with statistics two times larger than
the measured data. Moreover, for the acceptance and the analysis efficiency studies the dedicated $K_{S}\to 3\pi^{0}$
signal simulations were performed. The signal is generated taking into account the 
conditions of the KLOE experiment and assuming branching ratio equal to the best known upper limit~\cite{Matteo}
increased by a factor of 30.
Background to the searched $K_{S}\to 3\pi^{0} \to 6\gamma$ signal originates predominantly from the $K_S\to 2\pi^0$ events.
The four photons from this decay can be reconstructed as six due to fragmentation of the electromagnetic
showers (so called splitting). These events are characterized by two clusters with low energy and
position near the place where the true photon hits the calorimeter. The additional fake clusters can be generated also
by the accidental coincidence between the $\phi$ decay event and the DA$\Phi$NE background.
The other source of background are $\phi \to K_{S}K_{L} \to \pi^+\pi^-,3\pi^0$ events. Methods of suppressing
this kind of background will be discussed in the next section. 
\section{Discriminating variables and the signal region definition}
\subsection{Rejection of the $K_L\to 3\pi^0$ events}
The most dangerous source of background for our analysis are the
$\phi \to K_{S}K_{L} \to (K_S \to \pi^+\pi^-, K_L \to 3\pi^0)$ events.
In this case one of the charged pions can interact in the DA$\Phi$NE low -- beta insertion
quadrupoles producing neutral particles which may ultimately simulate the signal of $K_L$
interaction in the calorimeter, while the $K_L$ meson decays close to the interaction point
and produces six photons. To suppress this kind of background we first reject events
with charged particles coming from the vicinity of the interaction region.
Since $K_S$ decays near the interaction point we reject events with at least one track satisfying:
\begin{eqnarray}
\label{trv1}
\rho_{T}~=~\sqrt{x^{2}_{PCA} + y^{2}_{PCA}} < 4~\mathrm{cm}\\
|z_{PCA}| < 10~\mathrm{cm}~,
\end{eqnarray}
where PCA denotes the point of the closest approach of the reconstructed trajectory to the interaction region.
In principle one could reject all events with the track reconstructed in the drift chamber. However,
this would decrease the reconstruction efficiency of the $K_S \to 3\pi^0$ signal because of the relatively
high probability of an accidental coincidence of the real event with tracks generated by DA$\Phi$NE background.
The choice of conditions defined in Eq.~\ref{trv1} minimizes the loss of the signal events due to this effect.
\begin{figure}[h]
\centering
\includegraphics[width=0.45\textwidth]{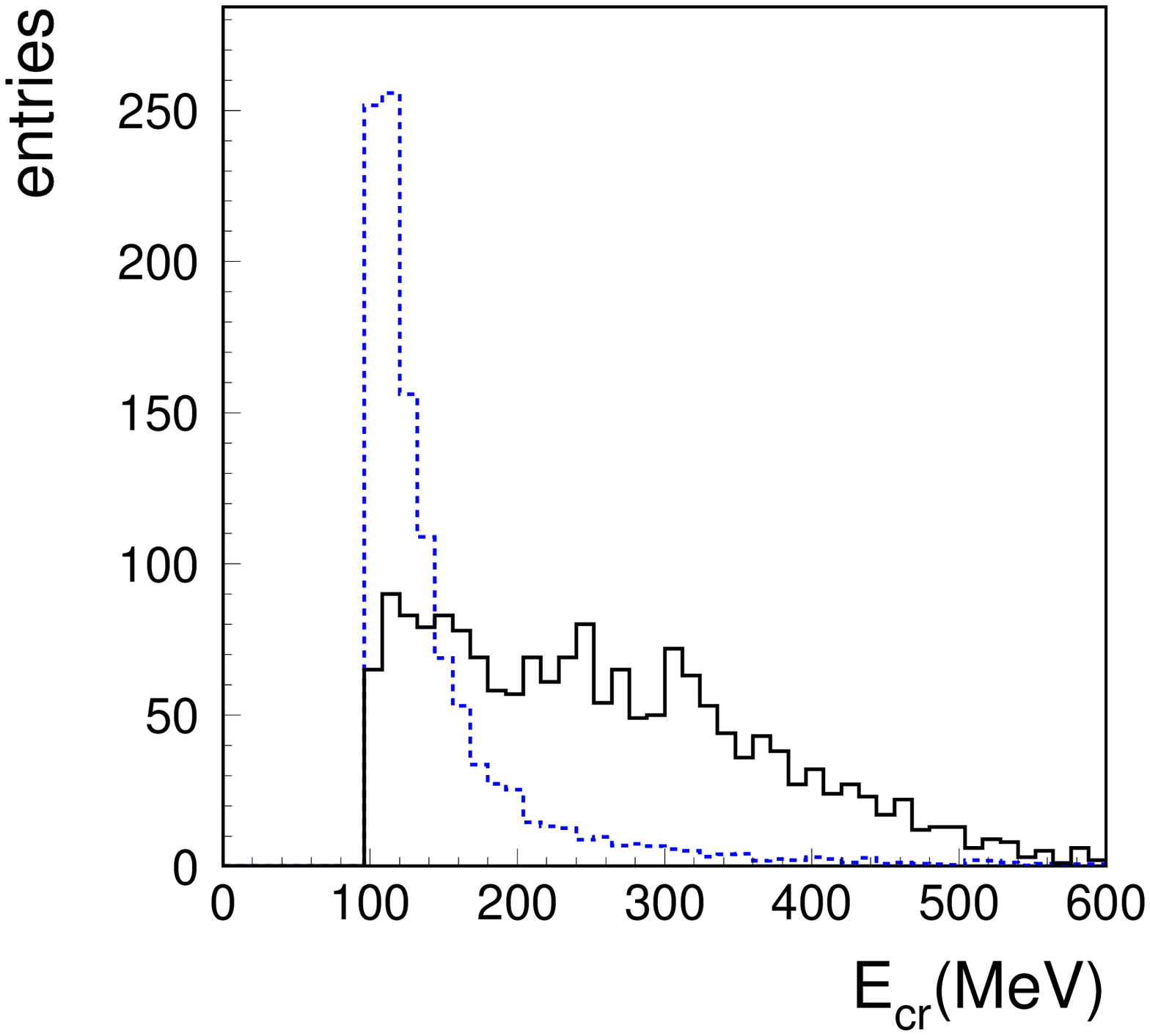}
\includegraphics[width=0.45\textwidth]{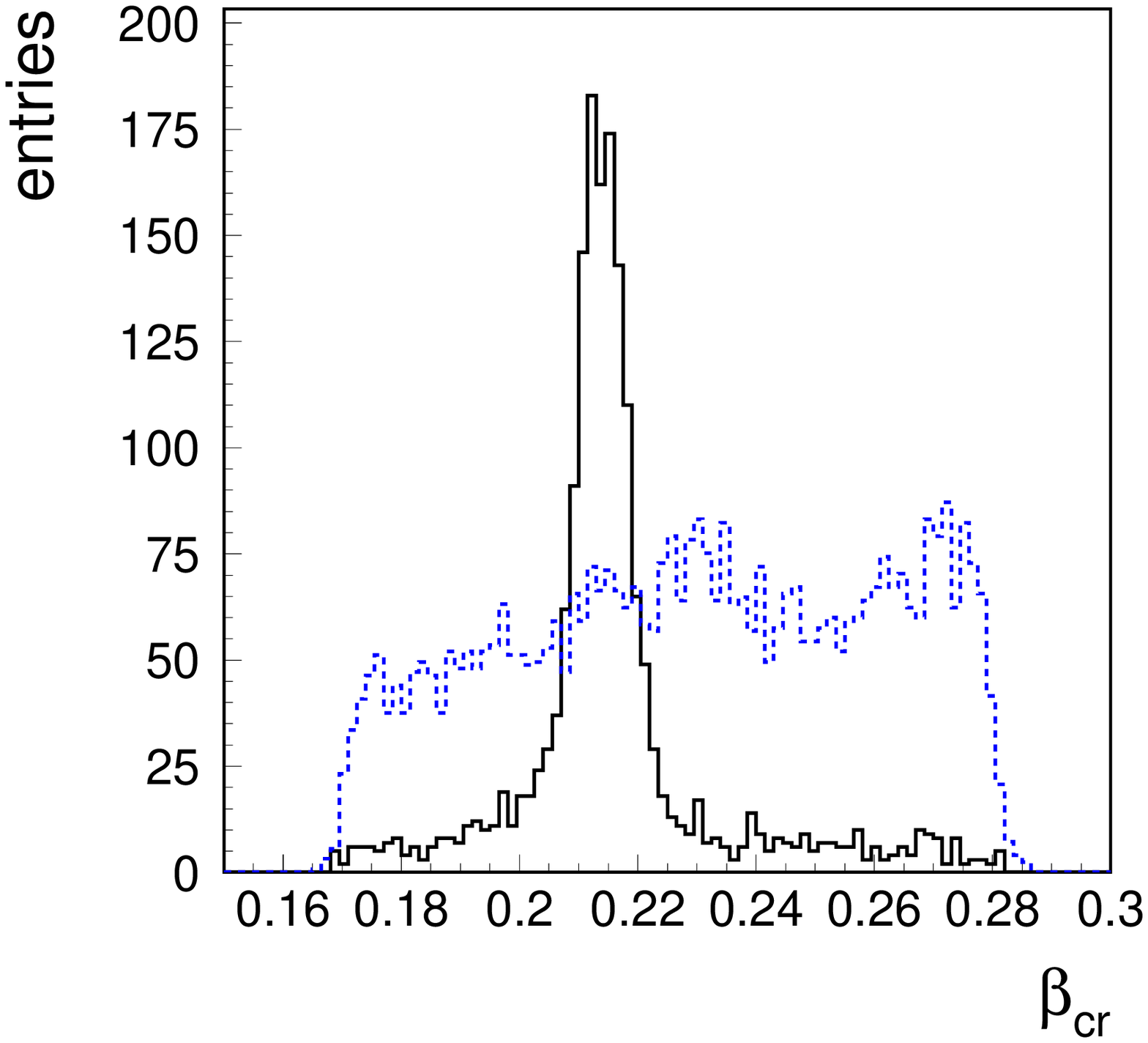}
\caption{Simulated distributions of the reconstructed $K_L$ energy $E_{cr}$ and velocity $\beta_{cr}$ for the
$\phi \to K_{S}K_{L} \to (K_S \to 3\pi^0, K_L) \to 6\gamma$
signal (solid) and $\phi \to K_{S}K_{L} \to (K_S \to \pi^+\pi^-, K_L \to 3\pi^0) \to \pi^+\pi^- 6\gamma$
background events (dashed) after the conservative cuts on reconstructed $\beta_{cr}$ and $E_{cr}$
described in Eq.~\ref{eq:consKlcr}.}
\label{hardKl}
\end{figure} 
\\As it is presented in Fig.~\ref{hardKl} the distribution of the velocity $\beta_{cr}$ of $K_L$ candidates 
reconstructed from signals in the calorimeter for events with fake ,,$K_L$ -- crash'' is relatively flat,
while there is a clear peak around 0.215 for the signal events. Similarly there is a substantial difference
in the reconstructed energy spectrum of the interacting $K_L$ mesons and fake ,,$K_L$ -- crash'' events.
Thus, tightening the cuts on $E_{cr}$ and $\beta_{cr}$ allows to reject almost all events belonging to this
background category:
\begin{figure}
\centering
\includegraphics[width=0.5\textwidth]{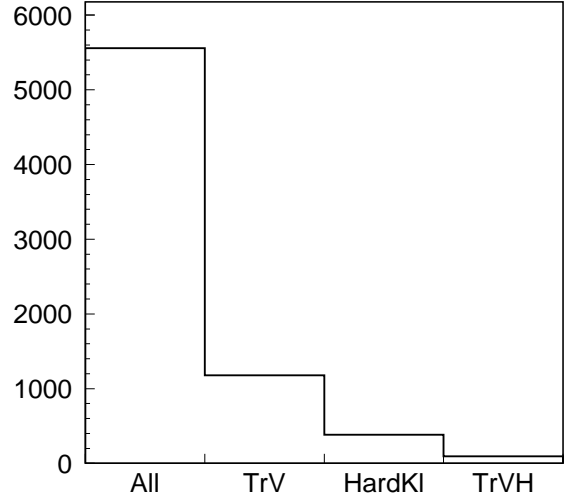}
\caption{The rejection efficiency for the $\phi \to K_{S}K_{L} \to \pi^+\pi^-,3\pi^0$ reaction chain
after cutting events with tracks (TrV), after the tight requirements for $K_L$ interaction in the calorimeter
(HardKl)
and both cuts (TrVH). The result was obtained based on the simulated background sample.}
\label{hardKlpor}
\end{figure}
\begin{eqnarray}
E_{cr} > 150~\mathrm{MeV}
\nonumber \\
0.200< \beta_{cr}< 0.225~.
\label{hardKlc}
\end{eqnarray}
These cuts reduce  the $K_S$ tagging efficiency from 34$\%$ to 23$\%$ but we gain a big reduction factor
(about 60) on the most important source of background (see Fig.~\ref{hardKlpor}).
\subsection{The kinematical fit}
\begin{figure}
\centering
\includegraphics[width=0.49\textwidth]{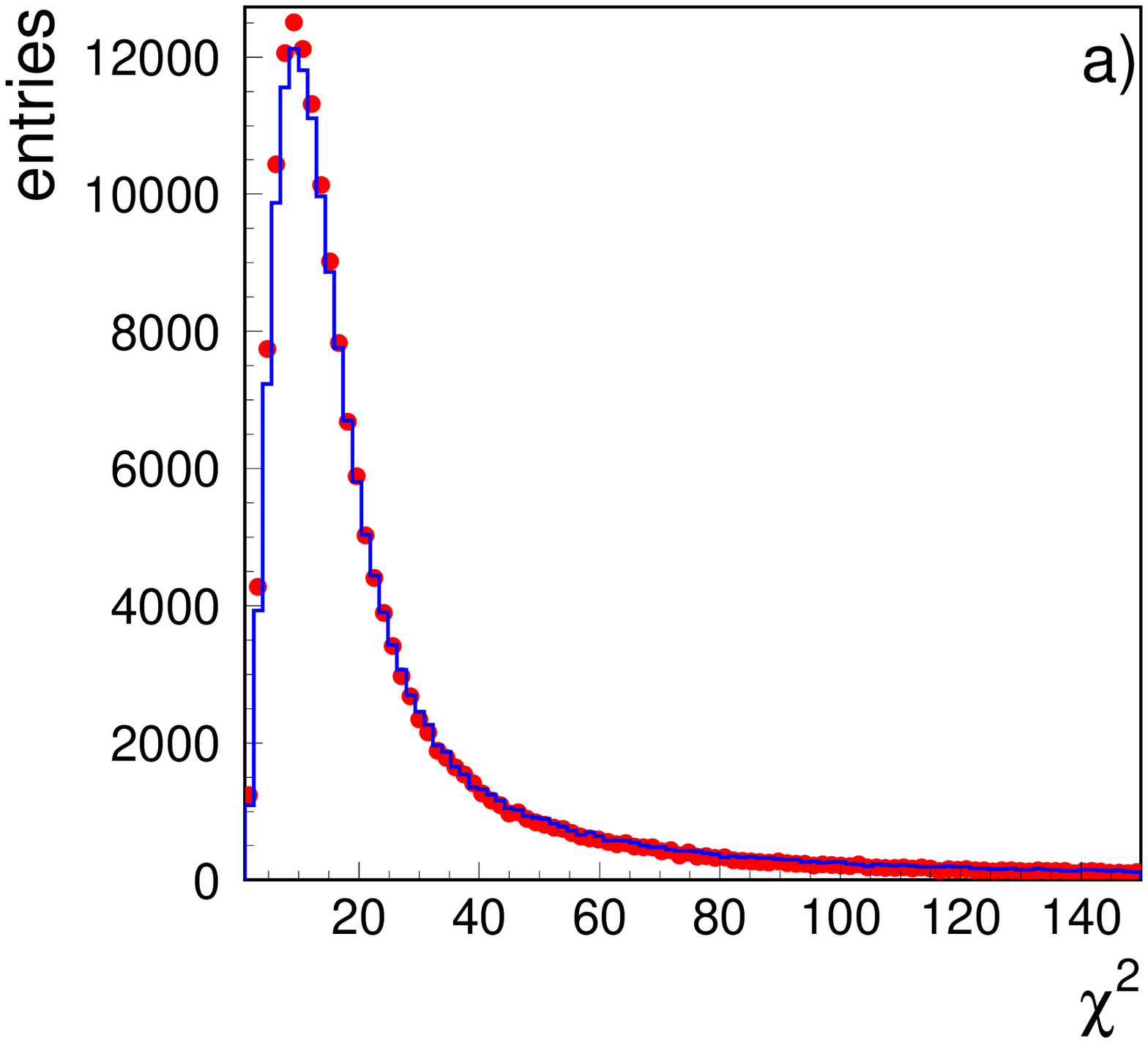}
\includegraphics[width=0.49\textwidth]{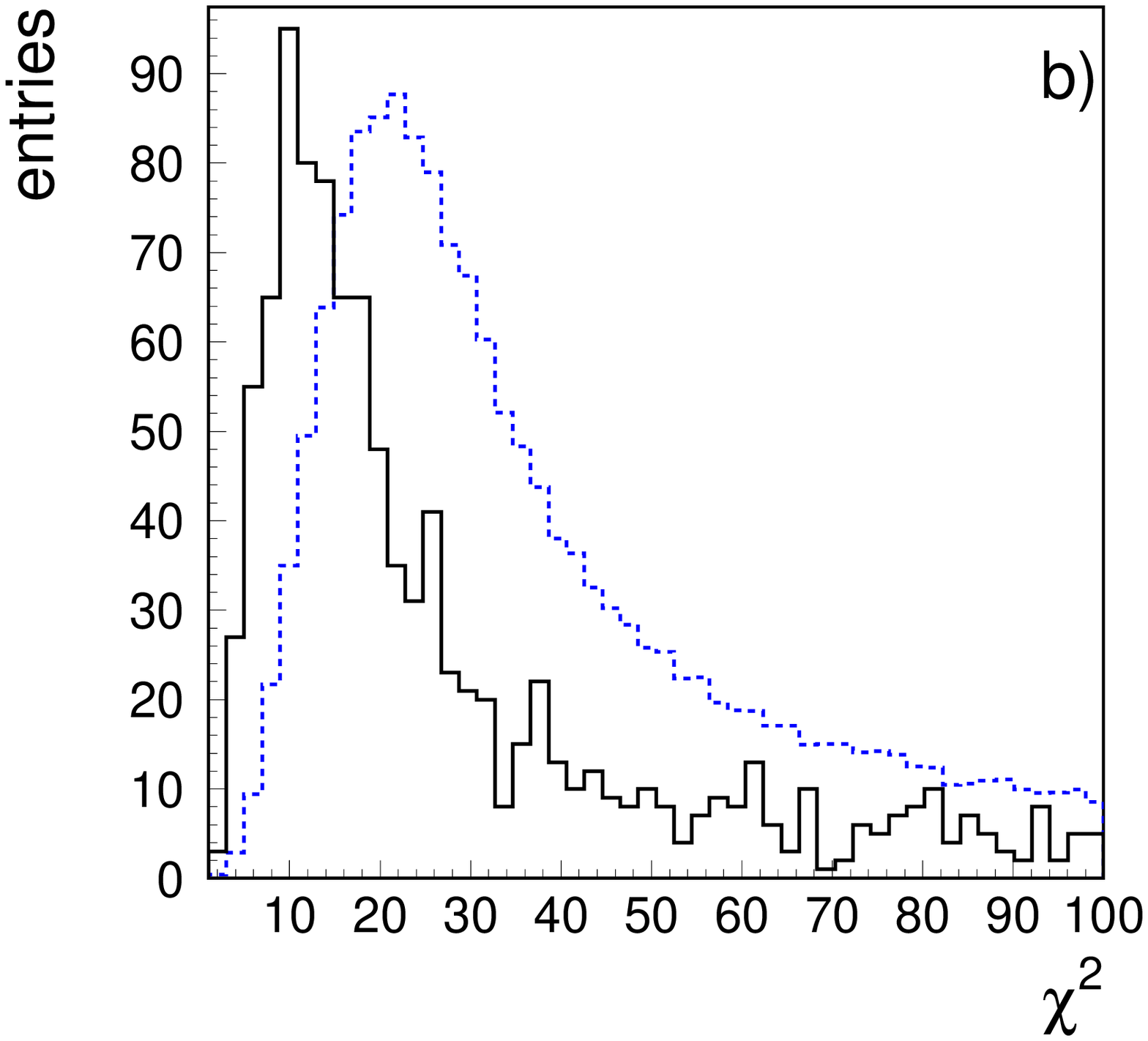}
\caption{a) $\chi^2$ distribution of the kinematical fit for 4$\gamma$ events of data (red pionts)
and simulations (blue histogram); b) The $\chi^2$ distributions from kinematical fit for simulated $K_S \to 3\pi^0$ signal
events (solid) and background (dashed).}
\label{kinfit2}
\end{figure}

In the next stage of the analysis we select only kinematically well defined events.
To this end we perform the kinematical fit procedure based on the least squares method
with the following set of variables as an input:
\begin{itemize}
 \item The total energy of the system $\sqrt{s}$
 \item The momentum $P_{\phi}$ and decay vertex of the $\phi$ meson
 \item The four -- momentum vector and decay vertex of $K_S$ meson
 \item Energy of each $\gamma$ quantum together with its time and position measured
in the calorimeter 
\end{itemize}
The initial value of the $K_S$ four -- momentum vector is determined for each
event using the position of the reconstructed $K_L$ cluster while for $\sqrt{s}$  and $P_{\phi}$
we use the mean values measured for each running period using the gathered sample
of $e^+e^-$ scattered at large angles. Using this set of variables
we can construct the following quadratic form:
\begin{equation}
X^{2}~=~(\mathbf{Y} -\mathbf{Y_{0}})^{T} \mathbf{V}^{-1} (\mathbf{Y} -\mathbf{Y_{0}})
+ \mbox{\boldmath{$\lambda$}}\cdot\mathbf{G(Y)}~,
\label{kinfit1}
\end{equation}
where $\mathbf{Y_0}$ is a vector of measured variables, \textbf{V} denotes covariance matrix, $\mathbf{Y}$ denotes the
vector of corrected variables fulfilling the kinematical constraints $\mathbf{G(Y)}$ and $\boldsymbol{\lambda}$ is a vector
of the Lagrange multipliers. The quadratic form is then minimized with respect to $\mathbf{Y}$ and $\boldsymbol{\lambda}$
leading to the determination of the best corrected variable values. The minimum value of $X^2$ can be treated
as the $\chi^2$ -- like variable with a probability function that can be used for hypothesis testing and evaluation of
the goodness of the fit. 
As the constraints we impose the total four -- momentum vector conservation, $K_S$ mass requirement and consistency of the
time of flight determined from the cluster position of each $\gamma$ quantum with its time reconstructed in the calorimeter.
The developed algorithm was first applied to the kinematically well defined
$K_S \to 2\pi^0 \to 4\gamma$ events to test how reliable is the fit procedure. The distributions for data and simulations for
this kind of events are shown in Fig.~\ref{kinfit2}a.
The same distributions for generated background events and $K_S \to 3\pi^0$ signal are shown
in Fig.~\ref{kinfit2}b.
It can be seen that applying a cut on $\chi^2$ around 40 we obtain a small background rejection, about 30$\%$,
with a good signal efficiency ($\sim$70$\%$). This does not improve significantly the signal to background
ratio but allows to reject bad quality reconstructed events.
\subsection{Testing the 3$\pi^0$ and 2$\pi^0$ hypotheses} 
\begin{figure}
\centering
\includegraphics[width=0.3\textwidth]{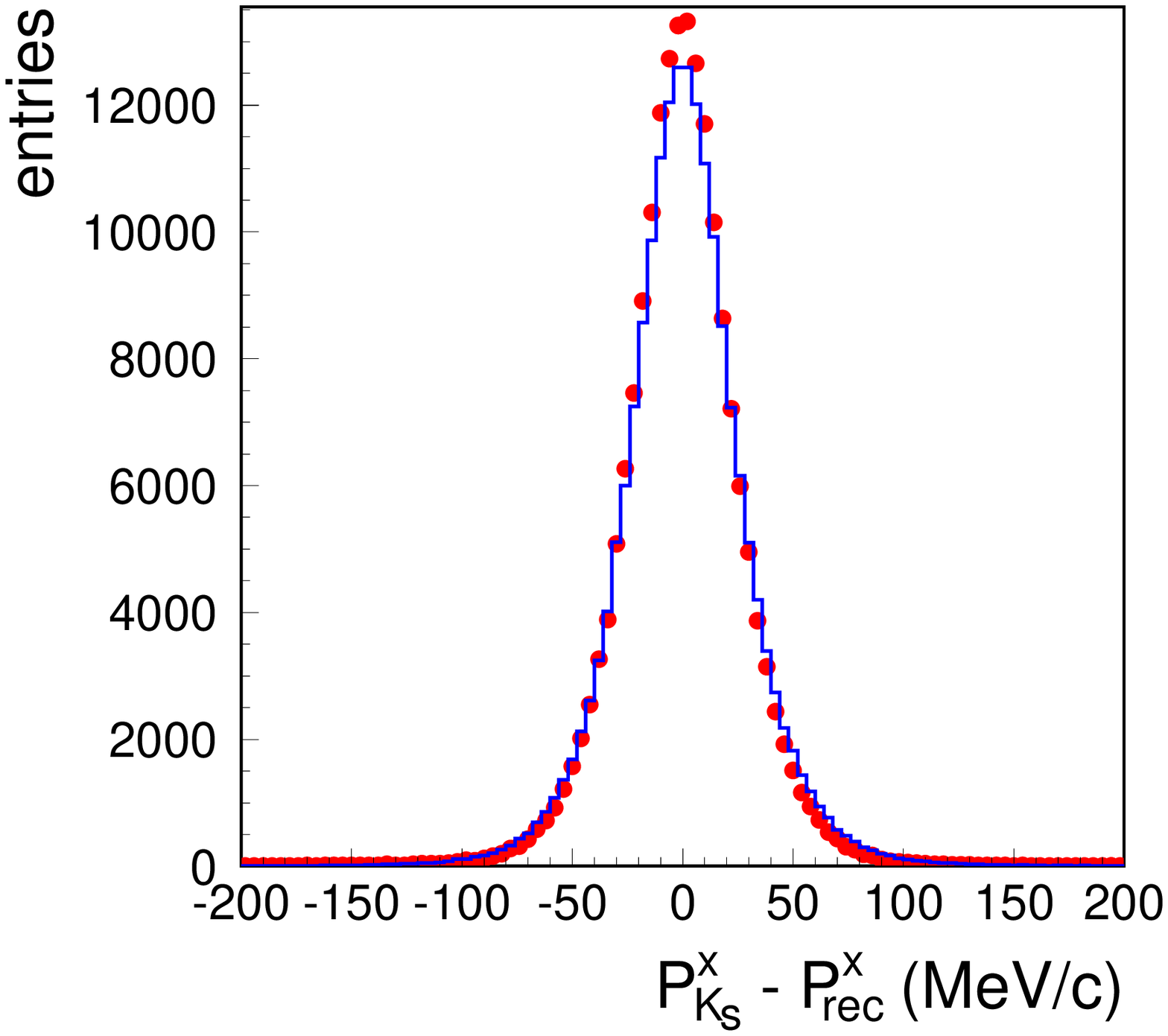}
\includegraphics[width=0.3\textwidth]{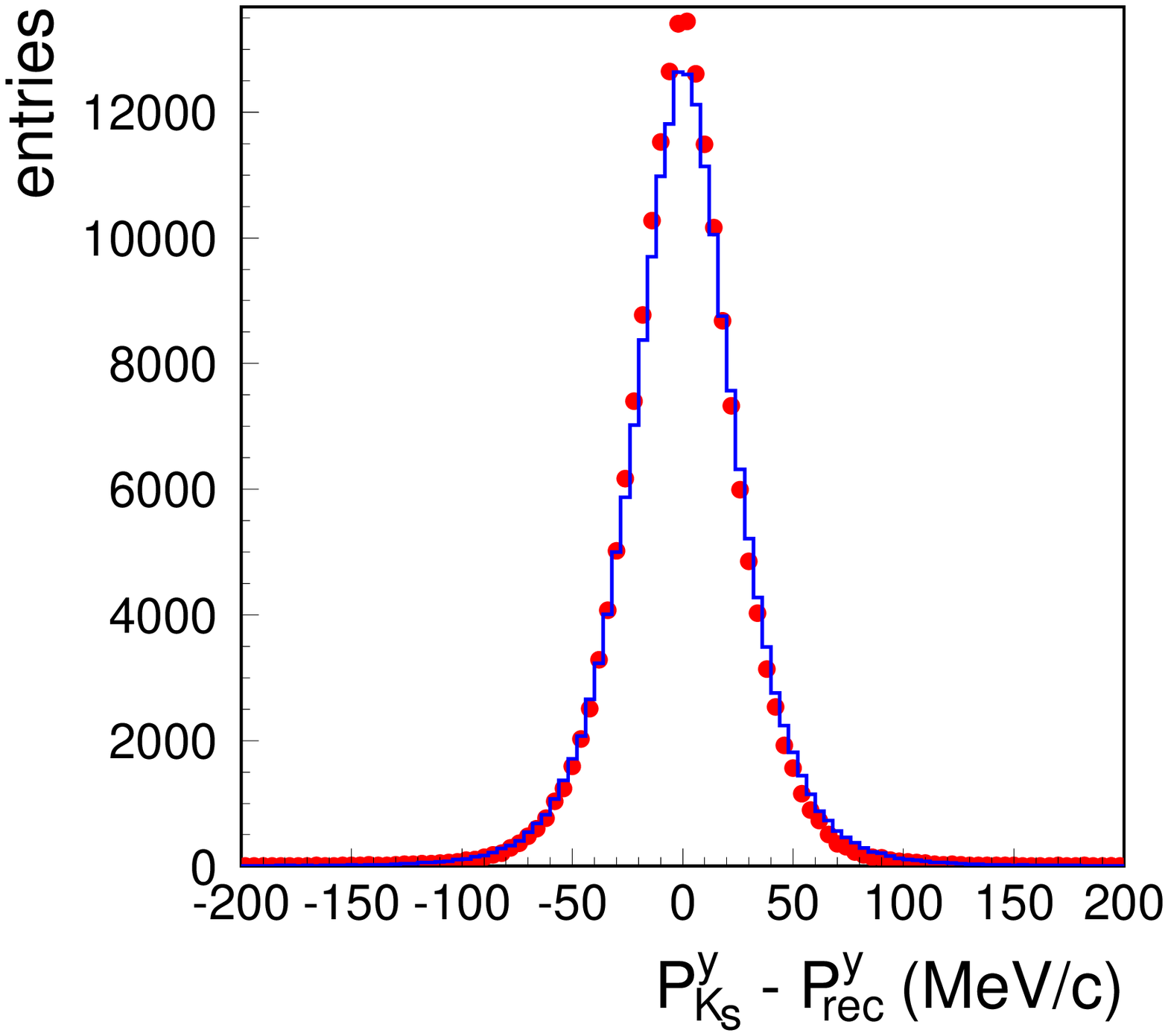}
\includegraphics[width=0.3\textwidth]{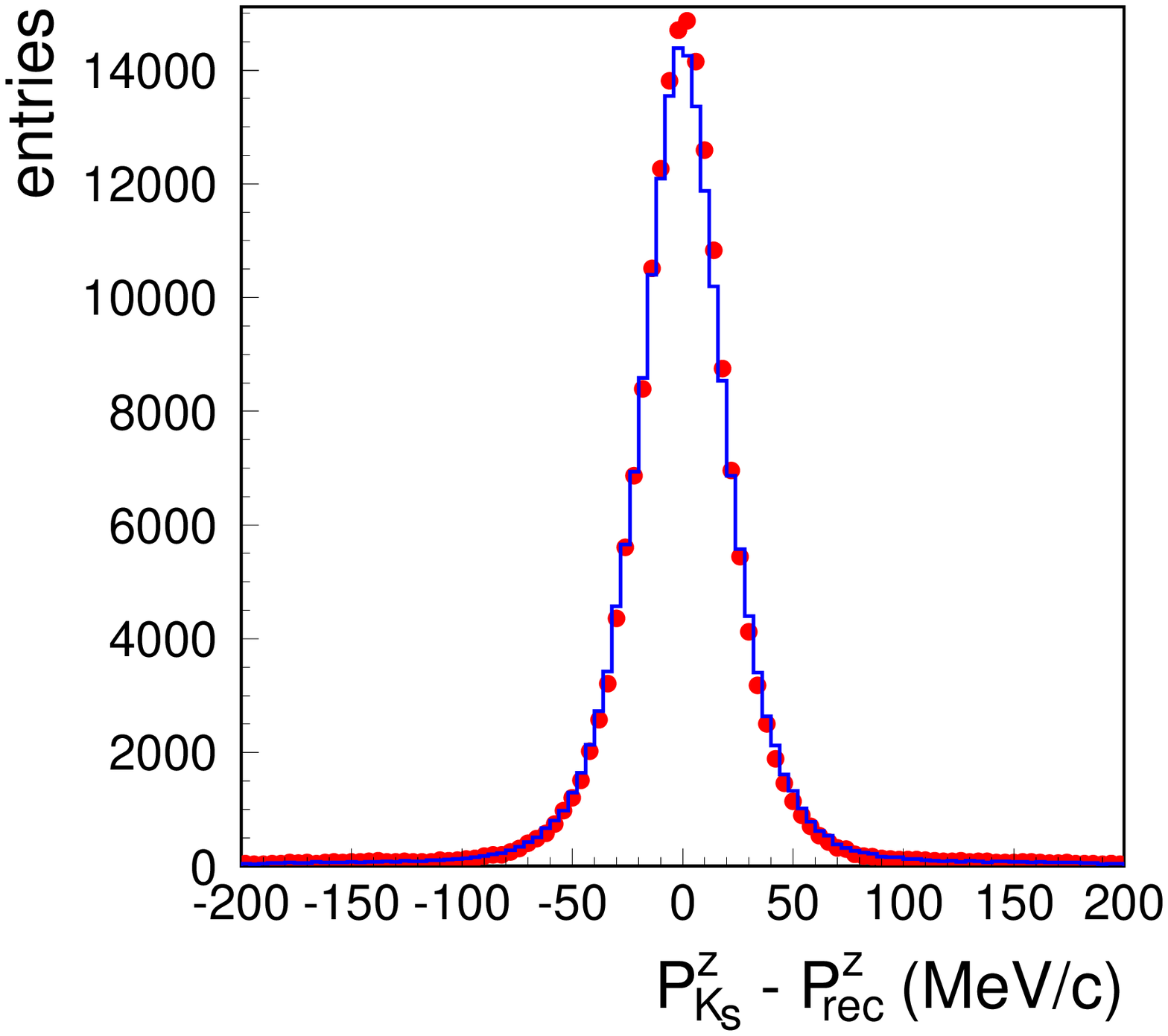}
\includegraphics[width=0.3\textwidth]{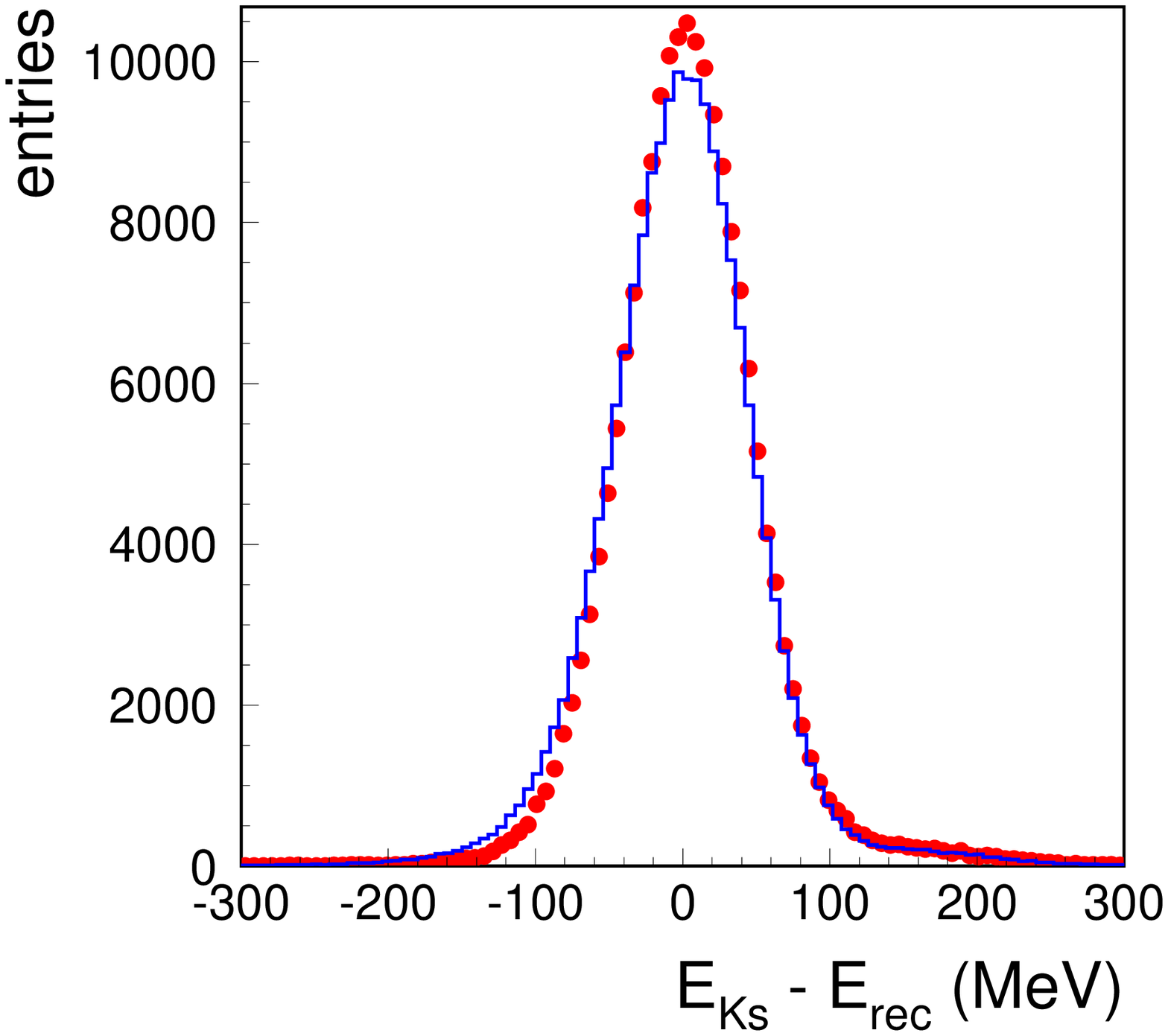}
\includegraphics[width=0.3\textwidth]{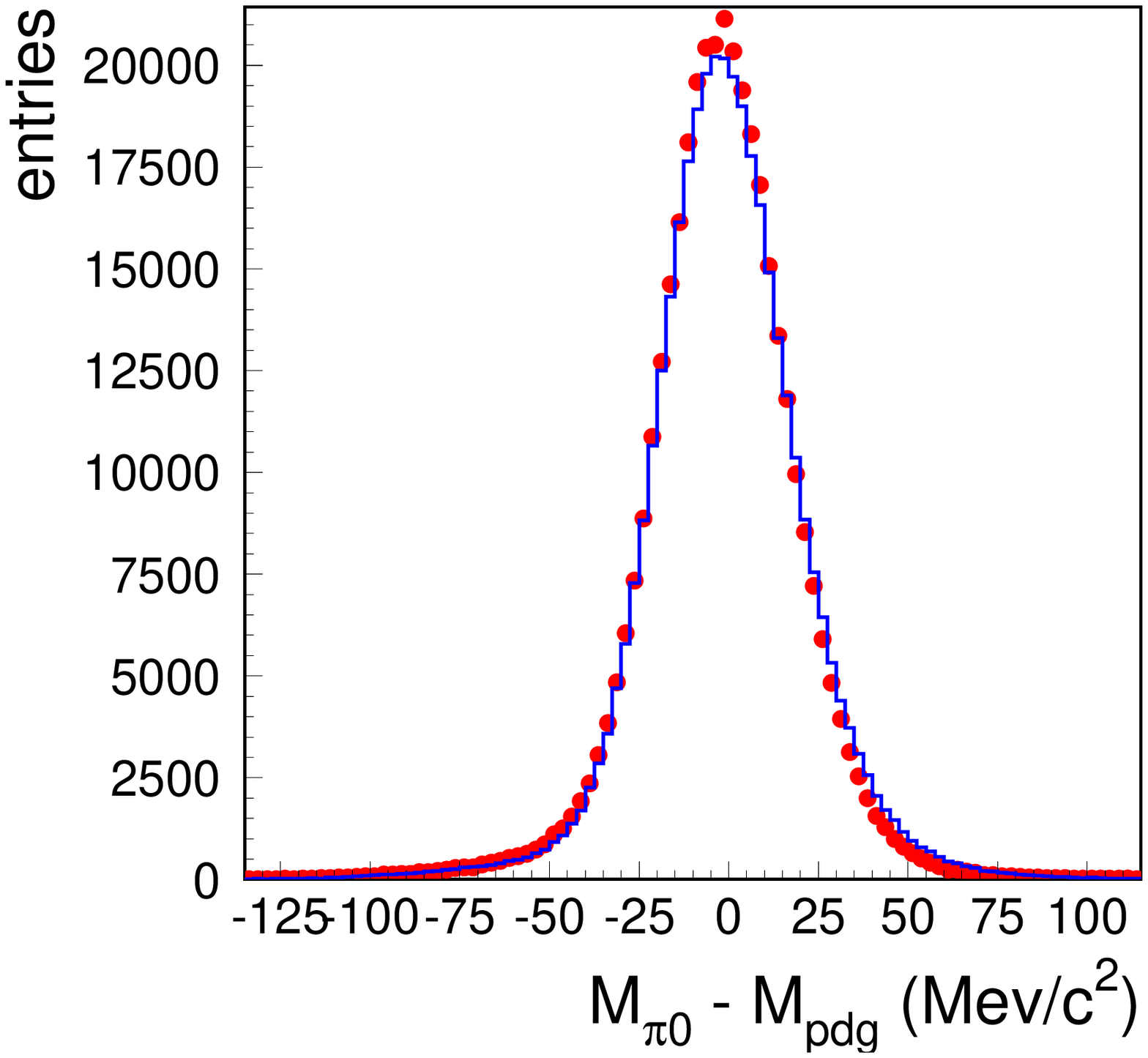}
\includegraphics[width=0.3\textwidth]{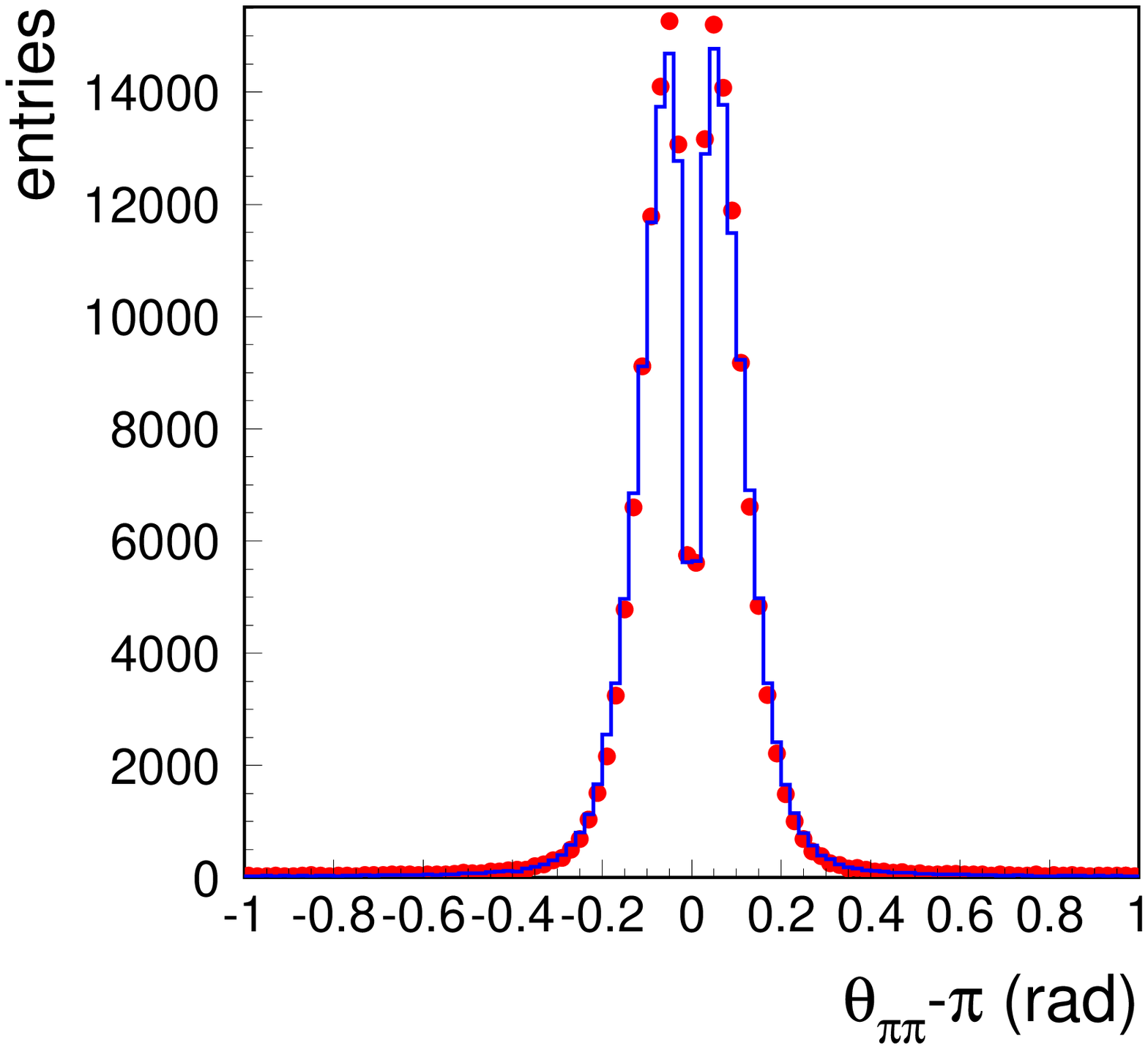}
\caption{Distributions of parameters used in $\chi^2_{2\pi}$ calculation for data (red points) and the simulations
of the $K_S \to 2\pi^0$ channel (blue histogram) after the energy scale correction described in Sec.~\ref{tlo}.}
\label{parametry}
\end{figure}

\begin{table}[p]
\begin{center}
\begin{tabular}{|c|c|c|}
\hline
\textbf{PARAMETER} & \textbf{DATA} & \textbf{SIMULATIONS}\\
\hline
$\boldsymbol{\sigma_{2\pi1}}$ & 18.66 $\pm$ 0.04& 19.24 $\pm$ 0.04\\
$\boldsymbol{\sigma_{2\pi2}}$ & 18.84 $\pm$ 0.04& 19.36 $\pm$ 0.04\\
$\boldsymbol{\sigma_{E_{K_{S}}}}$ & 44.07 $\pm$ 0.08& 46.83 $\pm$ 0.09\\
$\boldsymbol{\sigma_{P_x}}$ & 25.93 $\pm$ 0.05& 27.55 $\pm$ 0.06\\
$\boldsymbol{\sigma_{P_y}}$ & 26.12 $\pm$ 0.05& 27.55 $\pm$ 0.06\\
$\boldsymbol{\sigma_{P_z}}$ & 23.48 $\pm$ 0.05& 24.39 $\pm$ 0.06\\
$\boldsymbol{\sigma_{\theta_{\pi\pi}}}$ & 0.1238 $\pm$ 0.0002& 0.1257 $\pm$ 0.0002\\
\hline
$\boldsymbol{\sigma_{3\pi}}$ & 17.0 $\pm$ 0.5 & 17.0 $\pm$ 0.5\\
\hline
\end{tabular}
\end{center}
\caption{
\label{tabchipar}
List of parameters used in the calculation of $\chi^2_{2\pi}$ and $\chi^2_{3\pi}$.
The values were obtained by fitting Gaussian functions to the distributions presented in Fig.~\ref{parametry}
and to the distributions of pion masses reconstructed by minimization of the $\chi^2_{3\pi}$ value using
the simulated signal events.}
\end{table}
In order to reject events with split and accidental clusters we look at the correlation between
two $\chi^2$ -- like discriminating variables $\chi^{2}_{2\pi}$ and $\chi^{2}_{3\pi}$. 
$\chi^{2}_{2\pi}$ is calculated by an algorithm selecting four out of six clusters best satisfying
the kinematic constraints of the two-body decay, therefore it verifies the $K_S \to 2\pi^0 \to 4\gamma$
hypothesis. The pairing of clusters is based on the difference between reconstructed
$\pi^0$ masses $M_{\pi^{0}_1}$ and $M_{\pi^{0}_2}$ with respect to the PDG value $M_{PDG}$~\cite{pdg2010}
and on the opening angle of the reconstructed pions trajectories
in the $K_S$ center of mass frame $\theta_{\pi\pi}$
which should be equal to 180$^\circ$ for the $K_S \to 2\pi^0$ events.
Moreover, we check the consistency of the determination of the $K_S$ four -- momentum vector $\mathbb{P}_{K_S}$.
It is performed by comparing the $\mathbb{P}_{K_S}$
determined from the reconstructed four -- momentum of $K_L$
with the sum of the $\gamma$ quanta four -- momenta $\mathbb{P}_{rec} = \sum_{i=1}^4\mathbb{P}_{\gamma_{i}}$.
For every possible pairing choice the algorithm calculates the $\chi^{2}_{2\pi}$ defined as:
\begin{align}
\chi^{2}_{2\pi} &=~\frac{(M_{\pi^{0}_1} - M_{PDG})^2}{\sigma^{2}_{2\pi}}
+ \frac{(M_{\pi^{0}_2} - M_{PDG})^2}{\sigma^{2}_{2\pi}}
+ \frac{(\theta_{\pi\pi} - \pi)^2}{\sigma^2_{\theta_{\pi\pi}}}
\nonumber
+ \frac{\biggl(E_{K_{S}} - \displaystyle\sum_{i=1}^4 E_{\gamma_{i}}\biggr)^2}{\sigma^2_{E_{K_{S}}}}\\
&+ \frac{\biggl(P_{K_{S}}^x - \displaystyle\sum_{i=1}^4 P_{\gamma_{i}}^x\biggr)^2}{\sigma^2_{P_x}}
+ \frac{\biggl(P_{K_{S}}^y - \displaystyle\sum_{i=1}^4 P_{\gamma_{i}}^y\biggr)^2}{\sigma^2_{P_y}}
+ \frac{\biggl(P_{K_{S}}^z - \displaystyle\sum_{i=1}^4 P_{\gamma_{i}}^z\biggr)^2}{\sigma^2_{P_z}}~.
\label{chi2_2pi_def}
\end{align}
The minimization of the $\chi^2_{2\pi}$ gives the two photon pairs which out of the six clusters
fulfills best the criteria expected for the $K_S \to 2\pi^0 \to 4\gamma$ hypothesis.
The resolutions used in Eq.~\ref{chi2_2pi_def} were estimated using the
$K_S \to 2\pi^0 \to 4\gamma$ normalization sample.
The four -- momentum vectors of the reconstructed pions were used to make
distributions of the differences used for the $\chi^2_{2\pi}$ calculation
(see Fig.~\ref{parametry}), which were then fitted with the Gauss functions allowing
for the determination of the variances used in Eq.~\ref{chi2_2pi_def}.
All the values of the parameters are gathered in Tab.~\ref{tabchipar}.
\\The $\chi^{2}_{3\pi}$ instead verifies the signal hypothesis by looking on the reconstructed masses of three pions.
For every choice of cluster pairs we calculate the quadratic sum of the residuals
between the nominal $\pi^0$ mass and the invariant masses of three photon pairs~\cite{HyperF}:
\begin{equation}
\chi^{2}_{3\pi}~=~\frac{(M_{\pi^{0}_1} - M_{PDG})^2}{\sigma^{2}_{3\pi}}
+ \frac{(M_{\pi^{0}_2} - M_{PDG})^2}{\sigma^{2}_{3\pi}}
+ \frac{(M_{\pi^{0}_3} - M_{PDG})^2}{\sigma^{2}_{3\pi}}~.
\label{chi2_3pi_def}
\end{equation}
As the best combination of cluster pairs we take the configuration with lowest $\chi^{2}_{3\pi}$.
The resolution of pion mass $\sigma_{3\pi}$ was estimated applying the algorithm to the simulated signal events
(see Tab.~\ref{tabchipar}). In the definition of $\chi^{2}_{3\pi}$ we do not take into account the difference
between $\mathbb{P}_{K_S}$ and $\mathbb{P}_{rec} = \sum_{i=1}^6\mathbb{P}_{\gamma_{i}}$
because in this case it is the same for each combination of photon pairs.
\begin{figure}[h]
\includegraphics[width=0.5\textwidth]{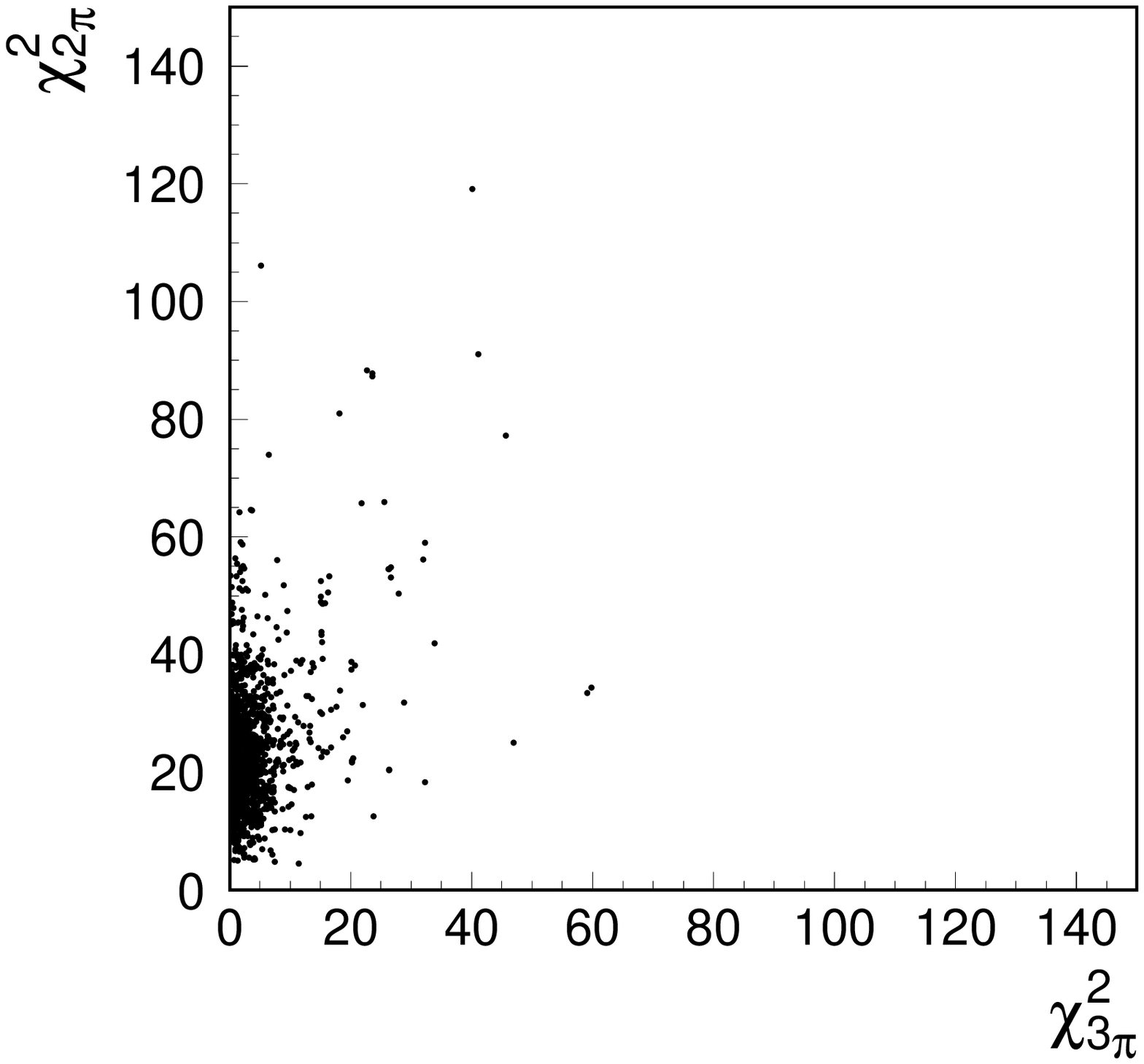}
\includegraphics[width=0.5\textwidth]{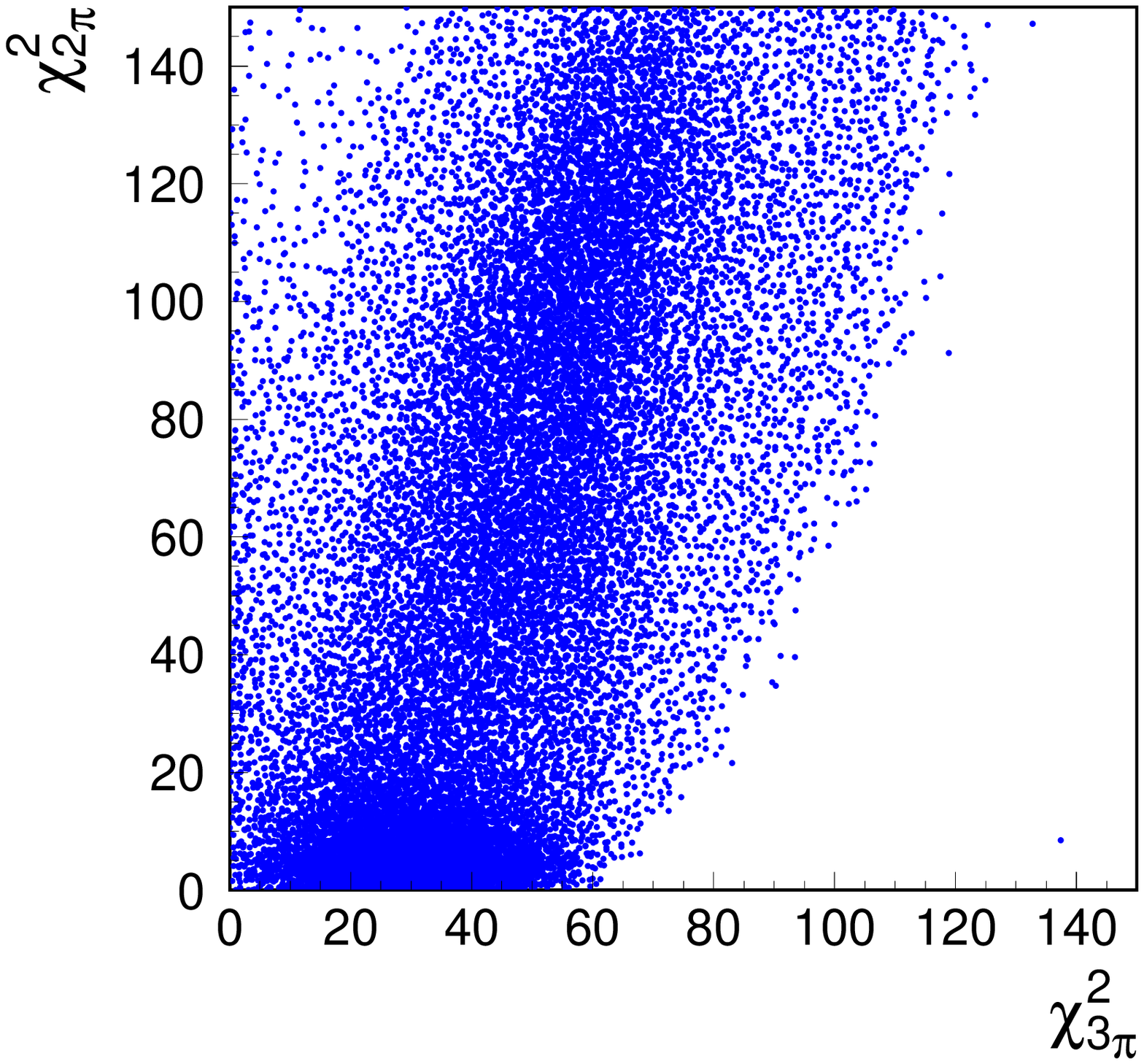}
\caption{Scatter plot of $\chi^2_{2\pi}$ versus $\chi^2_{3\pi}$ for simulated $K_S\to 3\pi^0$ signal (left)
and for the background after rejection of the $\phi \to K_{S}K_{L} \to \pi^+\pi^-,3\pi^0$ events (right).}
\label{picchi22pi3pi}
\end{figure}
\\As it is presented in Fig~\ref{picchi22pi3pi} the $K_S\to 3\pi^0$ signal is
characterized by low values of $\chi^2_{3\pi}$ and relatively high values
of $\chi^2_{2\pi}$. Background events are instead spread on the large area of 
$(\chi^2_{2\pi},\chi^2_{3\pi})$ plane with a maximum at low $\chi^2_{2\pi}$
being well distinguishable from the signal. Nevertheless in the region populated
by signal we find also some background events, mainly the $\phi \to K_{S}K_{L} \to \pi^+\pi^-,3\pi^0$
category even though it is already strongly suppressed by cuts defined with Eqs.~\ref{trv1} and~\ref{hardKlc}.
Further analysis dedicated to the rejection of the background due to the $K_S \to 2\pi^0$ decay is described
in the next section.
\subsection{Improvement of the $K_S \to 2\pi^0$ background suppression}
\begin{figure}
\includegraphics[width=0.5\textwidth]{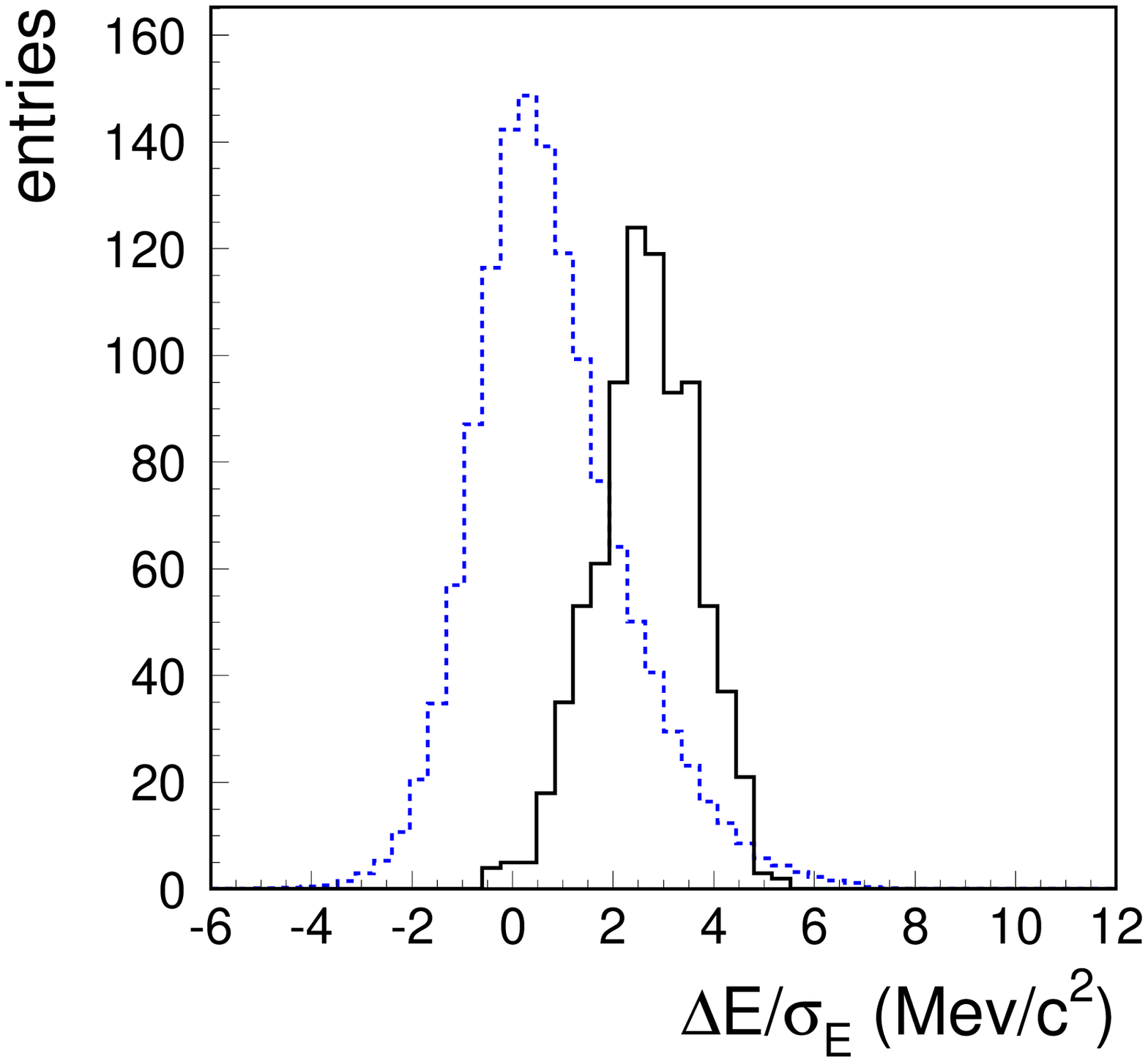}
\includegraphics[width=0.5\textwidth]{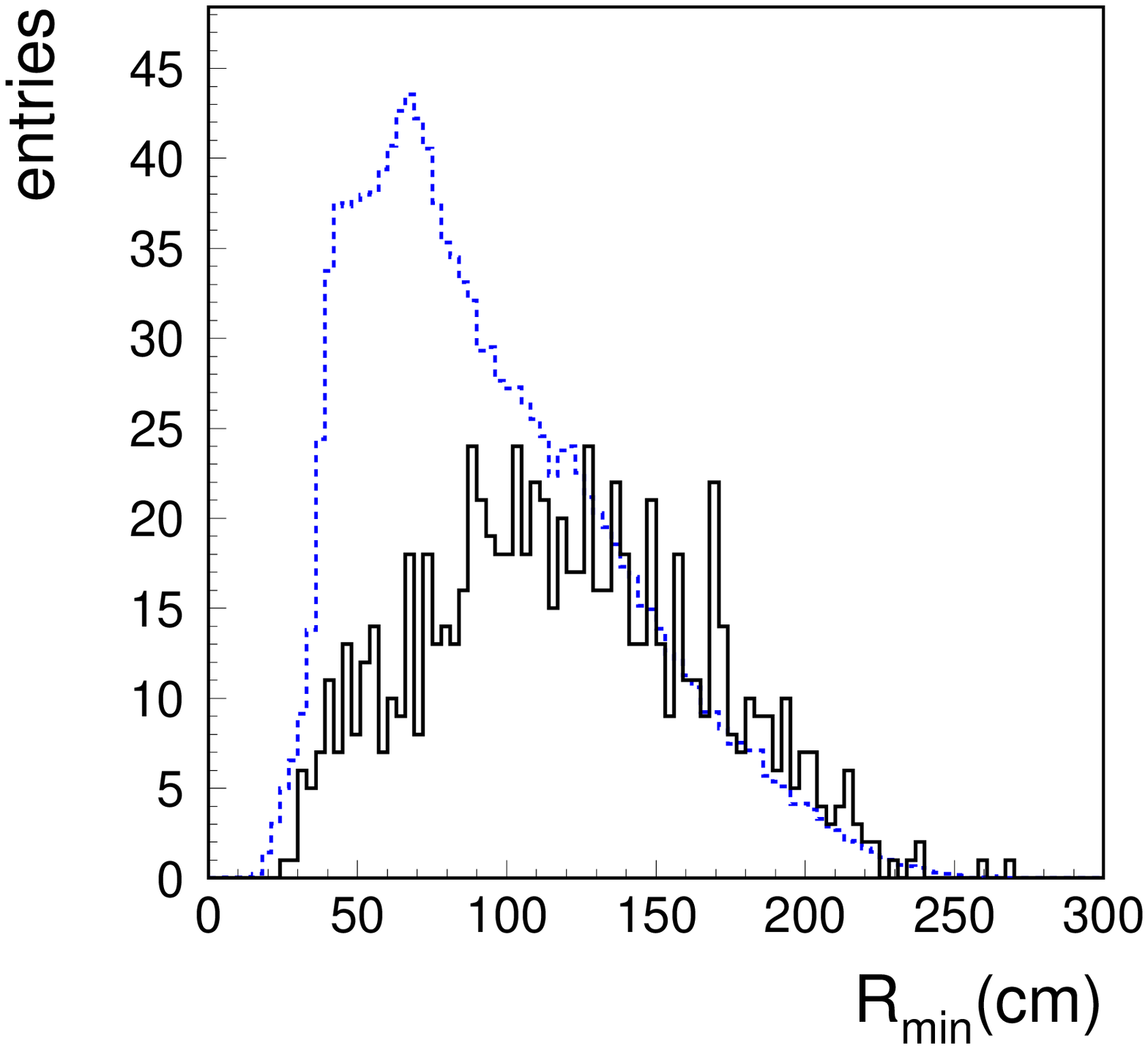}
\caption{Distributions of variables used to refine the rejection of the $K_S \to 2\pi^0$ events.
Dashed histogram indicates simulated background events from $K_S \to 2\pi^0$ and solid histogram
represents the Monte Carlo simulations of the signal. The distributions
are made for events with $\chi^2$ < 300. The variables are described in the text.}
\label{pic_rmin_deks}
\end{figure}
Since we are looking for a very rare decay and have to deal with a large background the rejection power
of the data analysis has to be as high as possible. The two additional four -- momentum vectors of photon candidates
reconstructed based on clusters originating from the machine background
or shower fragments results sometimes in an invariant mass close to the mass of $\pi^0$.
Thus, the $\chi^2_{3\pi}$ algorithm does not distinguish all the $K_S \to 2\pi^0$ decays
from the signal and we need another discriminant variable to refine the background rejection.\\
Events with two accidental clusters can be identified by measurement of the difference $\Delta E$
between the $K_S$ energy determined from the reconstructed $K_L$ four -- momentum and the sum of energies
of four gamma quanta selected by the $\chi^2_{2\pi}$ algorithm. For the $K_S \to 2\pi^0$
background this variable is close to zero since the event is kinematically closed. On the other hand
for the $K_S \to3\pi^0$ events $\Delta E$ should peak around 135 MeV since the rejected clusters
have an energy comparable to the pion mass.
In order to make the cuts as independent as possible of the energy resolution determination we use the normalized
$\Delta E$:
\begin{equation}
\Delta E/\sigma_{E}~=~\frac{\biggl(E_{K_{S}} - \displaystyle\sum_{i=1}^4 E_{\gamma_{i}}\biggr)}{\sigma_{E}}~, 
\label{eqdeks} 
\end{equation}
where the used value of $\sigma_E$ is equal to the one listed in Tab.~\ref{tabchipar}.
An example of the $\Delta E/\sigma_{E}$ distributions for simulated background and signal are presented
in the left panel of Fig.~\ref{pic_rmin_deks}. One can see that cutting around
1.8 allows to reject about 60$\%$ of background events keeping the signal efficiency at the level of around 80$\%$.\\
Further on, events with splitted clusters are suppressed with cut on the distance between center of
reconstructed clusters. Here we take advantage of the fact that the distance between splitted clusters
is on average smaller than the distance between clusters originating from $\gamma$ quanta of the
$K_S \to 3\pi^0$ decay.
For all possible pairs of clusters $(i,j)$ we calculate the distance:
\begin{equation}
R_{ij}~=~\sqrt{(x_i - x_j)^2 + (y_i - y_j)^2 + (z_i - z_j)^2}~,
\label{eqRmin}
\end{equation}
where $(x_i,y_i,z_i)$ and $(x_j,y_j,z_j)$ are the position coordinates of clusters
reconstructed in the calorimeter. We look then for the minimum of obtained values
$R_{min}~=~\mathrm{MIN}(R_{ij})$ and impose the following cut:
\begin{equation}
R_{min} > 65~\mathrm{cm}~.
\label{eqRmin1}
\end{equation}
This suppresses the background by about 30$\%$ retaining about 85$\%$ of signal events 
(see right panel of Fig.~\ref{pic_rmin_deks}).
The $\Delta E/\sigma_{E}$ and $R_{min}$ close the cuts sequence improving the background rejection
power of the analysis chain.
\section{Background estimation}
\label{tlo}
Search for the rare decays like $K_S \to 3\pi^0$ demand a precise knowledge of the background processes.
To this end we used realistic Monte Carlo simulations based on GEANT3 package.
As it was mentioned in chapter~\ref{rozdz5} there is a substantial discrepancy between the data
and results of the simulation for the six photons sample, which had to be fixed.
To this end we first divide the background simulations into categories using for each event
information about the decay chain and about particles contributing to each cluster.
All the categories are next used in the fitting procedure to the experimental data, which
allowed for the determination of the weighting factors for events belonging to each class of
simulated background. The procedure used to determine the event weights is described in the
following subsection, where we present also small corrections related to the simulated $R_{min}$
distribution and rejection of events with charged particles.
\subsection{Event weights determination}
\label{subs:wagi}
A below listed classes of background events were recognized\footnote{ The given fractions of events refer
to the simulated sample after the preselection described at the beginning of this chapter.}:
\begin{itemize}\itemsep1pt
\item {Fakes: the $\phi \to K_{S}K_{L} \to (K_S \to \pi^+\pi^-, K_L \to 3\pi^0)$ events together
with non -- $K_{S}K_{L}$ channels like $\phi \to K^{+}K^{-}$ or $\phi \to \pi^+\pi^-\pi^0$ (about 2$\%$)}
\item 2A+1A1S: events with two accidental clusters (about 24$\%$) or events with one accidental and one
splitted cluster (about 6$\%$)
\item 2S: events with two splitted clusters (about 60$\%$) or with more than two accidental or splitted
clusters, as well as $K_S \to 2\pi^0$ events with only one splitted or only one accidental cluster (about 8$\%$)
\end{itemize}
The category called \textit{Fakes} contains, apart from the standard
$\phi \to K_{S}K_{L} \to (K_S \to \pi^+\pi^-, K_L \to 3\pi^0)$ events, small admixture of the
non -- $K_SK_L$ background, which is however suppressed by the $\chi^2$ and
$\Delta E/\sigma_{E}$ cuts. As regards other classes of background it is relatively easy to justify
the presence of more than two clusters originating from the machine accidental activity
or from cluster splitting.\\
The $K_S \to 2\pi^0$ events with only one splitted or only one
accidental cluster in principle should not pass the requirement of the registration of six gamma
quanta. However, more detailed analysis revealed, that the additional real cluster can be generated by
the initial state radiation or by the wrongly reconstructed $K_L$
cluster\footnote{It has turned out, that for many events in this category the additional cluster is very close
to the place where $K_L$ interacted with calorimeter. Thus they can result from wrongly reconstructed $K_L$ showers.}.
\begin{figure}
\includegraphics[width=0.5\textwidth]{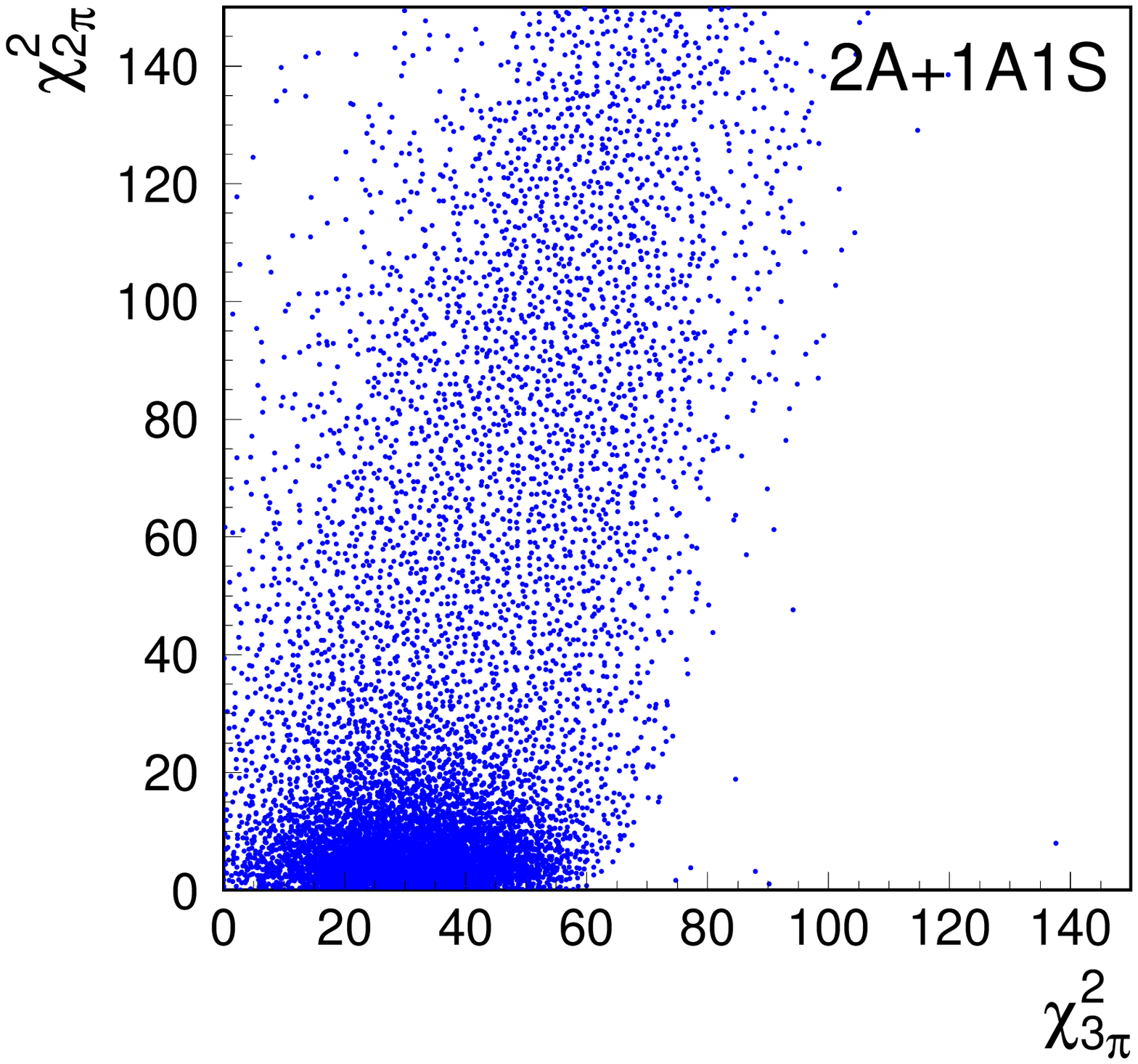}
\includegraphics[width=0.5\textwidth]{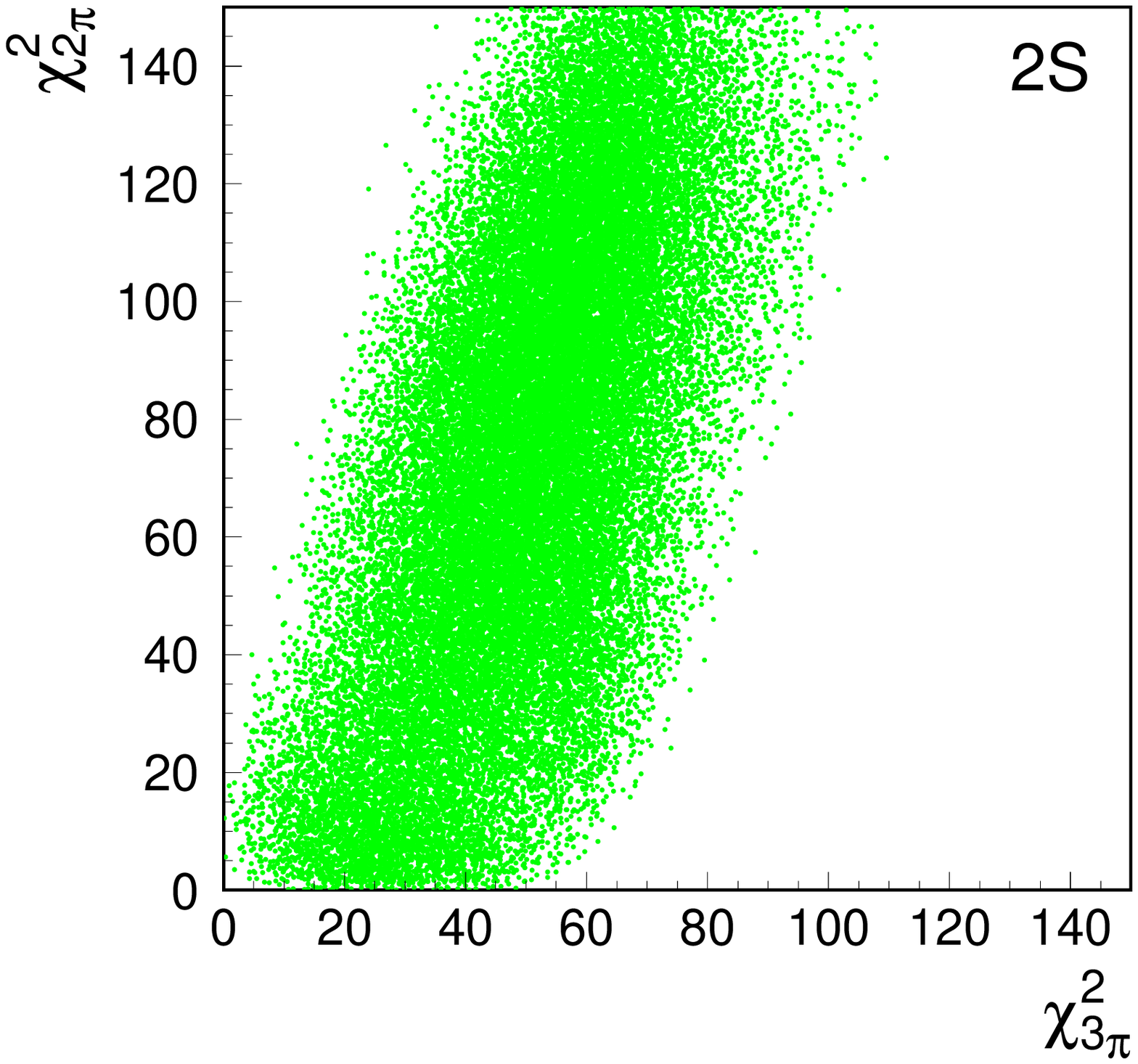}
\includegraphics[width=0.5\textwidth]{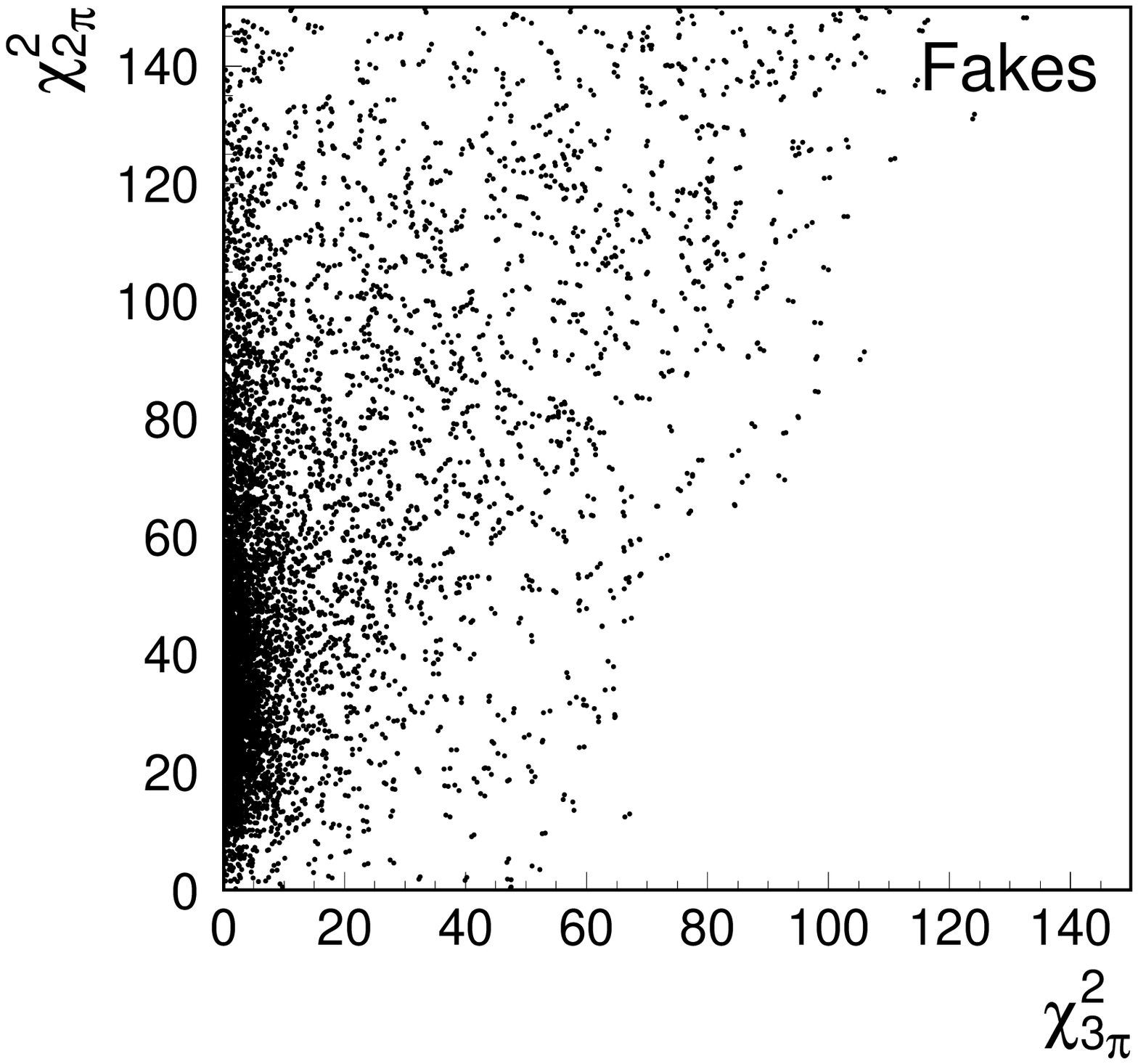}
\includegraphics[width=0.5\textwidth]{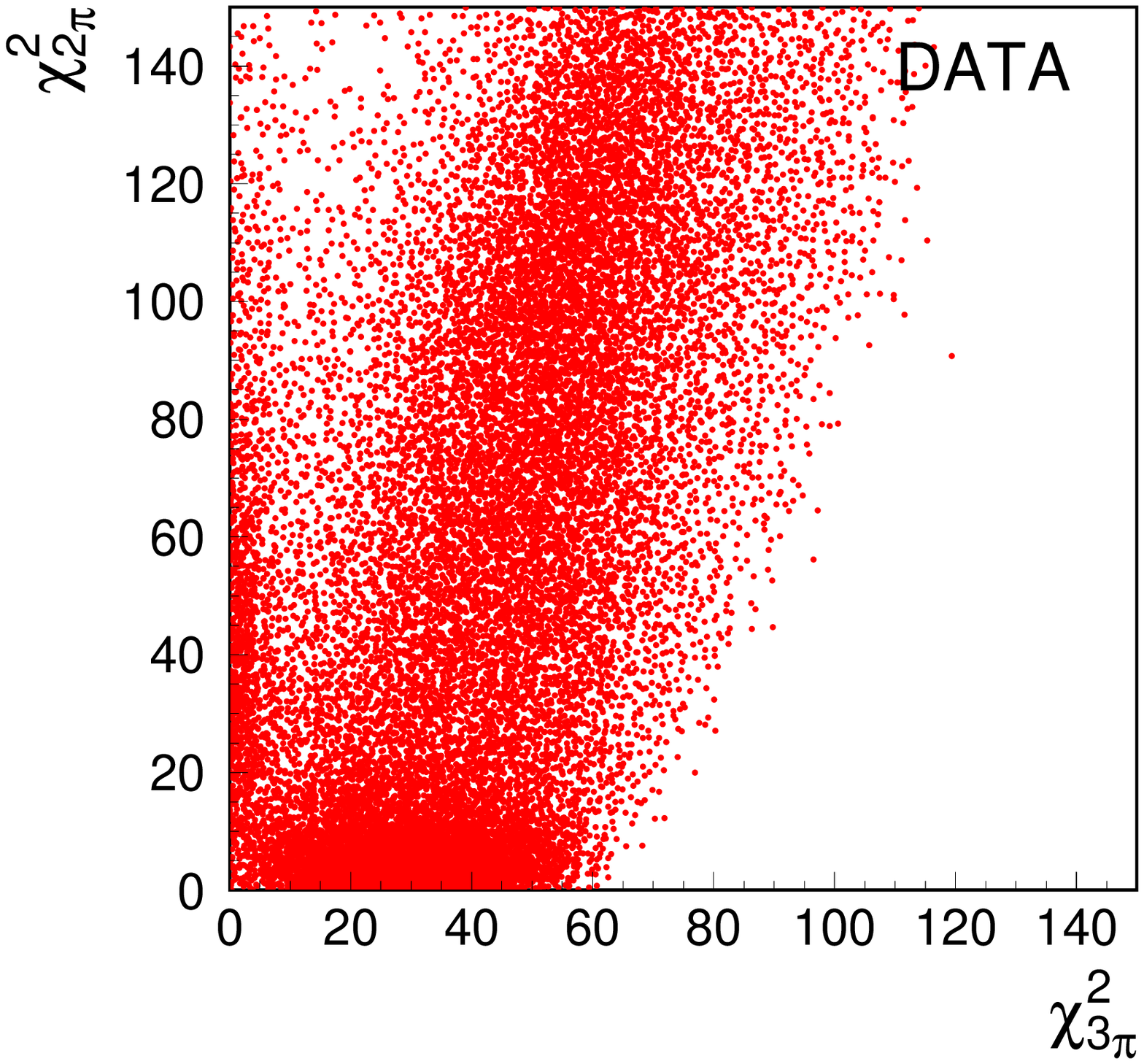}
\caption{The $\chi^2_{2\pi}$ versus $\chi^2_{3\pi}$ scatter plots for different background categories.
The label of each plot is defined in the text. The corresponding experimental spectrum is also shown.}
\label{figfit1}
\end{figure}
For both data and all the background categories we have made the ($\chi^2_{2\pi}$,$\chi^2_{3\pi}$)
scatter plots. 
\\The Monte Carlo distributions were then fitted to the data as a linear combination providing
scaling factors $W_T$ for each background category used next to weight events:
\begin{align}
\nonumber
 \mathrm{Data}(\chi^2_{2\pi},\chi^2_{3\pi}) &= W_{T}^{\mathrm{2S}}\cdot \mathrm{Sim}^{\mathrm{2S}}
(\chi^2_{2\pi},\chi^2_{3\pi})+ W_{T}^{\mathrm{2A+1A1S}}\cdot \mathrm{Sim}^{\mathrm{2A+1A1S}}
(\chi^2_{2\pi},\chi^2_{3\pi})\\
\nonumber
&+ W_{T}^{\mathrm{Fakes}}\cdot \mathrm{Sim}^{\mathrm{Fakes}}(\chi^2_{2\pi},\chi^2_{3\pi})~,
\end{align}
where $\mathrm{Data}(\chi^2_{2\pi},\chi^2_{3\pi})$ denotes the experimental $(\chi^2_{2\pi},\chi^2_{3\pi})$
scatter plot and \\$\mathrm{Sim}(\chi^2_{2\pi},\chi^2_{3\pi})$ stands for the simulated distribution
of each background category.
\begin{table}[h]
\begin{center}
\begin{tabular}{|c|c|c|c|}
\hline
\textbf{Category} & $\boldsymbol{N_{MC}}$ & $\boldsymbol{N_{fit}}$ & $\boldsymbol{W_T}$\\
\hline
\textbf{2A+1A1S} & 64544 $\pm$ 254 & 26346 $\pm$ 236 & 0.4082 $\pm$ 0.0040 \\
\textbf{2S} & 145996 $\pm$ 382 &  43446 $\pm$ 283 & 0.2976 $\pm$ 0.0021\\
\textbf{Fakes} & 5670 $\pm$ 75 &  6897 $\pm$ 148 & 1.216 $\pm$ 0.031\\
\hline
\end{tabular}
\end{center}
\caption{
\label{tab:tfit}
Scaling factors for Monte Carlo background categories used in the fit to the data.
$N_{MC}$ and $N_{fit}$ denote the number of events in each category before and after the fit, respectively.
}
\end{table}
The results of the fit are gathered in Tab.~\ref{tab:tfit}.
The quality of the procedure used to refine the simulations for the six -- photon sample can be
controlled by the comparison of simulated and experimental inclusive distributions of discriminating
variables described in previous sections just after preselection.
\begin{figure}
\centering
\includegraphics[width=0.45\textwidth]{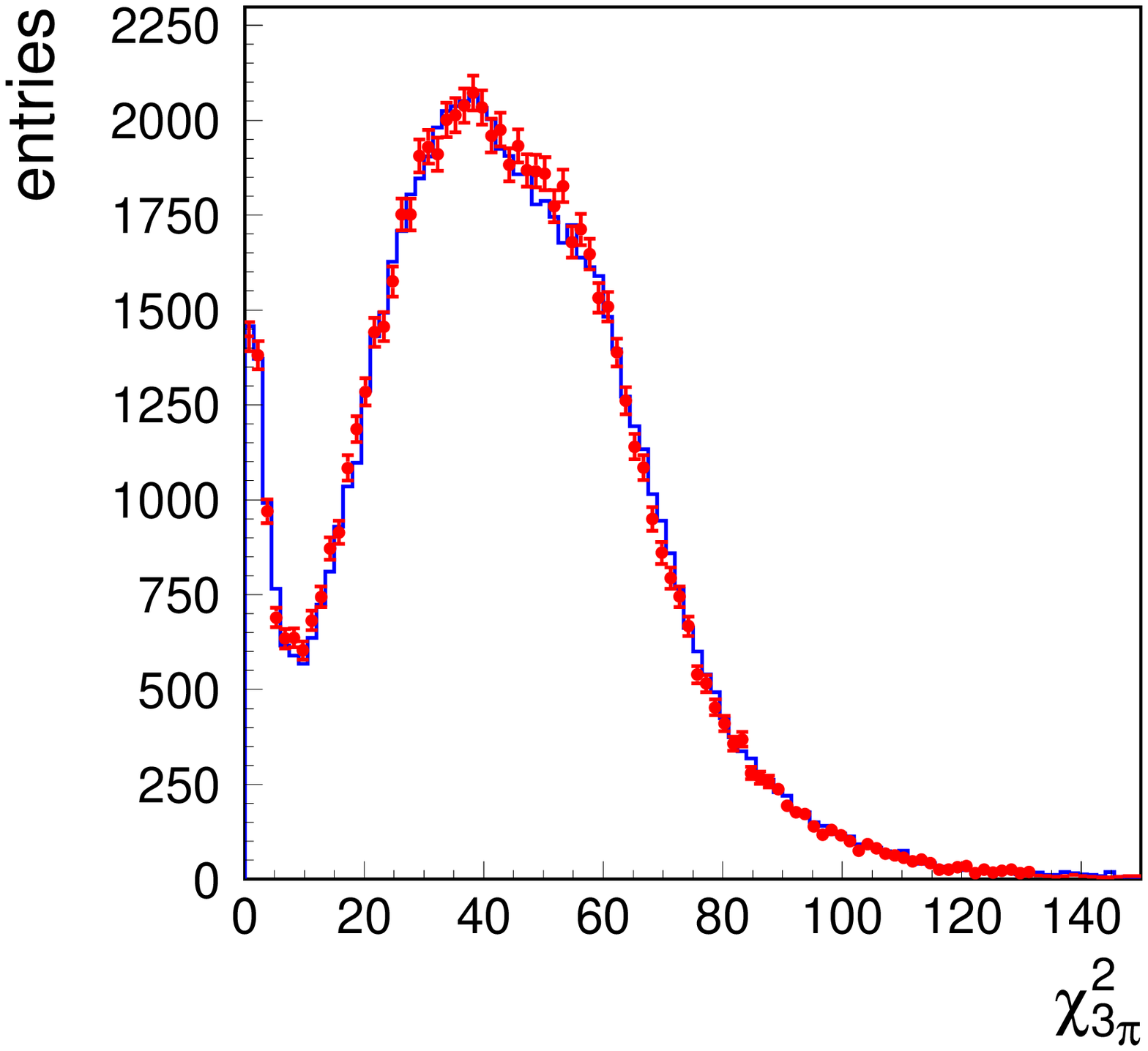}
\includegraphics[width=0.45\textwidth]{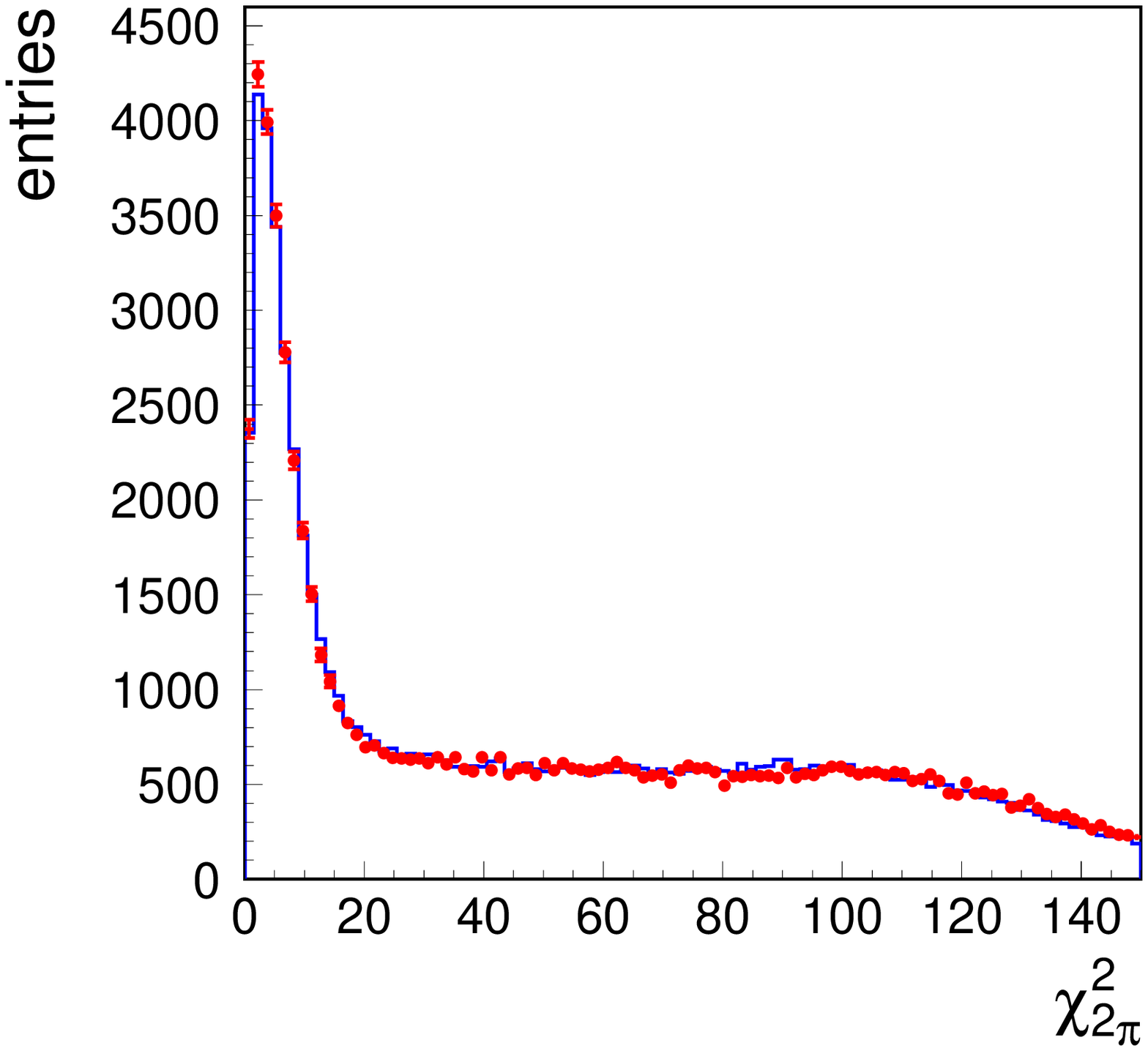}
\includegraphics[width=0.45\textwidth]{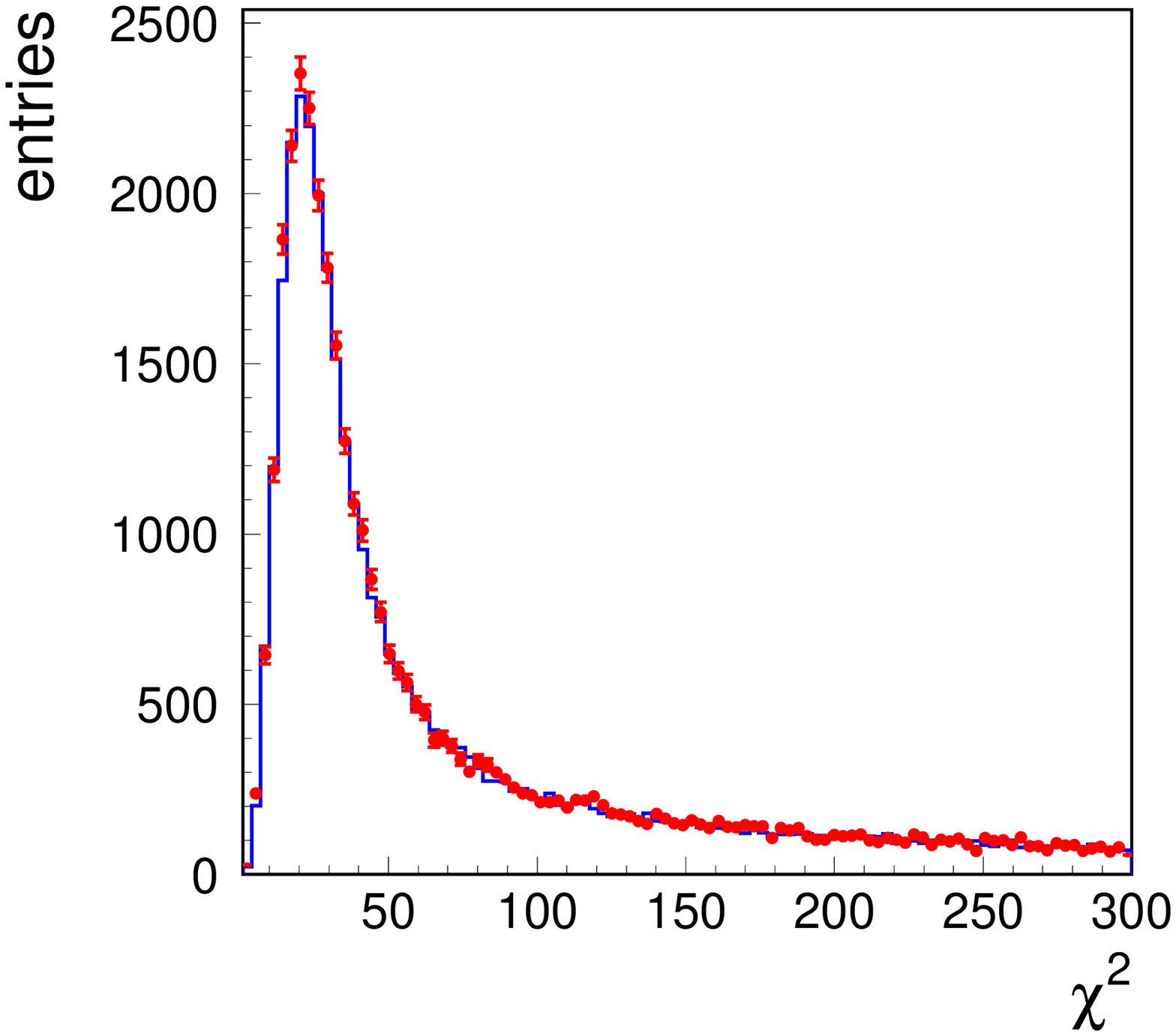}
\includegraphics[width=0.45\textwidth]{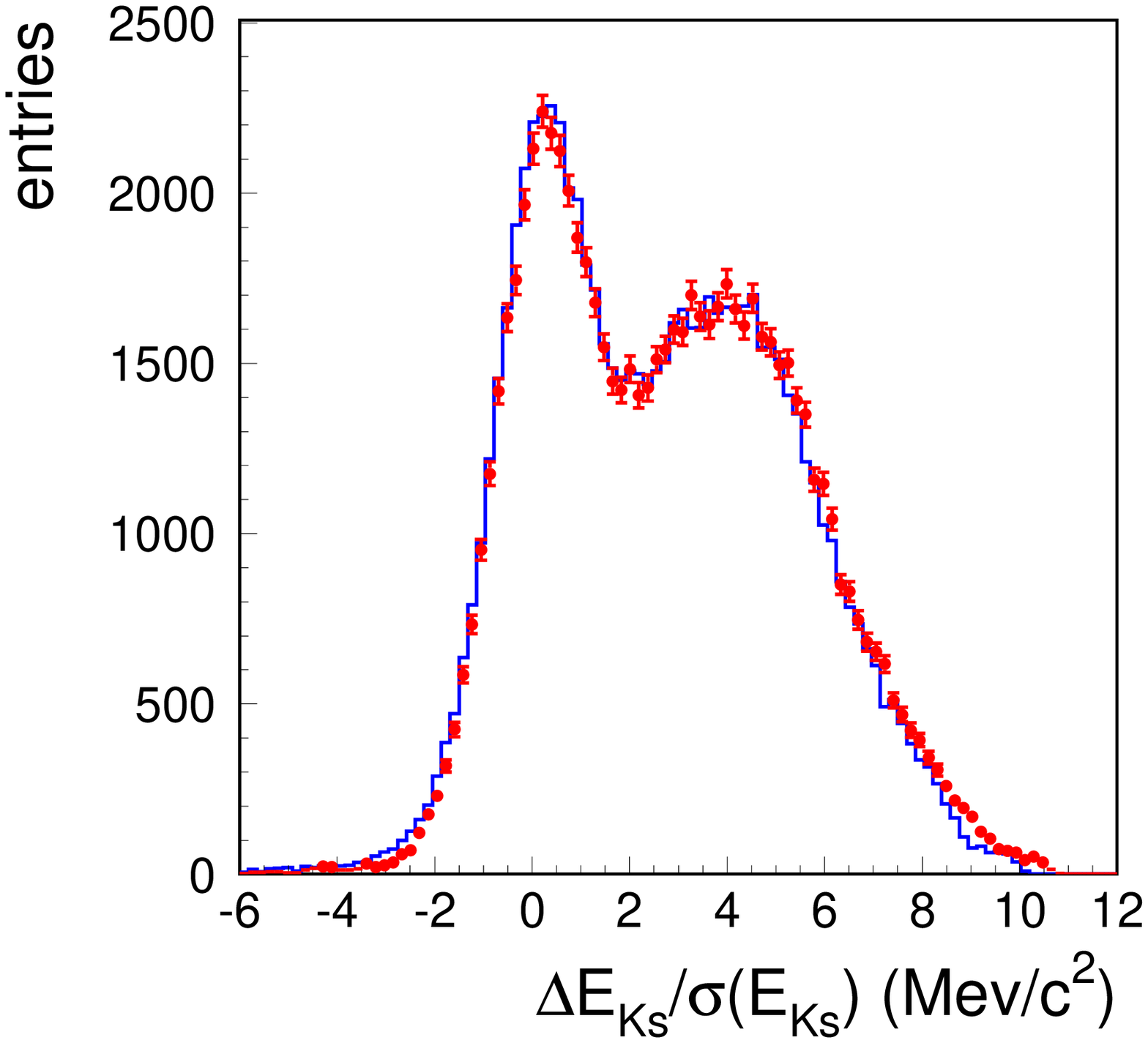}
\caption{Inclusive distributions of discriminating variables
for data (red points) and the simulations of the background (blue histograms) weighted with $W_T$ factors
as described in the text.}
\label{figfit2}
\end{figure}
This comparison is shown in Fig.~\ref{figfit2} and implies that the agreement between data and
background simulations is reasonable after the fit. Another check was done dividing the 
$(\chi^2_{2\pi},\chi^2_{3\pi})$ plane onto different regions to compare the experimentally
observed number of events with expectations based on the Monte Carlo simulations. Five control boxes were chosen
around the signal region defined with preliminary cuts on $\chi^2_{2\pi}$ and $\chi^2_{3\pi}$
(see Fig~\ref{fig:boxes}).
\begin{figure}
\centering
\includegraphics[width=0.6\textwidth]{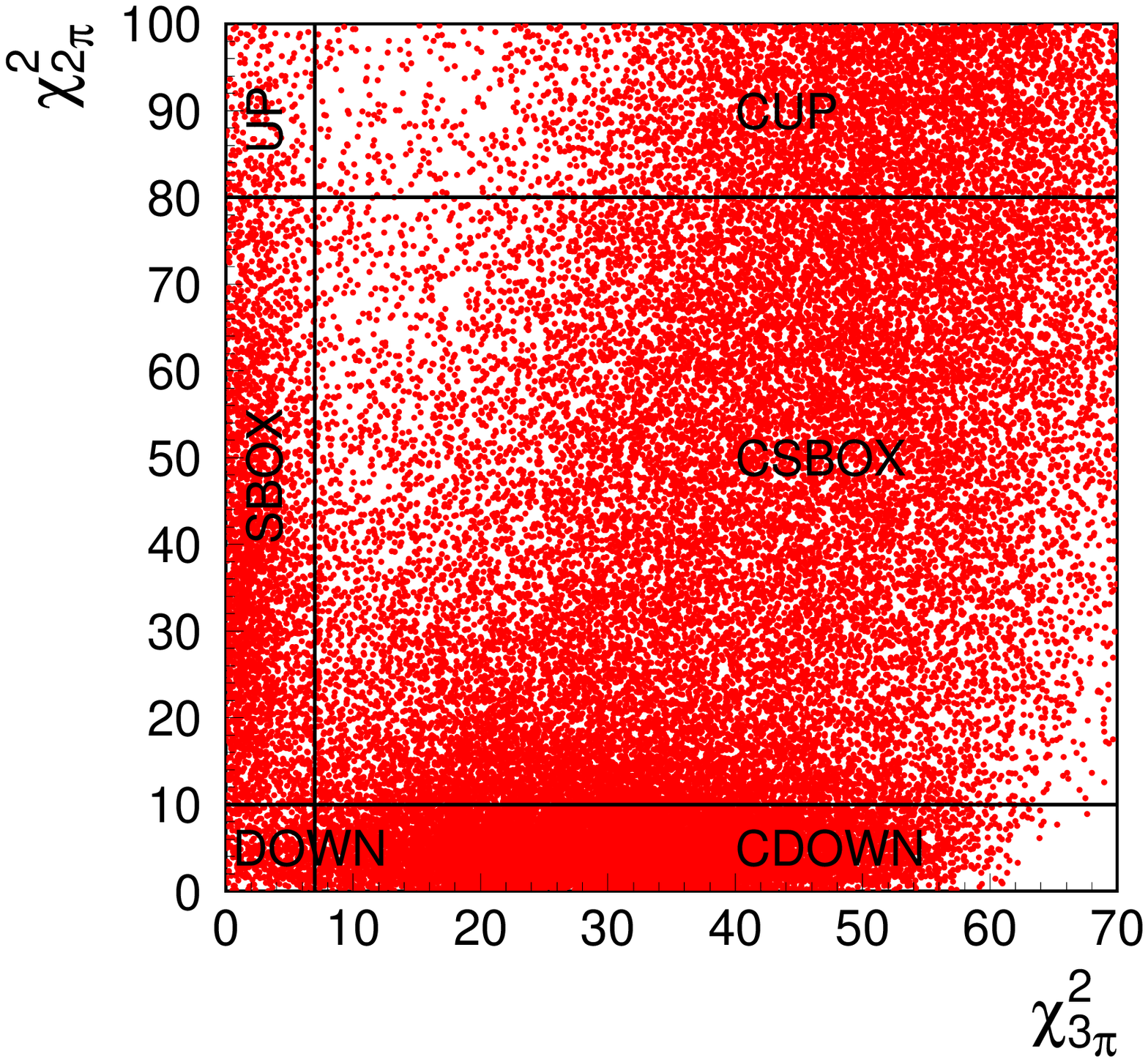}
\caption{Zoom of the $(\chi^2_{3\pi},\chi^2_{2\pi})$ distribution for data illustrating
the division of the plane into control boxes.}
\label{fig:boxes}
\end{figure}
The number of events registered in each box is reported in Tab.~\ref{tab:box},
where one can see that simulations results agree with experiment when taking
into account the statistical uncertainties calculated using
the standard deviations of the scaling factors (see Tab~\ref{tab:tfit}):
\begin{equation}
\Delta N_{box} = \sqrt{(W_{T}^{2S} \sqrt{N^{2S}})^2 +(W_{T}^{2A+1A1S} \sqrt{N^{2A+1A1S}})^2
+ (W_{T}^{Fakes} \sqrt{N^{Fakes}})^2}~,
\label{eq:wagiblad}
\end{equation}
where $N$ denotes the number of events belonging to each background category.
\begin{table}
\begin{center}
\begin{tabular}{|c|c|c|c|c|c|c|}
\hline
\textbf{} & \textbf{SBOX} & \textbf{DOWN} & \textbf{UP} & \textbf{CDOWN} & \textbf{CUP} & \textbf{CSBOX}\\
\hline
\textbf{DATA}& 200 $\pm$ 14 & 416 $\pm$ 21 & 7 $\pm$ 3 & 14385 $\pm$ 120 & 16321 $\pm$ 128 & 17634 $\pm$ 133\\
\hline
\textbf{MC} & 228 $\pm$ 10 & 313 $\pm$ 12 & 8 $\pm$ 3 & 14380 $\pm$ 134 & 16143 $\pm$ 124 & 17940 $\pm$ 123\\
\hline
\end{tabular}
\end{center}
\caption{
\label{tab:box}
The number of events populating control boxes in the $(\chi^2_{3\pi},\chi^2_{2\pi})$ plane defined in
Fig.~\ref{fig:boxes} after tight requirements for the reconstructed $K_L$ energy and velocity.}
\end{table}
\begin{figure}
\centering
\includegraphics[width=0.49\textwidth]{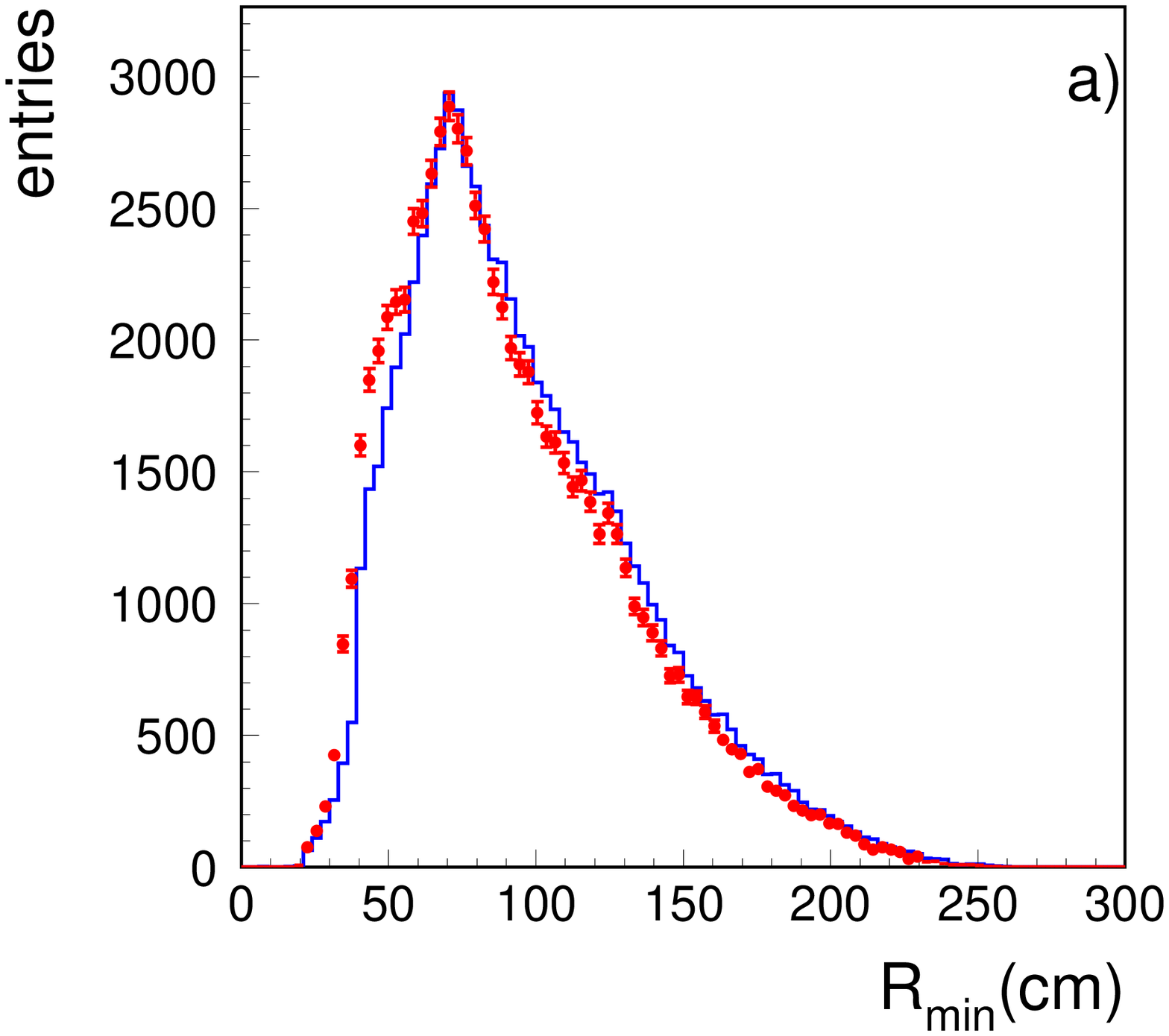}
\includegraphics[width=0.49\textwidth]{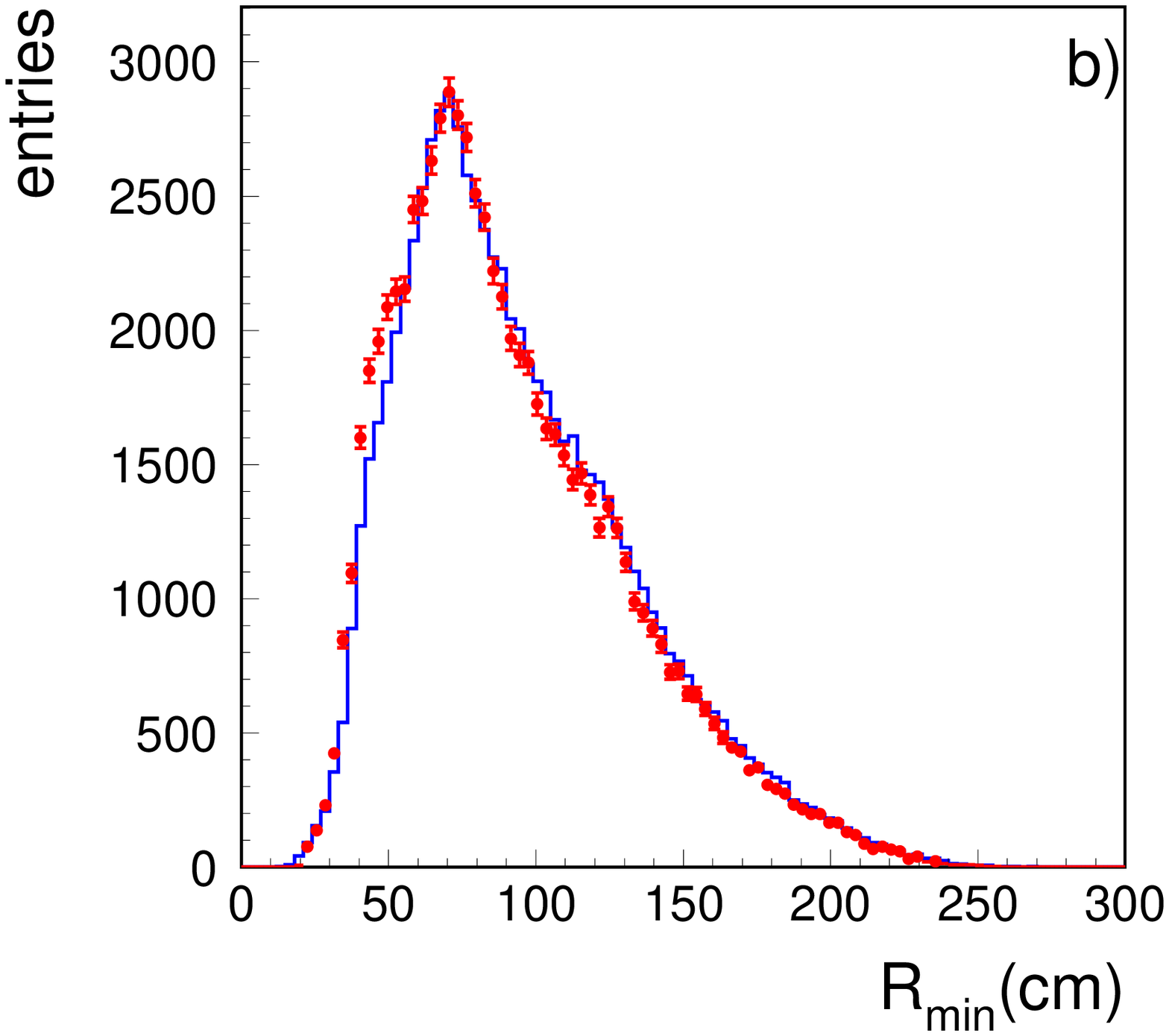}
\includegraphics[width=0.49\textwidth]{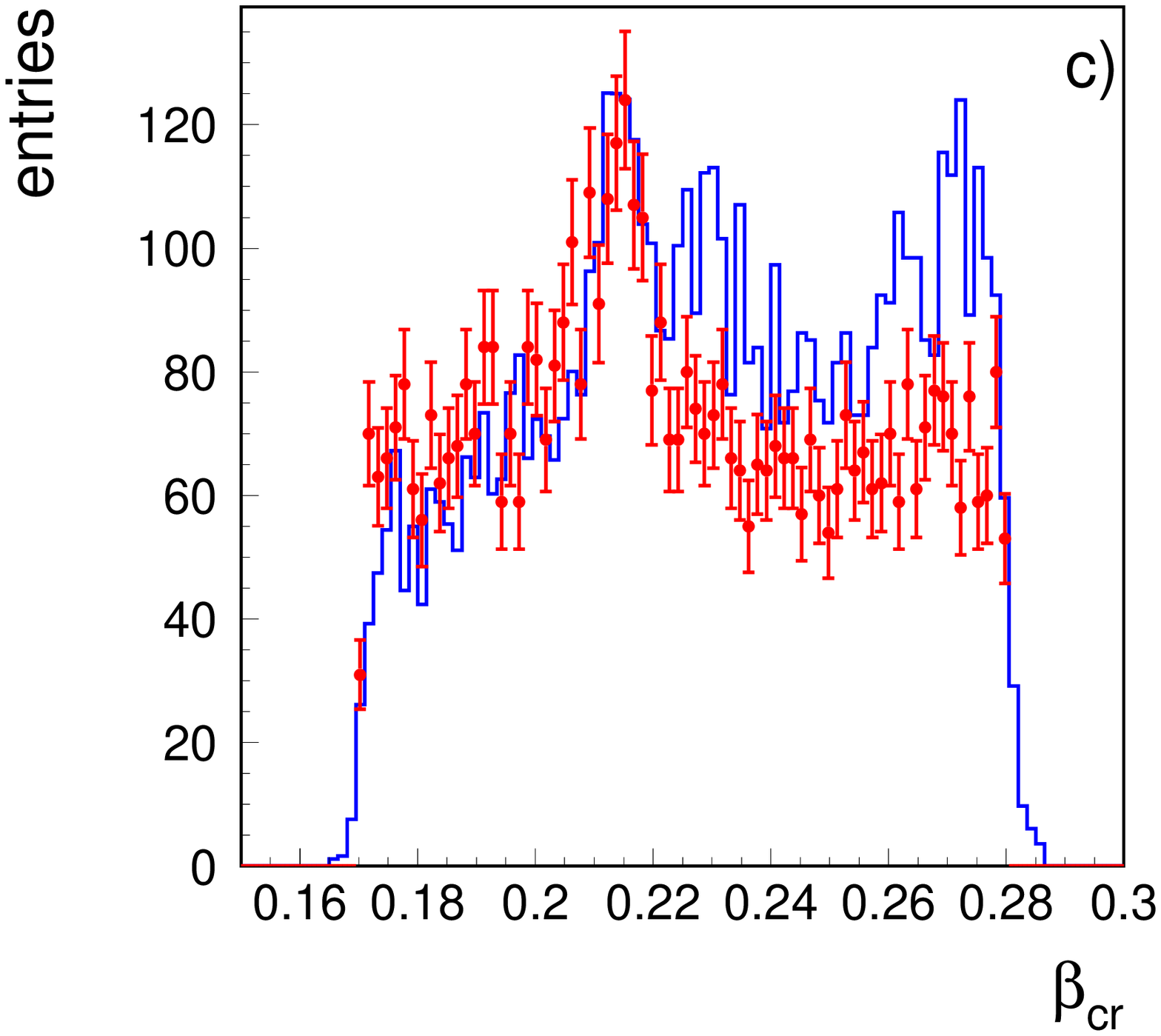}
\includegraphics[width=0.49\textwidth]{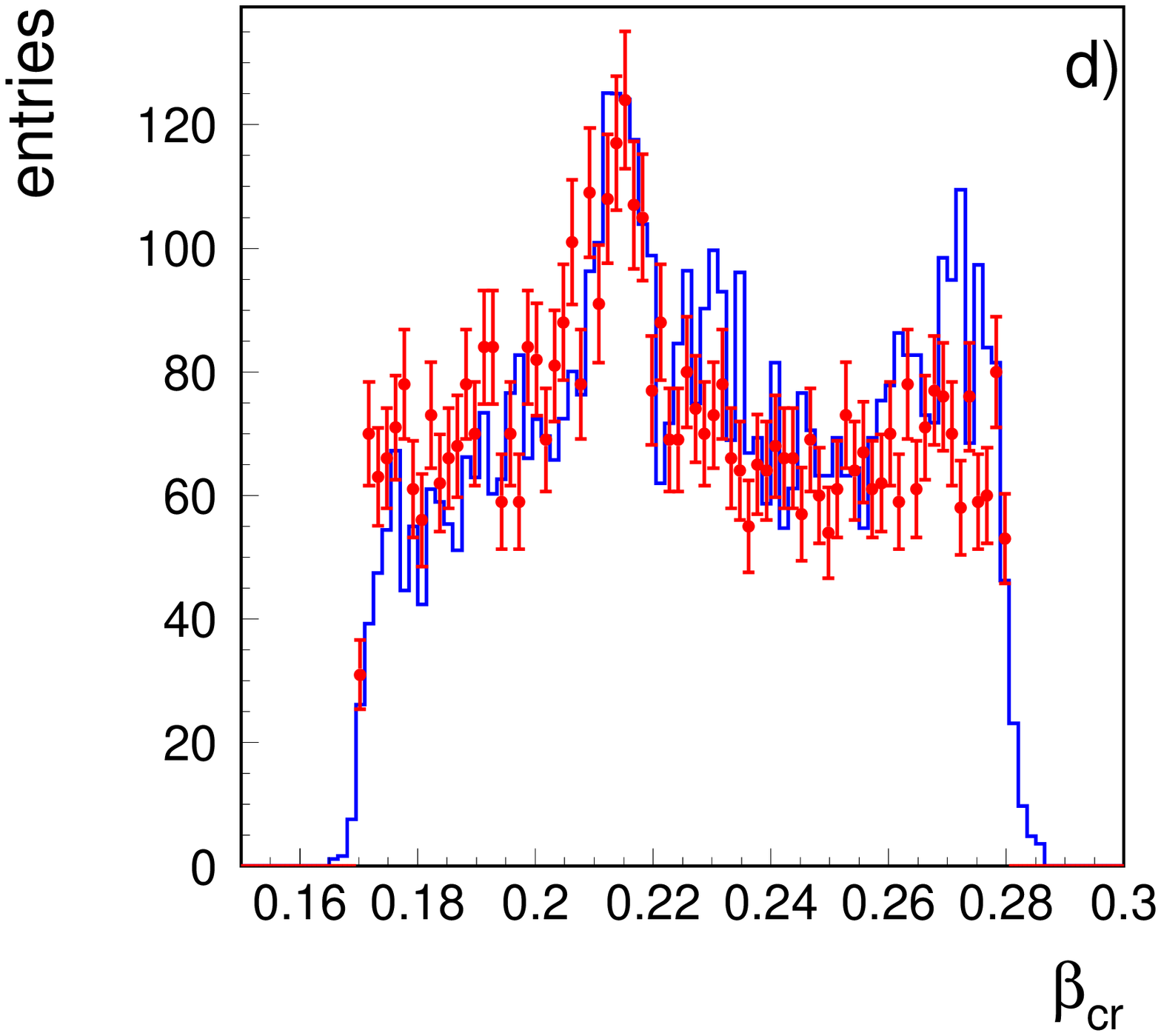}
\caption{The inclusive distributions of $R_{min}$ before (a ) and after (b ) correction. Plots c and d
present $\beta_{cr}$ spectrum for events with charged particles before and after the correction, respectively.
The red points represent the data, while blue histograms denote simulation results.}
\label{fig:Rminbeta}
\end{figure}
\\As it can be seen in Fig.~\ref{fig:Rminbeta}a, despite of application of the weighting factors for
different background categories there is still a small discrepancy between data and simulations
for the $R_{min}$ discriminant variable. Moreover it turns out, that we reject more events with charged
particles having $\beta_{cr}>0.220$ (see Fig.~\ref{fig:Rminbeta}c). Therefore, we introduce a small correction
shifting the $R_{min}$ by 2 cm smeared with a Normal-distributed random number. The $R_{min}$ distribution
after this correction is presented in Fig.~\ref{fig:Rminbeta}b where we observe much better agreement with data.
For the simulated events with $\beta_{cr} > 0.220$ belonging to the Fake category a correction based on
the ,,Hit \& Miss''
method was applied. For each rejected simulated event we draw a random uniformly-distributed number from the range
(0.; 1.), if this number was greater than 0.93 the event was passed for the further analysis. The $\beta_{cr}$
distribution for the rejected events after the correction is presented in Fig.~\ref{fig:Rminbeta}d.
\section{Optimization of selection criteria}
After all corrections described in last section as well as in chapter~\ref{rozdz5}, after application
of the scaling factors the Monte Carlo simulations provide a good description of the measured data.
This allows to determine the set of discriminant variables values which provide the best signal to
background ratio.
\\As a next step we optimize the event selection in order to reduce the background as strongly
as possible while keeping high signal efficiency. To this end the following
cuts were varied:
\begin{itemize}
\item the $\chi^2$ of the kinematical fit
\item the topological $\Delta E/\sigma_{E}$ cut
\item signal box definition in the $(\chi^2_{3\pi},\chi^2_{2\pi})$ plane
\item $R_{min}$~.
\end{itemize}
Each set of the cut values was applied to the simulated background and signal samples
excluding events with charged particles coming from the vicinity
of the interaction region and for several sets of tight cuts on the reconstructed $K_L$ energy and velocity.
This procedure allowed to determine the number of selected background events $B$ and
the signal efficiency $\epsilon_{3\pi}$ as a function of the five variables listed above:
\mbox{$B = B(\chi^2,\Delta E/\sigma_{E},\chi^2_{3\pi},\chi^2_{2\pi},R_{min})$} and 
\mbox{$ \epsilon_{3\pi} = \epsilon_{3\pi}(\chi^2,\Delta E/\sigma_{E},\chi^2_{3\pi},\chi^2_{2\pi},R_{min})$}.
Since the expected number of events at the end of the analysis chain is
small we define the following function:
\begin{equation}
f_{cut}~=~\frac{N_{up}(B)}{\epsilon_{3\pi}}~,
\end{equation}
where $N_{up}$ denotes the mean upper limit on the expected number of signal events calculated
at 90$\%$ confidence level assuming well -- known number of background events $B$\footnote{The
detailed description of the meaning of the mean upper limit and the statistical methods
for its estimation will be described in chapter~\ref{rozdz8}.}~\cite{lal92}.
The best choice of cut values is defined as the one which minimizes the $f_{cut}$ value.
As the result of the optimization we have obtained the following values of discriminant variables:
\begin{eqnarray}
\nonumber
\chi^2 < 57.2\\
\nonumber
\Delta E/\sigma_{E} \geq 1.88\\
\label{eqopcuts}
4.0 \leq \chi^2_{2\pi} \leq 84.9\\
\nonumber
\chi^2_{3\pi} \leq 5.2\\
\nonumber
R_{min} > 65~.
\end{eqnarray}
The signal efficiency corresponding to this set of cuts amounts to:\\
\begin{center}
$\epsilon_{3\pi} = 0.233 \pm 0.012_{stat}$.
\end{center}
\section{Counting of the $K_S\to 3\pi^0$ events}
\begin{figure}
\centering
\includegraphics[width=0.45\textwidth]{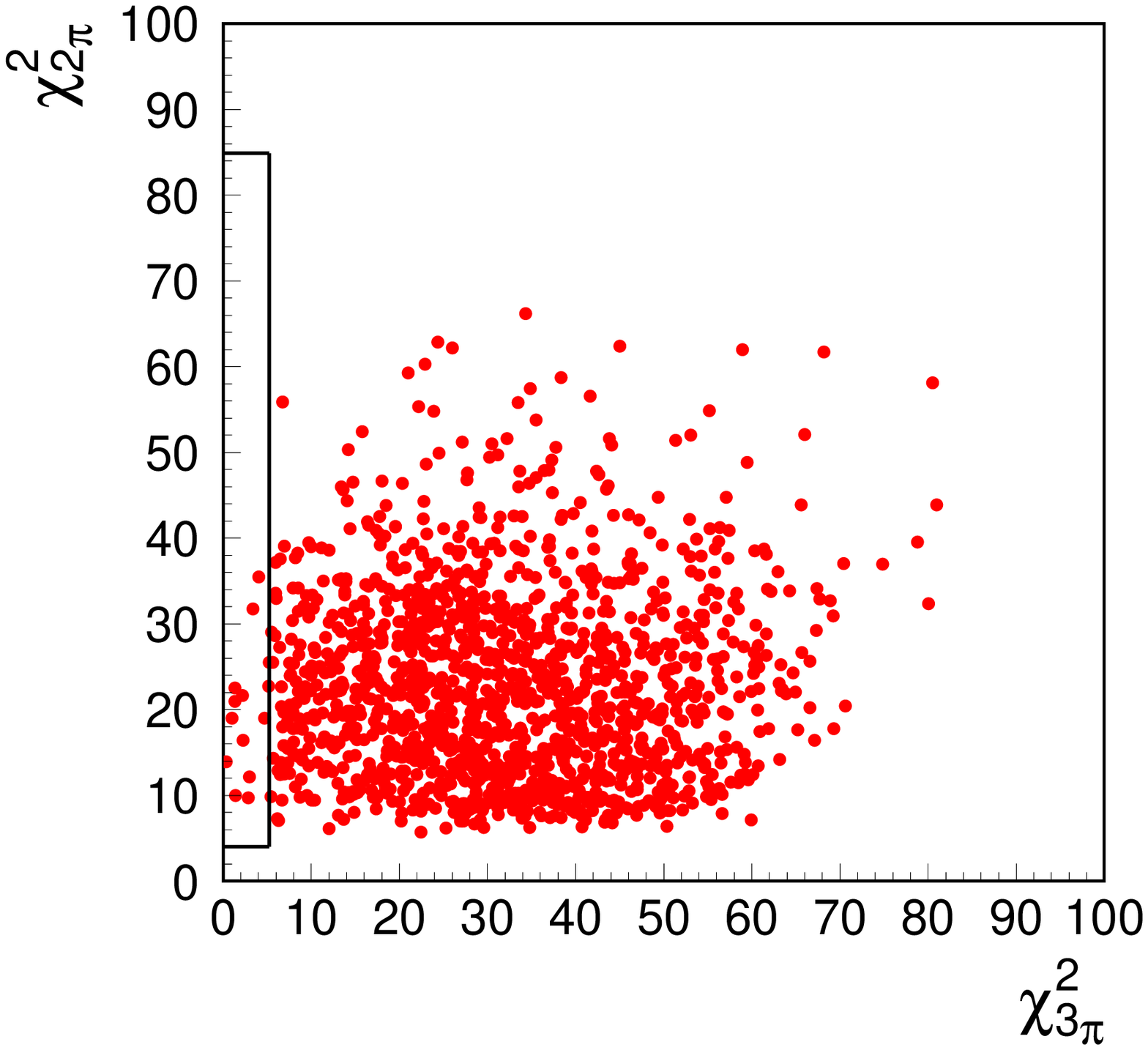}
\includegraphics[width=0.45\textwidth]{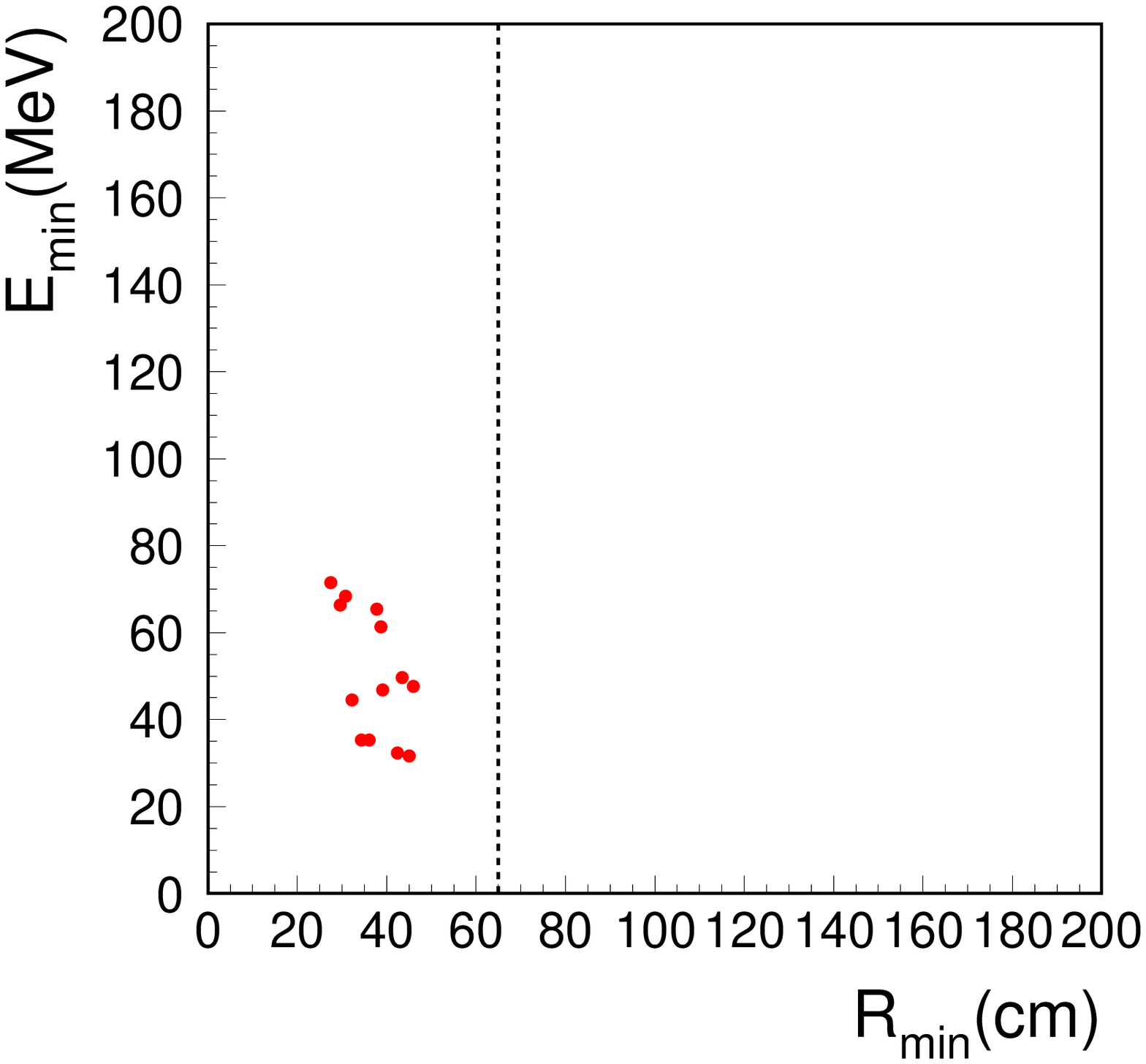}
\caption{Experimental $(\chi^2_{3\pi},\chi^2_{2\pi})$ distribution after
the $\chi^2$ and $\Delta E/\sigma_{E}$ cuts (left). The right panel presents the scatter
plot of $R_{min}$ versus the minimum energy of the cluster $E_{min}$ for events
in the Signal Box at the last stage of the analysis. The lines are described in the text.}
\label{RminEminChi2pi}
\end{figure}

After validation of the Monte Carlo simulations and determination of the optimal set of cut
values defined in Eq.~\ref{eqopcuts} we preform the discriminant analysis of the
experimental six gamma
sample preselected using the requirements for the $K_L$ mentioned before:
\begin{eqnarray}
E_{cr} > 150~\mathrm{MeV}
\nonumber \\
0.200 < \beta_{cr}< 0.225~,
\label{eq:hardKl}
\end{eqnarray}
The same analysis was also applied to the simulated background events with six reconstructed
gamma quanta fulfilling the tight requirements for $K_L$ listed above. This provided
the estimation of the expected background at the end of the analysis chain.
The experimental scatter plot of $(\chi^2_{3\pi}$ versus $\chi^2_{2\pi})$ after the $\chi^2$
and $\Delta E/\sigma_{E}$ cuts is presented in the left panel of Fig.~\ref{RminEminChi2pi}.
The solid lines show the signal region defined in Eq.~\ref{eqopcuts}.
As it can be seen in the right panel of Fig.~\ref{RminEminChi2pi} all events
selected by the Signal Box are characterized by $R_{min}$ less than 65 cm denoted by
the dashed line, thus at the end of the analysis chain we have found $N = 0$ of the $K_S \to 3\pi^0$
candidates in data. The expected background contribution amounts to $B_{exp} = 0$.
\begin{table}
\begin{center}
\begin{tabular}{|c|c|c|c|c|c|c|}
\hline
\textbf{} & \textbf{all 6 -- $\boldsymbol{\gamma}$ events} & \textbf{TRV} 
& $\boldsymbol{\chi^2_{fit}}$ & $\boldsymbol{\Delta E/\sigma_{E}}$ & \textbf{SBOX} & $\boldsymbol{R_{min}}$\\
\hline
\textbf{DATA}& 76689 $\pm$ 278 & 48963 $\pm$ 222 & 16501 $\pm$ 129 &  1400 $\pm$ 38 & 13 $\pm$ 4 & 0 $\pm$ 1\\
\hline
\textbf{MC} &  76721 $\pm$ 446 & 48984 $\pm$ 283 & 16230 $\pm$ 136 & 1210 $\pm$ 21 & 17 $\pm$ 3 & 0 $\pm$ 0.06\\
\hline
\end{tabular}
\end{center}
\caption{
\label{tab:cuts}
The number of events surviving each subsequent cut. Results for data are given in the first row (DATA)
and the second row shows the results for Monte Carlo simulations (MC). TRV denotes the
rejection of events with charged particles combined with the tight cuts on the energy and
velocity of $K_L$.}
\end{table}
\begin{table}
\begin{center}
\begin{tabular}{|c|c|c|c|c|c|c|}
\hline
\textbf{} & \textbf{SBOX} & \textbf{DOWN} & \textbf{UP} & \textbf{CDOWN} & \textbf{CUP} & \textbf{CSBOX}\\
\hline
\textbf{DATA}& 220 $\pm$ 15 & 137 $\pm$ 12 & 5 $\pm$ 3 & 6931 $\pm$ 83& 15179 $\pm$ 123 & 26491 $\pm$ 163\\
\hline
\textbf{MC} & 232 $\pm$ 11 & 100 $\pm$ 7 & 4 $\pm$ 3 & 6797 $\pm$ 76 &  14906 $\pm$ 116 & 26962 $\pm$ 169\\
\hline
\end{tabular}
\end{center}
\caption{
\label{tab:cuts1}
The population of control boxes in the $(\chi^2_{3\pi},\chi^2_{2\pi})$ plane right after rejection of events
with charged particles imposing the tight cuts on the reconstructed $K_L$ energy and velocity.}
\end{table}
\begin{table}
\begin{center}
\begin{tabular}{|c|c|c|c|c|c|c|}
\hline
\textbf{} & \textbf{SBOX} & \textbf{DOWN} & \textbf{UP} & \textbf{CDOWN} & \textbf{CUP} & \textbf{CSBOX}\\
\hline
\textbf{DATA}& 13 $\pm$ 4 & 0 $\pm$ 1 & 0 $\pm$ 1 & 0 $\pm$ 1 & 0 $\pm$ 1 & 1387 $\pm$ 37\\
\hline
\textbf{MC} & 17 $\pm$ 3 & 0 $\pm$ 0.06 & 0 $\pm$ 0.06 & 0 $\pm$ 0.06 & 0 $\pm$ 0.06 & 1194 $\pm$ 21\\
\hline
\end{tabular}
\end{center}
\caption{
\label{tab:cutsrminbox}
Population of control boxes in the $(\chi^2_{3\pi},\chi^2_{2\pi})$ plane defined in Subsec.~\ref{subs:wagi}
before the cut on $R_{min}$.}
\end{table}
\\For the final cross-check of the simulations credibility at each stage of the selection
the number of surviving events of both data and Monte Carlo samples were counted.
These numbers are reported in Tab~\ref{tab:cuts}. 
The statistical uncertainties for the results of simulations were estimated taking into account the scaling factors
and using the formula presented in Eq.~\ref{eq:wagiblad}. It can be seen that the simulated background
is consistent with data after each cut. Moreover, the agreement has been found also both at the beginning and
at the end of the analysis in all control boxes in the $(\chi^2_{3\pi},\chi^2_{2\pi})$ distribution
(see Tab.~\ref{tab:cuts1} and Tab.~\ref{tab:cutsrminbox}).
\\Since with the optimal cuts no events were observed only the upper limit on the $K_S \to 3\pi^0$ branching ratio
can be determined. The procedure used to estimate the upper limit taking into account the statistical
and systematical uncertainties is described in chapter~\ref{rozdz8}.
\chapter{Systematic error estimation}
\label{rozdz7}
In this chapter we present evaluation of the systematic uncertainties for the measurement
of the $K_S \to 3\pi^0$ branching ratio. They are related mainly to the determination
of the selection efficiencies for the signal and normalization sample $\epsilon_{2\pi}$
and $\epsilon_{3\pi}$, and estimation of the background and cuts used in the discriminant
analysis. Moreover we discuss small corrections due to the differences in the efficiencies
of the $K_S$ tagging and preselection with so called FILFO filter for the $K_S \to 3\pi^0$
and $K_S \to 2\pi^0$ decays. FILFO (\textit{FILtro FOndo}: background filter) is an off -- line procedure
identifying background events at a very early stage of the data reconstruction using only information
from the calorimeter. Events rejected by FILFO do not enter the track fitting and pattern
recognition algorithms which saves CPU-time during events reconstruction or reprocessing~\cite{memo288}.
The efficiency of the trigger and cosmic veto for both channels has been neglected due to the
fact, that they were found to be very close to 100$\%$ in previous KLOE analysis\footnote{
Since we have used the tighter ,,$K_L$ -- crash'' requirements the energy released in the calorimeter
by $K_L$ meson was much larger thus the efficiency of trigger is even higher than in the prior analysis.}~\cite{Matteo}.
We conclude giving a summary of the estimated systematic error affecting our result.
\section{Systematics related to acceptance for the $K_S \to 2\pi^0$ channel}
For the normalization sample we have considered the following list of systematic effects related
to the determination of the $\epsilon_{2\pi}$ efficiency:
\begin{itemize}
	\item \textbf{Difference in splitting and accidental probabilities between data and simulations}
	\\As it was mentioned in chapter~\ref{rozdz5}, the probability to find one or more accidental
	clusters in the acceptance is slightly different for data and simulations. To estimate
	the systematic uncertainty originating from this discrepancy we consider the fractions of reconstructed
	number of photons for the simulated $K_S \to 2\pi^0$ events
	$F^{true}_{k\gamma} = \frac{N^{rec}_{k\gamma}}{N_{tot}}$
	gathered in Tab.~\ref{tab:eff2pi} as the true ones. The presence of additional
	accidental clusters changes the true fraction $F^{true}_{4\gamma}$ (neglecting the second order effects)
	to:
	\begin{equation}
	F'_{4\gamma}~\approx~F^{true}_{4\gamma}\cdot(1-P_{A1}) + F^{true}_{3\gamma}\cdot P_{A1},
	\end{equation} 
	where $P_{A1}$ denotes the probability to find one accidental cluster in the event (see Tab.
	~\ref{tabprob}). This corresponds to the change
	$\Delta F_{4\gamma} = (F^{true}_{3\gamma} - F^{true}_{4\gamma})\cdot P_{A1}$.
	Therefore the systematic uncertainty originating from the difference
	of the probabilities $\Delta P_{A1}$ for data and Monte Carlo  amounts to:
\begin{center}
\mbox{$\Delta \epsilon_{2\pi}/\epsilon_{2\pi} = \Delta P_{A1} \cdot (F^{true}_{3\gamma} - F^{true}_{4\gamma})/
F^{true}_{4\gamma} = 7\cdot 10^{-4}$.}
\end{center}
Similarly we have calculated the systematic uncertainty due to different probabilities of shower
	fragmentation for data and simulations. In this case the variation of the true $F^{true}_{4\gamma}$
	amounts to:
	\begin{equation}
	\Delta F_{4\gamma}~\approx~(3\cdot F^{true}_{3\gamma} -4\cdot F^{true}_{4\gamma})\cdot P_{S1}~,
	\end{equation} 
	and the corresponding systematic uncertainty:
\begin{center}
\mbox{$\Delta \epsilon_{2\pi}/\epsilon_{2\pi} = \Delta P_{S1} \cdot (3\cdot F^{true}_{3\gamma} -4\cdot F^{true}_{4\gamma})/
F^{true}_{4\gamma} = 3\cdot 10^{-4}$,}
\end{center}
where $\Delta P_{S1}$ is the corresponding difference in the probabilities to generate one
	splitted cluster for data and Monte Carlo.
\item \textbf{Correction of the cluster reconstruction efficiency}\\
The systematic error due to this correction was estimated conservatively as the difference between
the true $f_{4\gamma}$ fractions evaluated with and without correction and amounts to:
$\Delta \epsilon_{2\pi}/\epsilon_{2\pi} = 5.2\cdot 10^{-3}$.
\item \textbf{Acceptance related effects}\\
The number of events counted as a normalization sample was determined in chapter~\ref{rozdz5} taking
into account only events with four reconstructed gamma quanta $N_{norm} = N_{2\pi}/\epsilon_{2\pi}$.
Assuming, that events with 3 -- 6 reconstructed photons originate from the $K_S \to 2\pi^0$ decay\footnote{
This assumption was checked looking to the true event decay chains for the simulations.}
$N_{norm}$ should be consistent with the result determined based on the number of events
with 3 -- 6 reconstructed photons: $N'_{norm} = N_{3-6\gamma}/\epsilon_{3-6\gamma}$\footnote{
The $\epsilon_{3-6\gamma}$ was determined using the same simulated $K_S \to 2\pi^0$ sample which was used
for the $\epsilon_{2\pi}$ determination (see Tab.~\ref{tab:eff2pi}).}.
Thus, the difference between these two numbers constitute the measure of the systematic
error amounting to $\Delta \epsilon_{2\pi}/\epsilon_{2\pi} = 1.5\cdot 10^{-2}$.
\item \textbf{FILFO preselection efficiency}\\
To reject the DA$\Phi$NE background before the track reconstruction a fast filter FILFO~\cite{memo288}
based only on information from calorimeter is applied. Further reconstruction is done only for events
which pass this filter. In the Monte Carlo we keep however all the simulated events which allows us to
estimate the efficiency. To this end we have considered the sample of $K_S \to 2\pi^0$
events simulated without ,,$K_L$ -- crash'' requirements retained and rejected by FILFO
(see Tab.~\ref{tab:filfo2pi}).
\begin{table}
\begin{center}
\begin{tabular}{|c|c|c|c|c|c|c|c|c|}
\hline
$\boldsymbol{k}$ & 0 & 1 & 2 & 3 & 4 & 5 & 6 & $\boldsymbol{\sum^{6}_{i=3}}$\\
\hline
$\boldsymbol{N^{k}_{A}}$ & 141 & 337 & 2916 & 16096 & 40048 & 649 & 39 & \textbf{56832}\\
\hline
$\boldsymbol{N^{k}_{R}}$ & 29 & 43 & 352 & 352 & 148 & 21 & 1 & \textbf{522}\\
\hline
$\boldsymbol{N^{k}_{A}/(N^{k}_{A} + N^{k}_{R}) [\%]}$ & 83 & 89 & 89.2 & 97.86 & 99.63 & 96.9 & 98 & \textbf{99.09}\\
\hline
\end{tabular}
\end{center}
\caption{
\label{tab:filfo2pi}
The number of simulated $K_S \to 2\pi^0$ events accepted ($N^{k}_{A}$) and rejected ($N^{k}_{R}$) by the FILFO
filter as a function of the number of reconstructed gamma quanta $k$. The numbers were obtained without any
cuts on the $K_L$ energy and velocity.}
\end{table}
\begin{table}
\begin{center}
\begin{tabular}{|c|c|}
\hline
\textbf{Source} & $\boldsymbol{\Delta \epsilon_{2\pi}/\epsilon_{2\pi}} [\%]$\\
\hline
\bf{Accidental} & 0.07\\
\hline
\bf{Splitting} &0.03\\
\hline
\bf{Accept. rel.} & 1.50\\
\hline
\bf{Clu. eff. corr} & 0.52\\
\hline
\bf{FILFO} & 0.46\\
\hline
\hline
\bf{TOTAL} & 1.65\\
\hline
\end{tabular}
\end{center}
\caption{
\label{tab:syssumm2pi}
Summary table of the systematic uncertainties on the selection efficiency for the $K_S \to 2\pi^0$
normalization sample.}
\end{table}
The efficiency is defined as a ratio of the number of events accepted by the filter with photon
multiplicities 3 -- 6 to the total number of events with gamma quanta in the same
range of multiplicity\footnote{We do not consider events with the multiplicities less than 3 because they originate
mainly from the wrong $T_0$ time assignment to the event.} (see Tab.~\ref{tab:filfo2pi}), and amounts to:
$\epsilon^{F}_{2\pi} = 0.9909 \pm 0.0004$. This value will be used in the final evaluation of the upper
limit presented in chapter~\ref{rozdz8}. As a systematic error contribution related to the preselection
with FILFO we take conservatively $\Delta \epsilon_{2\pi}/\epsilon_{2\pi} = (1-\epsilon^{F}_{2\pi})/2\epsilon_{2\pi} = 4.6\cdot 10^{-3}$.
\end{itemize}
Summary of the different contributions to the systematic uncertainty on $\epsilon_{2\pi}$ is presented
in Tab.~\ref{tab:syssumm2pi}, where the total error was evaluated adding all the contributions
in quadrature.\\
Finally, the estimated selection efficiency for the normalization sample amounts to:
\begin{equation}
 \epsilon_{2\pi} = (0.660 \pm 0.002_{stat} \pm 0.010_{sys})~.
\end{equation} 
\section{Systematics related to the selection efficiency and background for the $K_S \to 3\pi^0$ channel}
For the search of the $K_S \to 3\pi^0$ signal the main sources of systematic uncertainties originate from
the estimation of background and selection efficiency. As in the case of the $K_S \to 3\pi^0$ channel we have
considered also the systematic effects related to the acceptance.\\
For systematic study of the background we have repeated the analysis changing the parameters values used in
$\chi^2_{2\pi}$ and $\chi^2_{3\pi}$ calculation (see Tab.~\ref{tabchipar}) as well as varying all the corrections
applied to the Monte Carlo simulations, namely:
\begin{itemize}
 \item using the same resolutions for data and simulations in the $\chi^2_{2\pi}$ definition
 \item using different resolutions for data and simulations in the $\chi^2_{3\pi}$ definition obtained with
control sample consisting of events with charged particles (mainly $K_S \to \pi^+\pi^-; K_L \to 3\pi^0$)
\item removing correction on $R_{min}$ 
\item repeating  the analysis with different energy scale corrections
\item removing correction on the rejection of events with charged particles
\item varying $\sigma_{E}$ in the $\Delta E/\sigma_{E}$ definition ($\sigma_{E} \pm \delta(\sigma_{E})$, where
$\delta(\sigma_{E})$ denotes the standard deviation of $\sigma_{E}$)
\item varying the cuts on the reconstructed $K_L$ energy $E_{cr}$ and velocity $\beta_{cr}$ arbitrarly
by $\pm$ 5$\%$.
\end{itemize}
The full analysis was repeated in total twenty times performing each time one of the systematical checks
listed above. For all of the checks we have not observed any changes in the number of background events
at the end of the analysis chain.\\
As in the case of the $K_S \to 2\pi^0$ decay we have considered the following systematic effects for
the selection efficiency related to the acceptance cuts:
\begin{itemize}
 \item \textbf{Splitting and accidental probabilities for data and simulations}\\
Based on the same simulated sample of the $K_S \to 3\pi^0$ events which has been used for estimation of
the selection efficiency $\epsilon_{3\pi}$ we have determined the true fractions $F^{true}_{k\gamma}$ of
the reconstructed number of photons for signal (see Tab.~\ref{tab:eff3pi}).
\begin{table}
\begin{center}
\begin{tabular}{|c|c|c|c|c|c|c|c|c|c|}
\hline
$\boldsymbol{k}$ & 0 & 1 & 2 & 3 & 4 & 5 & 6 & 7 & 8\\
\hline
$\boldsymbol{N^{rec}_{k\gamma}}$ & 2 & 3 & 1 & 12 & 133 & 435 & 597 & 11 & 1\\
\hline
$\boldsymbol{F^{true}_{k\gamma}} [\%]$ & 0.17 & 0.25 & 0.08 & 1.0 & 11.1 & 36.4 & 50.0 & 0.92 & 0.09\\
\hline
\end{tabular}
\end{center}
\caption{
\label{tab:eff3pi}
The number of events $N^{rec}_{k\gamma}$ reconstructed with a multiplicity of clusters $k$
for $N_{tot}$ = 1195 of $\phi \to K_SK_L \to 6\gamma K_L$ events simulated with 0.200 < $\beta_{cr}$ < 0.225 and
$E_{cr}$ > 150 MeV. $F^{true}_{k\gamma}$ denotes the true fraction defined as $F^{true}_{k\gamma} =
N^{rec}_{k\gamma}/\sum^{8}_{k=0} N^{rec}_{k\gamma}$.}
\end{table}
The changes  of the true $F^{true}_{6\gamma}$ fraction introduced by the presence of accidental clusters amounts
to approximately: $\Delta F_{6\gamma} = (F^{true}_{5\gamma} - F^{true}_{6\gamma})\cdot P_{A1}$. Systematic uncertainty
originating form the difference of the probabilities for data and Monte Carlo simulations amounts to:
\begin{center}
$\Delta \epsilon_{3\pi}/\epsilon_{3\pi} = \Delta P_{A1} \cdot (F^{true}_{5\gamma} -
F^{true}_{6\gamma})/F^{true}_{6\gamma} = 7 \cdot 10^{-4} $.
\end{center}
Similarly the systematic uncertainty corresponding to the difference in the splitting probabilities
has been found to be:
\begin{center}   
\mbox{$\Delta \epsilon_{3\pi}/\epsilon_{3\pi} = \Delta P_{S1} \cdot (5 \cdot F^{true}_{5\gamma} -
6 \cdot F^{true}_{6\gamma})/F^{true}_{6\gamma} = 3 \cdot 10^{-4} $.}
\end{center}
\item \textbf{Correction of the cluster reconstruction efficiency}\\
The systematic error due to this correction was estimated as in the case of the $K_S \to 2\pi^0$ channel
and amounts to $\Delta \epsilon_{3\pi}/\epsilon_{3\pi} = 2 \cdot 10^{-3}$.
\item \textbf{Energy scale correction}\\
As it was described in chapter~\ref{rozdz5} we have modified the energy scale of the reconstructed gamma
quanta in the Monte Carlo simulations to provide a better agreement with data. Systematic error connected
with this correction has been estimated by changing the value of the energy shift from 2.2$\%$ to 2.6$\%$,
and amounts to $\Delta \epsilon_{3\pi}/\epsilon_{3\pi} = 1 \cdot 10^{-2}$.
\item \textbf{FILFO preselection efficiency for the $K_S \to 3\pi^0$}\\
The efficiency of FILFO filter was estimated analogously to the derivation presented in the previous
section. The distribution of the $K_S \to 3\pi^0$ events simulated without any requirements for $K_L$
energy and velocity surviving and rejected by FILFO is presented in Tab.~\ref{tab:filfo3pi}.
\begin{table}
\begin{center}
\begin{tabular}{|c|c|c|c|c|c|c|c|c|c|}
\hline
$\boldsymbol{k}$ & 0 & 1 & 2 & 3 & 4 & 5 & 6 & 7 & $\boldsymbol{\sum^{6}_{i=3}}$\\
\hline
$\boldsymbol{N^{k}_{A}}$ & 11 & 12 & 1 & 23 & 208 & 658 & 903 & 15 & 1792\\
\hline
$\boldsymbol{N^{k}_{R}}$ & 0 & 0 & 0 & 0 & 6 & 4 & 1 & 1 & 11\\
\hline
$\boldsymbol{N^{k}_{A}/(N^{k}_{A} + N^{k}_{R}) [\%]}$ &100 & 100 & 100 & 100 & 97.2 & 99.4 & 99.9 & 93.8& 99.4\\
\hline
\end{tabular}
\end{center}
\caption{
\label{tab:filfo3pi}
The number of simulated $K_S \to 3\pi^0$ events accepted ($N^{k}_{A}$) and rejected ($N^{k}_{R}$) by the FILFO
filter as a function of the number of reconstructed gamma quanta $k$. The numbers were obtained without any
cuts on the $K_L$ energy and velocity.}
\end{table}
The estimated efficiency of the filter for signal is equal to
$\epsilon^{F}_{3\pi} = 0.994 \pm 0.002$. This value will be used in the final evaluation of the upper
limit presented in chapter~\ref{rozdz8}. As a systematic error contribution related to the preselection
with FILFO we take conservatively $\Delta \epsilon_{3\pi}/\epsilon_{3\pi} = (1-\epsilon^{F}_{3\pi})/2\epsilon_{3\pi} = 3\cdot 10^{-3}$.
\end{itemize}
The last group of systematic uncertainties is connected with the cut sequence used in the discriminant
analysis:
\begin{itemize}
\item \textbf{Energy resolution}\\
The systematical uncertainty due to energy resolution was determined by estimation of the selection efficiency
 $\epsilon_{3\pi}$ with different $\sigma_{E}$ values in the $\Delta E/\sigma_{E}$ definition, which was varied
 as it was described in the case of systematics related to the background estimation.
 In this case it amounts to $\Delta \epsilon_{3\pi}/\epsilon_{3\pi} = 1.1 \cdot 10^{-2}$.
\item \textbf{$\boldsymbol{R_{min}}$ cut}\\
The systematic effects related to the $R_{min}$ cut was studied comparing the selection efficiencies evaluated
with and without the $R_{min}$ correction. Difference of these two values gives the systematic uncertainty
equal to $\Delta \epsilon_{3\pi}/\epsilon_{3\pi} = 9 \cdot 10^{-3}$.
\item \textbf{$\boldsymbol{\chi^2_{fit}}$ cut}\\
Systematic effects due to the $\chi^2_{fit}$ cut were investigated using the $K_S \to 2\pi^0$ events with four
reconstructed gamma quanta. Since the photon multiplicity and energy spectrum of the $K_S \to 2\pi^0$
events differs from the ones for signal we expect differences in the shape
of $\chi^2_{fit}/ndof$\footnote{
$ndof$ denotes the number of degrees of freedom which amounts to 11 for the $K_S \to 3\pi^0$ events
and 9 for the $K_S \to 2\pi^0$ channel.} for the two samples. To estimate the systematic uncertainty
related to this difference we have compared the simulated $\chi^2_{fit}/ndof$ distributions
for the $K_S \to 2\pi^0$ and $K_S \to 3\pi^0$ events. For both distributions the cumulative curves
$f^{MC}_{2\pi}$ and $f^{MC}_{3\pi}$ were determined (see Fig.~\ref{fig:cumu}a).
The ratio $f^{MC}_{2\pi}/f^{MC}_{3\pi}$ which is shown in Fig.~\ref{fig:cumu}b constitutes
the estimation of the contribution to the systematic error. For the cut value used in the analysis
($\chi^2_{fit}/ndof = 5.2$) it corresponds to $1.22 \cdot 10^{-2}$. Also for the measured $K_S \to 2\pi^0$ events
we have constructed a cumulative curve $f^{Data}_{2\pi}$ presented in Fig.~\ref{fig:cumu}c.
The ratio $f^{Data}_{2\pi}/f^{MC}_{2\pi}$ gives us the second part of the systematic error equal
to $8 \cdot 10^{-3}$ (see Fig.~\ref{fig:cumu} d ). Adding the two contributions in quadrature
we obtain the total systematic uncertainty on the selection efficiency related to the $\chi^2_{fit}$ cut
equal to $\Delta \epsilon_{3\pi}/\epsilon_{3\pi} = 1.46 \cdot 10^{-2}$. 
\begin{figure}
\centering
\includegraphics[width=0.4\textwidth]{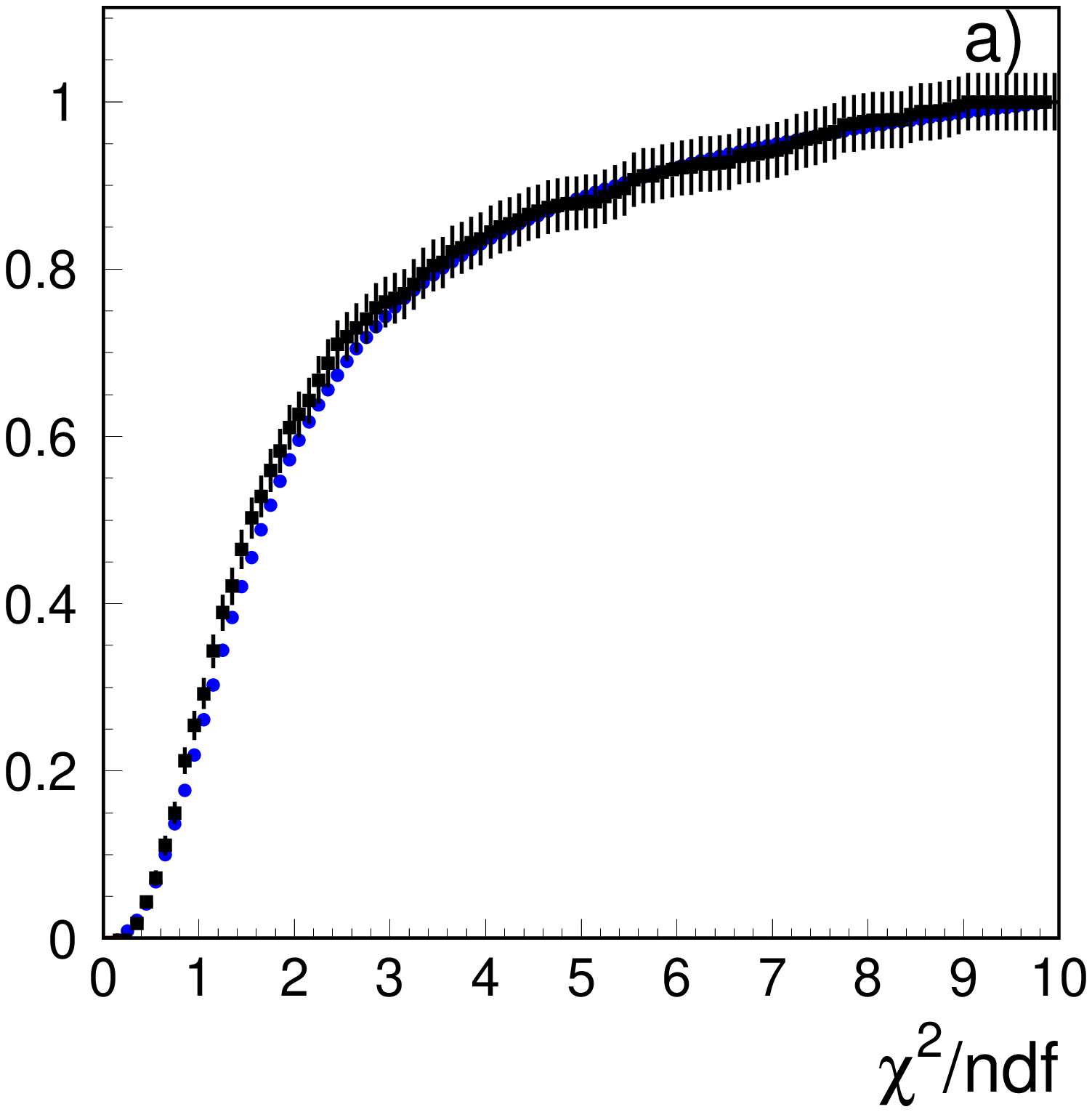}
\includegraphics[width=0.4\textwidth]{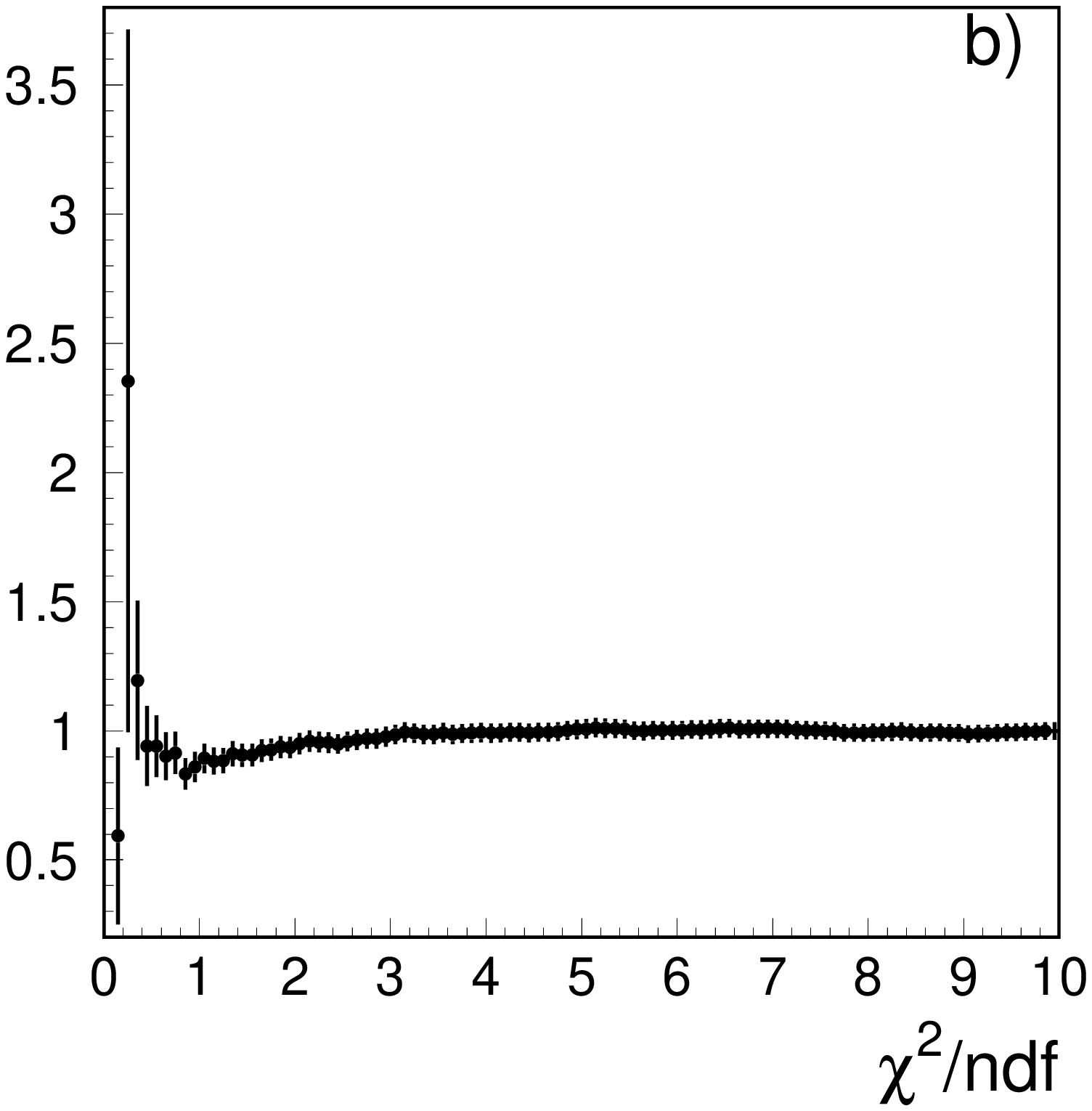}
\includegraphics[width=0.4\textwidth]{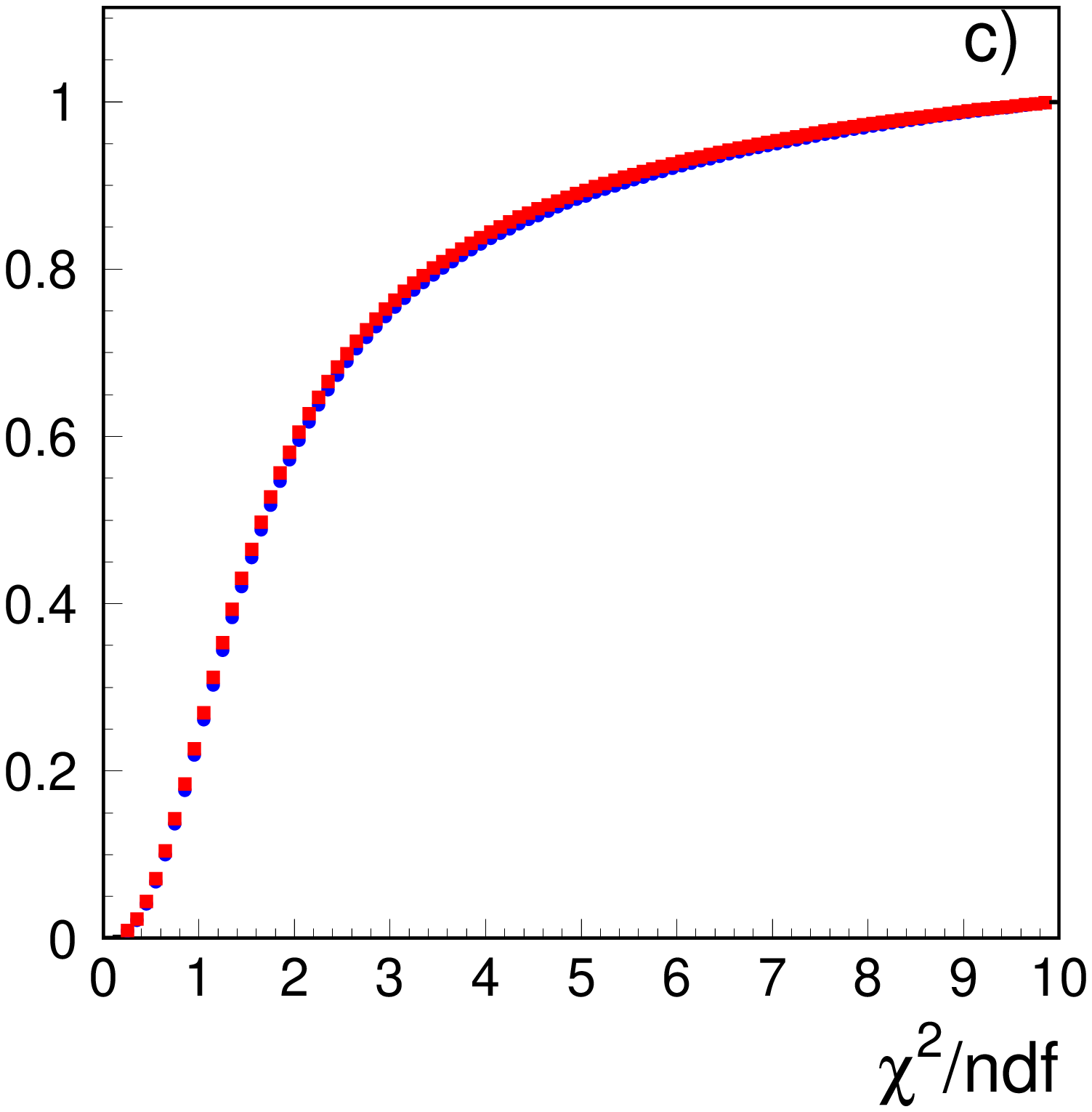}
\includegraphics[width=0.4\textwidth]{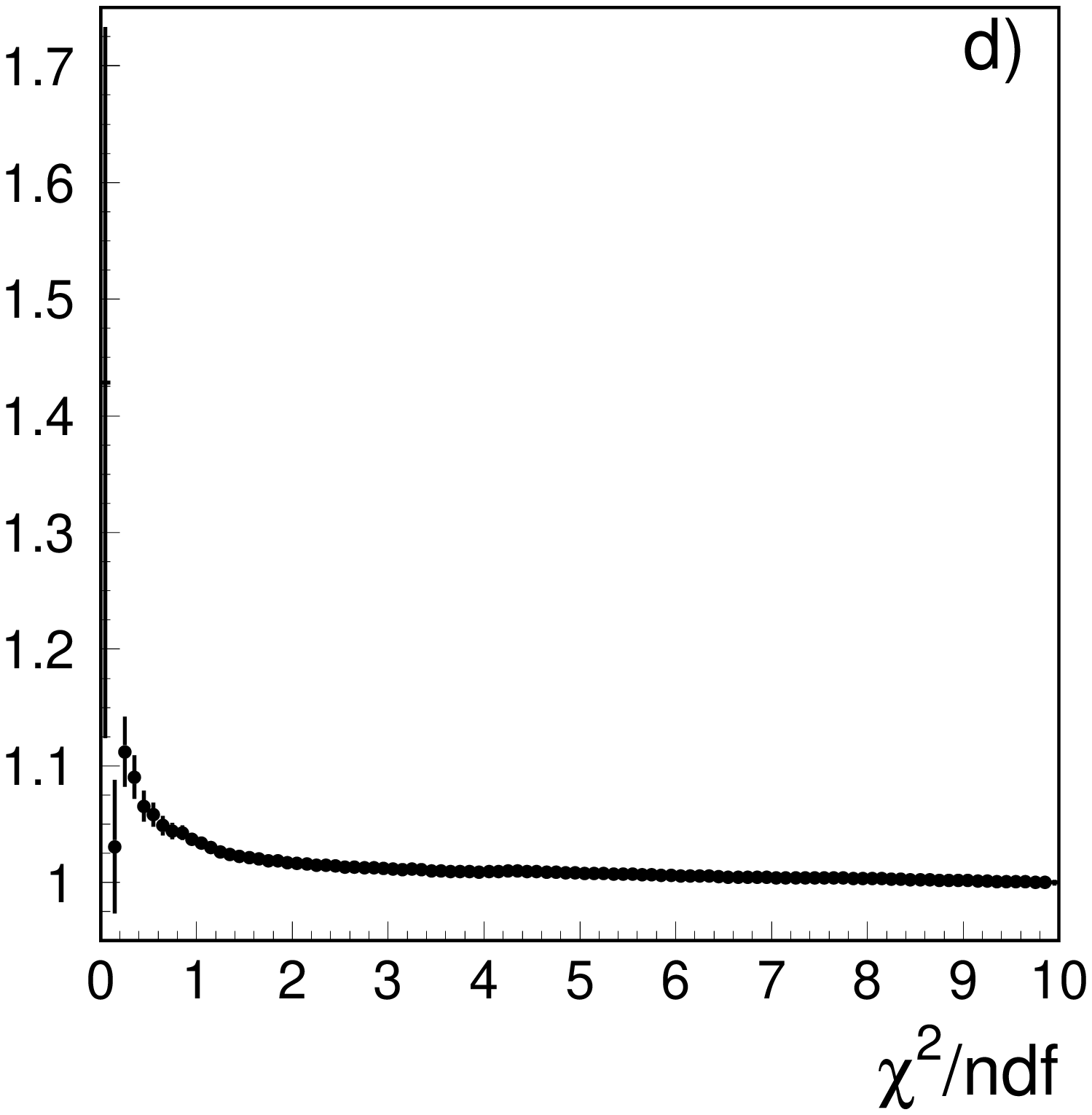}
\caption{a) The cumulative curves of the $\chi^2_{fit}/ndof$ distributions for the simulated
$K_S \to 3\pi^0$ (black squares) and $K_S \to 2\pi^0$ (blue circles) samples and  b) their ratio;
c) Comparison of $\chi^2_{fit}/ndof$ cumulative curves obtained with data (red squares) and
simulations (blue circles) for the four gamma events and d) their ratio. 
}
\label{fig:cumu}
\end{figure}
\end{itemize}
\begin{table}
\begin{center}
\begin{tabular}{|c|c|}
\hline
\textbf{Source} & $\boldsymbol{\Delta \epsilon_{3\pi}/\epsilon_{3\pi}} [\%]$\\
\hline
\bf{Accidental} & 0.03\\
\hline
\bf{Splitting} & 0.02\\
\hline
\bf{Energy scale} & 1.00\\
\hline
\bf{Clu. eff. corr} & 0.20\\
\hline
\bf{FILFO} & 0.30\\
\hline
$\boldsymbol{\chi^2_{fit}}$ & 1.46\\
\hline
\bf{Energy resolution} & 1.10\\
\hline
$\boldsymbol{R_{min}}$ & 0.90\\
\hline
\hline
\bf{TOTAL} & 2.30\\
\hline
\end{tabular}
\end{center}
\caption{
\label{tab:syssumm3pi}
Summary table of the systematic uncertainties on the selection efficiency for the $K_S \to 3\pi^0$ decay.}
\end{table}
Different contributions to the systematic error on $\epsilon_{3\pi}$ are summarized
in Tab.~\ref{tab:syssumm3pi}, where the total systematic uncertainty was evaluated adding all of them
in quadrature.\\
The final value of the selection efficiency for the $K_S \to 3\pi^0$ decay amounts to:
\begin{equation}
 \epsilon_{3\pi} = 0.233 \pm 0.012_{stat} \pm 0.006_{sys}~. 
\end{equation}
\section{Correction for the different $K_S$ tagging efficiencies for the $K_S \to 3\pi^0$ and $K_S \to 2\pi^0$ decays}
The difference in the kinematics  and in the photon multiplicity between the $K_S \to 3\pi^0$ and $K_S \to 2\pi^0$
decays creates a small difference in the $K_S$ tagging efficiencies for these channels. This may be result, for
example of accidental or splitting clusters which can modify spectrum of the reconstructed $K_L$ energy.
To take into account this small effect we have determined the $K_S$ tagging efficiencies independently
for each channel using the appropriate simulated sample of events. The estimated efficiencies amounts to
\begin{center}
$\epsilon^{2\pi}_{cr} = (23.65 \pm 0.12) \cdot 10^{-2}$ for the $K_S \to 2\pi^0$, 
\end{center}
\begin{center}
and $\epsilon^{3\pi}_{cr} = (23.90 \pm 0.90) \cdot 10^{-2}$ for the $K_S \to 3\pi^0$.
\end{center}
The ratio $R_{cr} = \epsilon^{3\pi}_{cr}/\epsilon^{2\pi}_{cr} = 1.01 \pm 0.04$ constitutes a correction which will be used
in the final evaluation of the upper limit on the $K_S \to 3\pi^0$ branching ratio.
\chapter{Upper limit on the $K_S \to 3\pi^0$ branching ratio and $|\eta_{000}|$}
\label{rozdz8}
\section{Upper limit on the measured number of $K_S \to 3\pi^0$ decays}
The search for the $K_S\to 3\pi^0$ decay presented in this work failed to detect a signal
of sufficient statistical significance. Therefore, one can only determine an upper
limit on the branching ratio for this decay at a chosen confidence level.
A limit on a physical quantity at a given confidence level is usually set by comparing
a number of detected events with the expected number of background events in
the signal region. The expected background depends strongly on the systematic uncertainties
existing in the measurement, therefore they should be taken into consideration in the limit
or confidence belt calculation~\cite{Zhu}.\\
In the framework of frequentist statistics confidence limits are set using a Neyman
construction~\cite{Neyman}. This method suffers however from so-called undercoverage
when the observable is close to the physics boundary (the actual coverage is less than the
requested confidence level). Moreover, constructed confidence intervals may be unphysical
or empty, in particular in the case when no events have been observed this method gives no answer
for the confidence interval~\cite{Zhu}.\\
Feldman and Cousins~\cite{Feldman} proposed a new method of confidence interval estimation based
on likelihood ratios which automatically provides a central confidence interval
or an upper confidence limit decided by the observed data itself (so called unified approach)
~\cite{Zhu}. However, if the observable is a Poisson distributed variable, there is a background
dependence of the upper limit in the case of fewer events observed than expected background.
This gives a smaller upper limit for measurements with higher background, which is clearly
undesirable. A solution to this problem was proposed by Roe and Woodroofe taking advantage of
a fact that given an observation $n$, the background $b$ cannot be larger than $n$~\cite{Roe}.
Therefore, the usual Poisson pdf (probability density function) used to construct the confidence
interval should be replaced by a conditional pdf. This approach solves the background dependence
of the upper limit, however, does not satisfy all the requirements of proper coverage and has
problems when applied to the case of a Gaussian distribution with boundaries~\cite{Zhu}.\\
All approaches described above do not take into account the systematic uncertainties of estimated
background and signal efficiency. Several methods have been developed to incorporate the systematic
errors to the calculation of upper limits. An entirely frequentist approach has been proposed 
for the uncertainty in the background rate prediction~\cite{Rolke}. It is based on
a two-dimensional confidence belt construction and likelihood ratio hypothesis testing.
This method treats the uncertainty in the background as a statistical uncertainty rather than
as a systematic one~\cite{Zhu}.\\
Several methods combine classical and Bayesian elements (so called semi-Bayesian approaches)
for example incorporating systematic uncertainty by performing average over the probability of the
detection efficiency~\cite{Cousins}. This method is however of limited accuracy in the limit
of high relative systematic uncertainties~\cite{Zhu}.
Conrad \textit{et al.} extended the method of confidence belt construction proposed by
Feldman and Cousins. This approach includes systematic uncertainties of both the signal efficiency
and background prediction by integration over a pdf parameterizing our knowledge
about the sources of the uncertainties~\cite{Conrad}.\\
In the framework of Bayes statistics the systematic error is included by modification of the
usual conditional pdf with additional probability density functions for the background
expectation and signal efficiency~\cite{Zhu,Narsky}.\\
Summarizing, there are many approaches for upper limit estimation and still there is
a lot of effort made towards improving these methods and understanding their practical implications.
All the approaches mentioned before give in principle different confidence intervals and one has to
choose the method depending on the relevance of systematic uncertainties and personal opinion
about the Bayesian and classical statistics. The analysis described in this thesis resulted in
$n=0$ observed events with the number of expected background events $b=0$. In this case all
the methods reveal that the upper limit is almost insensitive to the
systematic errors of the background estimation and signal efficiency~\cite{Cousins,Rolke,Zhu,Narsky}.
Moreover, in the case of $n = b = 0$ the upper limit on the number of signal events at the 90$\%$ confidence level amounts to 
2.21~\cite{Rolke} or 2.33~\cite{Cousins,Zhu,Conrad,Narsky} for the systematic uncertainties on
background and signal efficiency less than 10$\%$. As it was shown in chapter~\ref{rozdz7} in our
search the systematical errors do not exceed 5$\%$, therefore we assume the upper limit on the number
of the $K_S\to 3\pi^0$ events at the 90$\%$ confidence level amounting to  $N^{up}_{3\pi}(0.9) = 2.33$.
This number will be used in the next section for the calculation of the upper limit on $K_S\to 3\pi^0$
branching ratio.
\section{Determination of the upper limits on the $K_S \to 3\pi^0$ branching ratio and $|\eta_{000}|$}
From the limit on the number of expected signal events one can calculate a limit on the branching ratio.
Recalling Eq.~\ref{eq:1roz5} and taking into account the corrections for the difference in tagging
and FILFO preselection efficiencies for signal and the $K_S \to 2\pi^0$ normalization sample described in
chapter~\ref{rozdz7} we obtain the following expression for the branching ratio:
\begin{equation}
BR(K_S \to 3\pi^0) = \frac{1}{R_{cr}}\cdot \frac{\epsilon^{F}_{2\pi}}{\epsilon^{F}_{3\pi}}\cdot
\frac{N^{up}_{3\pi}(0.9)/\epsilon_{3\pi}}{N_{2\pi}/\epsilon_{2\pi}}\cdot BR(K_S \to 2\pi^0)~,
\end{equation}
where $R_{cr}$ denotes the ratio of tagging efficiencies for signal and the $K_S \to 2\pi^0$
normalization sample amounting to $R_{cr}=1.01 \pm 0.04$. $\epsilon^{F}_{3\pi}=0.994 \pm 0.002$
and $\epsilon^{F}_{2\pi}=0.9869 \pm 0.0006$ are the FILFO preselection efficiencies determined
in chapter~\ref{rozdz7}.
The branching ratio for the $K_S \to 2\pi^0$ channel is equal to:
$BR(K_S \to 2\pi^0) = 0.3069 \pm 0.0005$~\cite{pdg2010}. Taking into account value of the upper
limit on the number of expected signal: $N^{up}_{3\pi}(0.9) = 2.33$, the selection efficiency for the
$K_S \to 3\pi^0$ channel: $\epsilon_{3\pi} = 0.233 \pm 0.012_{stat} \pm 0.005_{sys}$ and the total
number of $K_S \to 2\pi^0$ events: $N_{2\pi}/\epsilon_{2\pi} = (1.142 \pm 0.005) \cdot 10^8$,
we have obtained the upper limit on the $K_S \to 3\pi^0$ branching ratio at the 90$\%$ confidence level:
\begin{equation}
\nonumber
BR(K_S \to 3\pi^0) \leq 2.7\cdot 10^{-8}~.
\label{eq:brvalue}
\end{equation}
This value is almost five times lower than the latest result $BR(K_S \to 3\pi^0) \leq 1.2\cdot 10^{-7}$
published by KLOE~\cite{Matteo}.
The upper limit on the $K_S \to 3\pi^0$ branching ratio can be translated into a limit on the
$|\eta_{000}|$ at the 90$\%$ confidence level~\cite{Thomson}:
\begin{equation}
\nonumber
|\eta_{000}| = \left|\frac{A(K_S \to 3\pi^0)}{A(K_L \to 3\pi^0)}\right| = \sqrt{\frac{\tau_L}{\tau_S}
\frac{BR(K_S \to 3\pi^0)}{BR(K_L \to 3\pi^0)}} \leq 0.009~,
\label{eq:etavalue}
\end{equation}
which corresponds to an improvement of the $|\eta_{000}|$ uncertainty by a factor of two with respect to
the latest direct search~\cite{Matteo}.
\chapter{Summary and outlook}
Kaons have been a remarkably important particles in the development of the Standard Model.
The kaon system, being a relatively simple one, played the key role in the discovery of
such phenomena as parity and $\mathcal{CP}$ violation, the GIM mechanism and the existence
of charm, and was central in the investigation of lepton flavour and $\mathcal{CPT}$
symmetries~\cite{Sozzi:2011yt}.\\
$\mathcal{CP}$ violation is deeply related to such fundamental issues as the microscopic
time reversibility of physical laws and the origin of the baryonic asymmetry of the
universe. It is also the only known phenomenon which allows an absolute distinction
between particles and antiparticles~\cite{Sozzi:2003ve}.
Since the first discovery in the neutral kaon system the $\mathcal{CP}$ symmetry
breaking has been a very active field of research. Although at present the main
experimental effort is focused on the neutral B and D meson system studies, there are
still several interesting open issues in the kaon physics demanding investigation.
One of them is the $K_S \to 3\pi^0$ decay which still remains undiscovered.
The best published upper limit on its branching ratio $BR(K_S \to 3\pi^0) < 1.2\cdot 10^{-7}$
is still two orders of magnitude larger than the predictions based on the Standard Model.
Therefore, the complete understanding of the $\mathcal{CP}$ violation in the neutral kaon
system demands a new high -- precision experiments which will contribute also to the
$\mathcal{CPT}$ conservation tests.\\
This work presents the search of the $K_S \to 3\pi^0$ decay based on the data sample
gathered in 2004 -- 2005 with the KLOE detector operating at the $\Phi$ -- factory
DA$\Phi$NE in the Italian National Center for Nuclear Physics in Frascati. DA$\Phi$NE
is an $e^+$ and $e^-$ collider optimized to work at the center of mass energy $\sqrt{s} = 1019.45$ MeV.
In the two storage rings of DA$\Phi$NE 120 bunches of electrons and positrons are stored. Each bunch
collides with its counterpart once per turn, minimizing the mutual perturbations of
colliding beams.
The $e^+e^-$ collisions result in the $\phi$ meson creation which is almost at rest and
decay predominantly to kaon pairs.
The decay products are registered using the KLOE detection setup surrounding the $e^+e^-$ interaction
point. KLOE consists of large
cylindrical drift chamber surrounded by the electromagnetic calorimeter.
The detectors are immersed in the axial magnetic field generated by
superconducting solenoid. The $K_S$ mesons were identified via registration of these
$K_L$ mesons which crossed the drift chamber without decaying
and then interacted with the KLOE electromagnetic calorimeter. The $K_S$
four -- momentum vector was then determined using the registered position of the $K_L$ meson
and the known momentum of the $\phi$ meson. The search for the $K_S \to 3\pi^0 \to 6\gamma$ decay
was then carried out by the selection of events with six gamma quanta which momenta were
reconstructed using time and energy measured by the electromagnetic calorimeter.
Background for the searched decay originated mainly from the $K_S \to 2\pi^0$ events
with two spurious clusters from fragmentation of the electromagnetic showers or accidental
coincidences with signals generated due to particle loss of DA$\Phi$NE beams, or from
false $K_L$ identification.
To increase the signal over background ratio after identification of the $K_S$ meson and 
requiring six reconstructed photons a discriminant analysis was performed.
It started from rejection of events with charged particles coming from the vicinity of
the interaction region which suppress
the $\phi \to K_{S}K_{L} \to (K_S \to \pi^+\pi^-, K_L \to 3\pi^0)$ background events.
The further analysis was based on kinematical fit, testing of the signal and background
hypotheses and exploiting of the differences in kinematics of the $K_S$ decays into $2\pi^0$ and
$3\pi^0$ states.\\
As a result of the conducted analysis no events corresponding to the $K_S \to 3\pi^0$ decay
have been identified. Hence, we have obtained the upper limit on the
$K_S \to 3\pi^0$ branching ratio at the 90$\%$ confidence level:
\begin{equation}
BR(K_S \to 3\pi^0) \leq 2.7\cdot 10^{-8}~.
\label{eq:brvalu}
\end{equation}
This value is almost five times lower than the latest published result.
This upper limit can be translated into a limit on the $|\eta_{000}|$ at the 90$\%$
confidence level:
\begin{equation}
|\eta_{000}| = \left|\frac{A(K_S \to 3\pi^0)}{A(K_L \to 3\pi^0)}\right| = \sqrt{\frac{\tau_L}{\tau_S}
\frac{BR(K_S \to 3\pi^0)}{BR(K_L \to 3\pi^0)}} \leq 0.009~,
\label{eq:etavalu}
\end{equation}
which corresponds to an improvement of the $|\eta_{000}|$ uncertainty by a factor of two with
respect to the latest direct measurement~\cite{Matteo}.\\
However, the upper limit on the $K_S \to 3\pi^0$ branching ratio determined in this work is still
about one order of magnitude larger than the prediction based on the Standard Model. Thus, the picture
of $\mathcal{CP}$ symmetry violation in the neutral kaon system remains incomplete.
Therefore, among several other experiments aiming in the precise measurements of rare and ultra -- rare kaon
decays~\cite{Sozzi:2011yt}, we are continuing the research of
the $K_S \to 3\pi^0$ decay by means of the KLOE--2 detector. The KLOE--2 collaboration is
continuing the physics program of its predecessor. 
In the last years, a new machine scheme based on the Crab -- waist optics and a
large Piwinsky angle~\cite{Zobov} has been proposed and tested to improve the DA$\Phi$NE luminosity.
The test has been successful and presently DA$\Phi$NE can reach a peak luminosity of a factor of three
larger than previously obtained.
The next data taking campaign during 2013 -- 2015 will be conducted with a goal to collect total
integrated luminosity amounting to about 20~fb$^{-1}$, which corresponds to one order of magnitude
higher statistics with respect to what was used in this search.
For the forthcoming run the KLOE performance has been improved by adding new
subdetector systems: the tagger system for studies of the meson production in the $\gamma\gamma$
reactions, the Inner Tracker
based on the Cylindrical GEM technology and two calorimeters in the final focusing 
region~\cite{BranchiniIT,BranchiniCC}. These new calorimeters will increase the acceptance of the
detector, while the new inner detector for the determination of the $K_S$ vertex will significantly
reduce the contribution of the background processes involving charged particles.
Increasing the statistics and acceptance of the detector while significantly reducing the background gives
the realistic chances to observe the $K_S \to 3\pi^0$ decay for the first time in the near future.
\cleardoublepage
\thispagestyle{plain}
\thispagestyle{plain}
\cleardoublepage
\pdfbookmark[-1]{Last but not least}{}
\phantomsection
\addcontentsline{toc}{chapter}{Acknowledgments}
\pagestyle{empty}
\begin{center}
{\bf Acknowledgements}
\end{center}
\bigskip
I would like to express my highest gratitude to all the people without
whom this thesis would not come into being.\\
First of all, I am deeply indebted to my supervisor, prof. Pawe{\l} Moskal,
for his invaluable help during my work in his research group. I am very
grateful for his patience in correcting all my mistakes and for the time
that he spent correcting this thesis ( often devoting his free time).
I admire his knowledge and attitude towards young people who want to be
scientists.\\
I am very grateful to dr. Stefano Miscetti for being my KLOE supervisor,
for his assistance, guidance and patient correction  of my mistakes
during the data analysis.\\

I would like to express my appreciation also to:
\begin{itemize}
	\item dr. Fabio Bossi and dr. Caterina Bloise
for making it possible for me to work in the KLOE experiment and to visit Frascati;
\item Prof. Bogus{\l}aw Kamys for all suggestions concerning this work  and support during
my PhD studies;
\item Prof. Lucjan Jarczyk for all his comments and questions during my PhD seminars;
\item dr. Matteo Martini and dr. Matteo Palutan for all very useful
fortran code and an introduction to the KLOE data analysis;
\item all my KLOE -- 2 colleagues, especially to: dr. Erika De Lucia, dr. Gianfranco Morello,
dr. Antonio De Santis and dr. Salvatore Fiore for all their help and valuable (not only)
scientific discussions,for introducing me to the Italian cuisine (especially porchetta in  Ariccia);
\item dr. Eryk Czerwi{\'n}ski, Jaros{\l}aw Zdebik and Izabela Pytko for the nice and stimulating
atmosphere of work, and all the time we spent together discovering Italy;
\item Marcin Zieli{\'n}ki for his help in sending grant applications and all the valuable informations
concerning PhD studies;
\item all my colleagues from the 03A room: Magdalena Skurzok, Tomasz Bednarski, Tomasz Twar{\'o}g,
Szymon Nied{\'z}wiecki, Wiktor Bardan, Wiktor Parol, Kacper Topolnicki, Krzysztof Kacprzak and
Andrzej Pyszniak for a nice atmosphere of work;\\

Cha{\l}bym r{\'o}wnie{\.z} serdecznie podzi\k{e}kowa{\'c} swoim Rodzicom za to wszysko cze\-go nauczy{\l}em
si\k{e} od nich i dzi\k{e}ki nim, za ich nieustaj\k{a}ce wsparcie i mi{\l}o{\'s}{\'c}. Dzi\k{e}\-ku\-j\k{e}
mojej siostrze, kt{\'o}ra zawsze by{\l}a przy mnie i pomaga{\l}a w trudnych chwilach. Jestem r{\'o}wnie{\.z}
wdzi\k{e}czny reszcie mojej Rodziny za wsparcie i wszel\-k\k{a} pomoc w~czasie studi{\'o}w doktoranckich.
Na koniec, chcia{\l}bym podzi\k{e}kowa{\'c} Kasi Grzesik, kt{\'o}rej cierpliwo{\'s}{\'c}, nieocenione wsparcie i
mi{\l}o{\'s}{\'c} do\-da\-wa\-{\l}y mi si{\l} w czasie prowadzenia bada{\'n} i pisania tej pracy.
\end{itemize}

\pagestyle{fancy}
\backmatter
\phantomsection
\addcontentsline{toc}{chapter}{Bibliography}

\end{document}